\documentclass[prd,aps,tightenlines,floatfix,nofootinbib]{revtex4}
\usepackage{epsfig,graphicx}

\newcommand{\dE}{\dot{E}}
\newcommand{\dL}{\dot{L}_{z}}
\newcommand{\dQ}{\dot{Q}}
\newcommand{\ddp}{\dot{p}}
\newcommand{\de}{\dot{e}}
\newcommand{\di}{\dot{\iota}}
\newcommand{\be}{\begin{equation}}
\newcommand{\ee}{\end{equation}}
\newcommand{\bear}{\begin{eqnarray}}
\newcommand{\eear}{\end{eqnarray}}

\begin{document}

\title{Extreme Mass Ratio Inspirals: LISA's unique probe of black hole gravity.}

\author{Kostas Glampedakis} 
\affiliation{School of Mathematics, University of Southampton, Southampton 
SO17 1BJ, UK}

\date{\today}

\begin{abstract}

In this review article I attempt to summarise past and present-ongoing-work on the problem of
the inspiral of a small body in the gravitational field of a much more massive Kerr black hole. 
Such extreme mass ratio systems, expected to occur in galactic nuclei, will constitute prime sources of 
gravitational radiation for the future LISA gravitational radiation detector. The article's main goal is to provide a survey of 
basic celestial mechanics in Kerr spacetime and calculations of gravitational waveforms and backreaction on 
the small body's orbital motion, based on the traditional `flux-balance' method and the Teukolsky black hole 
perturbation formalism.

\end{abstract}
\maketitle


\section{Introduction}
\label{intro}

The launch and subsequent operational deployment of the LISA space-based gravitational wave detector \cite{LISA},\cite{webpage}
is an event eagerly awaited by the relativistic astrophysics community. Unlike the various ground-based 
detectors (both existent and scheduled for the near future) which will be capable of `scoring' only marginal detections 
of gravitational radiation sources, LISA is designed as a prime instrument with the expectation to 
provide us with unprecedented data on strong-gravity spacetime dynamics. 

Mounting astronomical observations point to the existence of massive `dark objects' in the majority of galactic nuclei 
(see Ref.~\cite{MBHs_review} for an excellent review). Moreover, accretion disc models crucially depend on the existence 
of an event-horizon in order to explain the `dimness' of galactic nuclei such as our own Milky Way \cite{adaf}. It is
universally accepted that in reality these objects are supermassive Kerr black holes. 

Such black holes would strongly interact with their surrounding `cusp' stellar population, and occasionally capture
small-mass $\mu \sim 1-10~M_{\odot}$ compact bodies (black holes, neutron stars and white dwarfs) that happened to
scatter towards their vicinity \cite{rates2}. Following the capture, the small body will suffer loss of energy and angular 
momentum by emitting gravitational radiation and slowly inspiral until its final plunge into the black hole. The dynamics of such 
extreme-mass-ratio-inspiral (EMRI) is (almost) purely gravitational, accurately modelled as a `test-particle' moving
in a Kerr spacetime. Due to the extreme mass ratio (the massive black hole should have a mass $M \sim 10^6 M_\odot $
in order for the emitted signal to fall in LISA's bandwidth), the evolution of the system is adiabatic; for many orbital periods, 
the small body moves in an almost geodesic trajectory of the background spacetime. Radiation reaction effects become noticeable at
much longer timescales ($ \sim M^2/\mu  $) and the precise calculation of this motion is one of the main current 
research objectives.    

EMRIs, being relatively simple systems, will perhaps prove to be among the `cleanest' sources for LISA, making them unique probes for
studying black hole spacetimes and, indeed, proving without any dispute the Kerr identity of these massive objects. 
Estimates for the event rate of these inspirals \cite{rates2},\cite{rates} are still uncertain, suggesting that LISA could detect a number
of events ranging from few, to hundreds or even over a thousand, during a $3-5$ years mission. The detection of numerous events will 
potentially improve our understanding of massive black holes demography and population synthesis \cite{demography}. For those rare 
cases where the black hole's environment is `dirty' due to pronounced gas accretion (quasars, active galactic nuclei), an EMRI signal 
would provide valuable  information on the structure of the disc itself (see Refs.~\cite{chakrabarty},\cite{drag}).         
Due to their nature, EMRI systems can be accurately studied using well-established black hole perturbation theory \cite{chandra},
and after a substantial  amount of work over the last decade or so, they can now be considered among the most well-understood sources 
of gravitational radiation. 

It is the purpose of this article (as part of a special Volume dedicated to gravitational radiation reaction) to review the main body 
of this work, in particular the computation of gravitational waveforms and the backreaction to the small body's geodesic motion. 
Deliberately, we only discuss results obtained by the `flux-balance' method: the amounts of energy and angular momentum radiated away 
at infinity and the hole's horizon equal the amounts removed from the orbit. Unfortunately, this prescription only works for special 
classes of Kerr orbits: equatorial or circular-inclined. The orbits expected to occur in reality  have no reason to belong to any of 
these cases. On the contrary, they are expected to be generic: with non-zero eccentricity and inclination. For these orbits, only two
(energy and angular momentum along the black hole's spin axis) out of the three orbital `constants' can be evolved
using flux-balance. The third one, the so-called Carter constant, requires knowledge of the local gravitational self-force.
The description of the inspiral in terms of this force involves a more sophisticated machinery than the simple flux-balance
method, but at the end of the day, this higher precision will be required for constructing waveform templates for LISA. 
The self-force formalism is thoroughly reviewed in other contributions of the present Volume.      
In the meantime, flux-balance calculations provide rigorous results for non-generic orbits, while certain approximations
are still possible for generic orbits. In the future they will also serve as benchmarks for self-force results. 
We should warn the reader that due to space limitation and this Volume's particular subject, we do not discuss any data-analysis 
related issues. This omission should by no means reflect on the importance of data analysis for the LISA project; 
on the contrary, we believe that this crucial component deserves its own review exposition in the near future.  

The remaining of this article is organised as follows. In Sections~\ref{geod} \& \ref{HJ} we summarise previous well-known, 
as well as  more recent material on Kerr geodesic motion. Next, in Section~\ref{EMRI} we review the Teukolsky-Sasaki-Nakamura 
formalism on which all rigorous EMRI calculations are based. That Section includes presentation of frequency-domain results 
(equatorial-eccentric and circular-inclined orbits) as well as (more recent) time-domain calculations. Section~\ref{hybrid}
is devoted to the `hybrid' method, an approximate scheme for generating generic inspirals and waveforms. The topic of 
Section~\ref{mapping} is somewhat unorthodox as it addresses the issue of possible non-Kerr identity of the central massive
body. A direct astrophysical application of EMRI calculations -- estimates of `recoil' velocities for merging massive
black hole binaries -- is presented in Section~\ref{recoils}. Our concluding discussion, with a brief summary and possible 
directions for future work can be found in Section~\ref{conclusions}. A number of Appendices can be found at the end of the
article. Throughout the article we use geometric units $G=c=1$.


\section{Geodesic motion in the Kerr field}
\label{geod}

Not surprisingly, geodesic motion in Kerr spacetime is a well-studied subject since 
the discovery of the Kerr metric itself; an exhaustive discussion can be found
in Chandrasekhar's classic textbook \cite{chandra}. Adopting a Boyer-Lindquist coordinate frame, 
the well known equations of motion are (with $\lambda $ denoting the affine parameter, related 
to proper time as $\lambda = \tau/\mu  $),
\bear
\Sigma\, \frac{dr}{d\lambda} &=& \pm R^{1/2}, \qquad R = T^2 -\Delta\,[\mu^2\,r^2 +(L_z -a E)^2 +
Q], \quad T = E\,(r^2+a^2) -L_z\,a
\nonumber \\
\nonumber \\
\Sigma \, \frac{d\theta}{d\lambda} &=& \pm \Theta^{1/2}, \qquad
\Theta = Q -\cos^2\theta\, [ a^2 (\mu^2 -E^2) + L_z^2/\sin^2\theta ]
\nonumber \\
\nonumber \\
\Sigma\, \frac{d\phi}{d\lambda} &=& -\left (a\, E -\frac{L_z}{\sin^2\theta} \right )
+ \frac{a T}{\Delta}
\nonumber \\
\nonumber\\
\Sigma \frac{dt}{d\lambda} &=& -a\, (aE \sin^2\theta -L_z) + (r^2+a^2)\, T/\Delta
\label{geod1}
\eear  
where $ z^a_{\rm geod} = [\,t(\tau),r(\tau),\theta(\tau),\phi(\tau)\,] $ is the test-body's 
worldline and $\Delta = r^2 -2Mr + a^2,~ \Sigma = r^2 + a^2\,\cos^2\theta   $. The energy $E$,
angular momentum along the rotation axis $L_z$, and the so-called Carter constant $Q$ constitute
the integrals of motion. They naturally emerge as separation constants in the solution of
the Hamilton-Jacobi equation in Kerr spacetime (see Section~\ref{HJ}).

Generic Kerr geodesics can be parameterised by a triplet of constant
orbital elements: the semi-latus rectum $p$, the eccentricity $e$, and
the inclination angle $\iota$. For weak-field orbits ($p \gg M$) these elements reduce to 
familiar Keplerian notions. The elements $p$ and $e$ define the
orbit's radial turning points, the apastron and periastron:
\be
r_a= \frac{p}{1-e}, \qquad r_p= \frac{p}{1+e}\;.
\label{turn_points}
\ee
The radial orbital period $T_r$ is defined as the time required for the body to move from apastron
to periastron and back to periastron.

In the strong field of a Kerr black hole, there are many ways that one
could define an `inclination angle' --- for example, the turning
points of the orbit's latitudinal motion, or the angle at which the
small body crosses the equator as seen by distant observers.  We use
the following definition:
\be
Q = L_z^2\, \tan^2\iota
\label{ioteq}
\ee
This definition does not correspond to either of these examples, but is very convenient: it 
depends simply on orbital constants and has a useful intuitive description, suggesting that 
the Carter constant $Q$ is essentially just the square of the angular momentum projected into
the equatorial plane (this description is in fact exactly correct for Schwarzschild black holes; 
for non-zero spin it is not quite correct, but is good enough to be useful.  We discuss this issue in 
more detail in Appendix {\ref{app:almostsphere}}). The orbital elements can be written as functions of 
$ \{E, L_z, Q\}$, and vice versa. These functions are obtained by solving for the radial motion's 
turning points, $ R(r) = 0$. This equation is in general a quartic polynomial of $r$. Only the two largest 
roots are relevant for bound stable orbits, and they correspond to $r_p,r_a$. When $\iota=0$ one of the 
roots is trivially zero and we are left with a cubic equation. 

The actual turning points of the body's latitudinal motion are found by solving 
$\Theta(\theta) =0$, which becomes,
\be
\beta\, z^2  -\left (\beta +\frac{1}{\cos^2\iota} \right )\, z + \tan^2\iota = 0
\ee 
where $ z = \cos^2\theta $ and $\beta = a^2\,(\mu^2-E^2)/L_z^2 $. From the two roots $z_{-} < z_{+} $
the relevant one is the smallest and leads to the turning points $\theta_n$ and $\theta_s = \pi -\theta_n $ 
which are related to $\iota$ as,
\be
\theta_n = \iota -\frac{\pi}{2} + {\cal O}(a^2) 
\ee
The latitudinal period $T_\theta $ is defined as the time required for the body to travel from $\theta_n$ 
to $\theta_s$ and back to $\theta_n$.

\begin{figure}
\centerline{\includegraphics[height=7cm,clip]{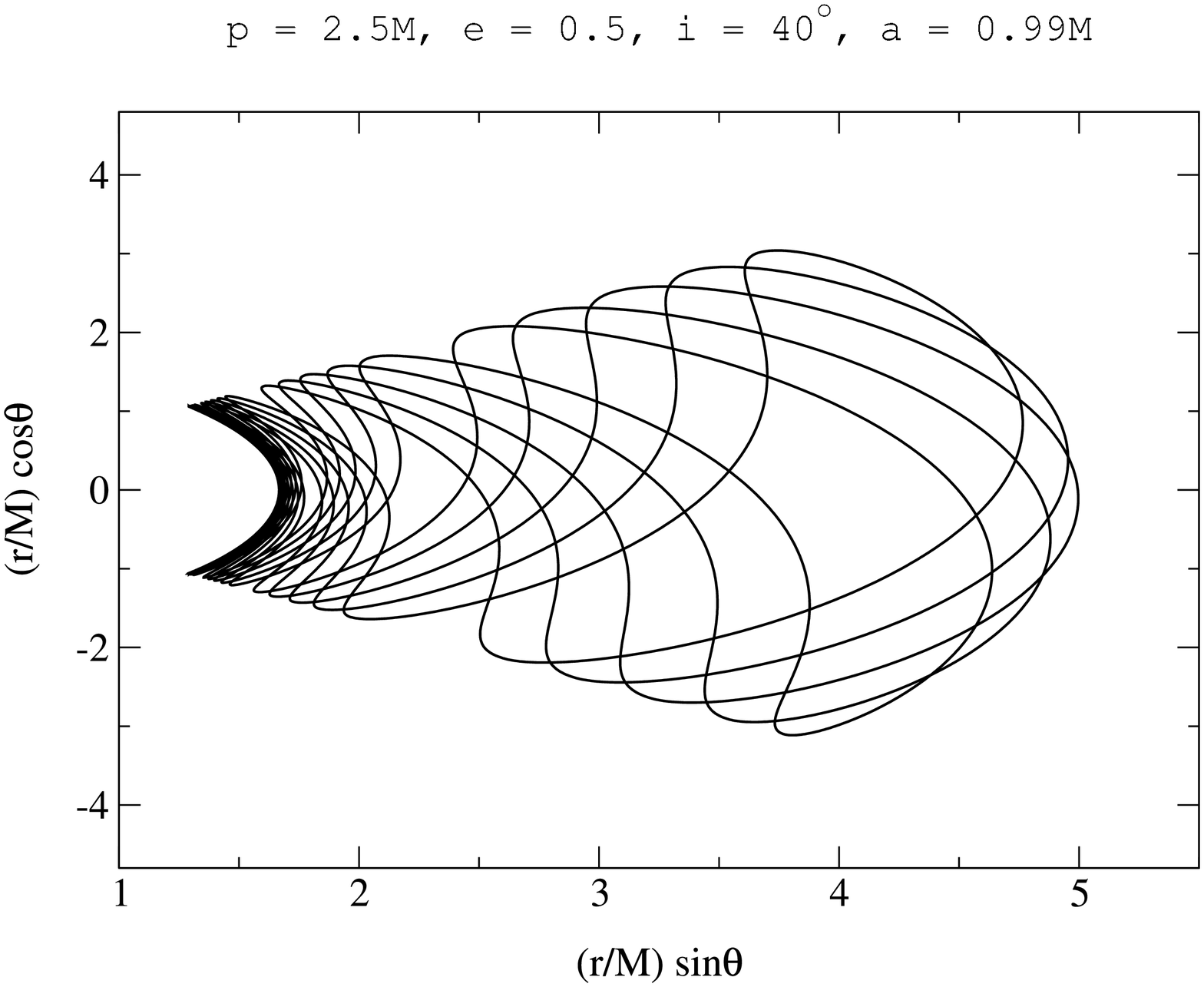}}
\caption{Generic Kerr orbit with $ p = 2.5M,~e=0.5,~\iota = 40^\circ $ around a rapidly spinning hole $a=0.99M$. The Figure displays a
Boyer-Lindquist $r-\theta $ coordinate slice of the orbit with the $\phi$-motion suppressed.}
\label{fig_orbit}
\end{figure}

Loosely speaking, Kerr bound orbits look similar to Keplerian ellipses especially when $r_p \gtrsim 15-20 M $.
The main qualitative orbital features originate from the incommensurate nature of the three fundamental periods 
$T_r, T_\theta,T_\phi $. In the weak-field Keplerian limit $p \gg M $ these periods almost coincide leading to 
familiar elliptical orbits. At closer distances $ T_r - T_\phi  $ grows resulting in the famous periastron advance. 
In addition, provided that $a \neq 0 $, we also  have $ T_\theta \neq T_\phi $ which is responsible for the well known 
Lense-Thirring precession of the orbital plane. Both these effects become important in the strong-field region and any 
resemblance to a quasi-elliptical orbit is lost. An example of a strong-field generic orbit is given in Fig.~\ref{fig_orbit}

\begin{figure}
\centerline{\includegraphics[height=6cm,clip]{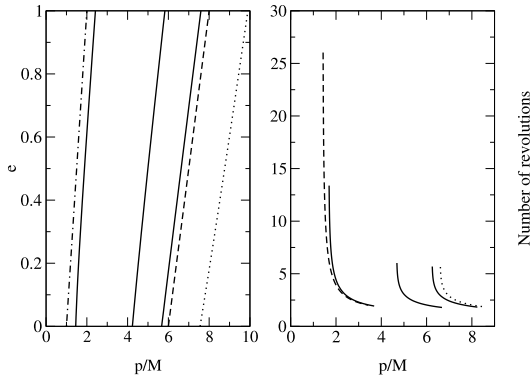}}
\caption{Left panel: Kerr equatorial separatrices, (from right to left): $a= 0.5 M$ (retrograde, dotted curve), $a=0 $
(dashed curve), $a= 0.1 M,~0.5 M,~ 0.99 M $ and $ a= M $ (dotted-dashed curve). For inclined orbits the
behaviour of the separatrix is qualitatively the same. Right panel: number of revolutions per one radial period
$T_r$ for zoom-whirl orbits with $ e =0.3 $ and black hole spin (from right to left), $a=0$ (dotted curve), 
$a = 0.1 M, 0.5 M, 0.99 M $ and $a= 0.999 M $ (dashed curve). Each of these curves terminate at a fixed distance 
$\delta p = 0.01 M$ from the respective separatrix.}
\label{fig_sepax}
\end{figure}

A point of the parameter space $\{p,e,\iota \}$ does not necessarily correspond to a stable bound orbit.
A separatrix surface $ F_{\rm s}(p,e,\iota) = 0$ delimits stable from unstable (plunging) orbits. 
This surface is determined by requiring a double root for $R(r) = 0$, specifically $r_2 = r_p $ 
where $r_2$ is the third largest root. For the case of Schwarzschild orbits the separatrix takes the 
simple form \cite{cutler},
\be
F_{\rm s} = p - M(6 + 2e) = 0 
\label{sepax_s}
\ee
The limits $e=0$ (circular orbits) and $e=1$ (parabolic orbits) correspond to the familiar values
$r_p = 6M$ and $r_p = 4M$ for marginally stable and marginally bound orbits \cite{chandra},\cite{bardeen}. 
For Kerr it is not generally possible to write down a simple expression as the functions $E(p,e,\iota)$ etc.
are quite cumbersome. For example, for equatorial orbits we obtain from $r_2 = r_p$ \cite{kgdk},
\be
F_{\rm s} = p^2 - (L_z -a\,E)^2\,(1+e)\,(3-e) = 0
\label{sepax_k}
\ee 
but this is not as informative as (\ref{sepax_s}).  For the special case of prograde orbits around an 
extreme Kerr black hole, eqn.~(\ref{sepax_k}) simplifies to $F_{\rm s} = p -M(1+e) $, which gives $r_p = M$ 
for all eccentricities. For all other cases,  the separatrix is calculated numerically, see relevant plots 
in \cite{kgdk}, \cite{scott_circ} and \cite{wolfram}. In Fig.~\ref{fig_sepax}, we illustrate the location 
of the separatrix for equatorial Kerr orbits as a function of the black hole spin. When plotted on the 
$p-e$ plane, the most prominent feature of the separatrix is that its location shifts towards smaller
(larger) $p$ with increasing black hole spin for prograde (retrograde) orbits.

Of particular importance are orbits with $p \approx p_s(e,\iota)$ (where $p_s$ solves $F_s = 0$ ). These
are called `zoom-whirl' orbits, heuristically named after their particular properties \cite{kgdk}. A body 
in a zoom-whirl orbit will spend a considerable amount of its orbital `life'  close to the periastron. 
An approximation for $T_r$, as $p \to p_s$, gives \cite{cutler},\cite{kgdk},
\be
T_{\rm r} \sim \ln(p-p_s) 
\label{nearTr}
\ee
which shows that the radial period will grow (and eventually diverge) as the separatrix is approached. 
In that region, the particle will trace a quasi-circular path before being reflected back to the apastron. 
Such behaviour will be particularly prominent for high eccentricity orbits: the particle will `zoom in' from 
its apastron position, and perform a certain number of quasi-circular revolutions (`whirls') reaching the 
periastron (which should have a value close to $p_s(e)/(1+e) $). Finally, the particle will be reflected 
and `zoom out' towards the apastron again. Zoom-whirl orbits resemble a set of orbits known in the literature 
as homoclinic orbits \cite{homo}. They occur in both Kerr and Schwarzschild geometries, and
their potential significance for the detection of gravitational waves by space-based
instruments was first pointed out some years ago by Cutler, Kennefick \& Poisson \cite{cutler} 
who concluded that the small number of whirls in the Schwarzschild case made the phenomenon less interesting 
for non-spinning black holes. The zoom-whirl behaviour is much more pronounced for prograde orbits around 
rapidly spinning holes, as illustrated in Fig.~\ref{fig_sepax}

The number of revolutions plotted in Fig.~\ref{fig_sepax} is defined as $N_r = \Delta\phi/2\pi$, where 
$\Delta\phi = \phi(t+T_r) - \phi(t) $, the accumulated azimuthal periastron advance during one complete 
radial period. Using (\ref{nearTr}), we find that $N_r$ exhibits the same logarithmic divergence as $p\to p_s$. 
According to Fig.~\ref{fig_sepax}, for small and moderate spins $N_r$ stays close to the corresponding
Schwarzschild  value, but it grows rapidly as $a \to M$, basically due to the intense 
`frame-dragging' induced by the black hole's rotation in the very strong field region close to the horizon 
which can be reached by bodies in prograde orbits. Note that the spin dependence enters (\ref{nearTr}) 
in the proportionality coefficient which we omitted. The overall behaviour can be understood as an extreme 
example of perihelion advance (as in the celebrated case of the planet Mercury).

In principle, as (\ref{nearTr}) suggests, the number of revolutions can be made arbitrarily large irrespective 
of the black hole spin, provided the body approaches sufficiently close to the separatrix. In practice however,
this limit is immaterial as sufficiently close to the separatrix, radiation reaction makes a significant correction 
to the body's motion in each orbital period, and therefore we can no longer speak of slow, adiabatic, orbital evolution. 
The motion during the stage of separatrix crossing has been discussed in detail in Refs.~\cite{transition1},
\cite{transition2}.

In a realistic scenario, we should not expect to find (apart from chance cases where the particle 
enters a near-separatrix orbit as a result of its initial scattering) very high eccentricity zoom-whirl orbits, 
as it is well known that eccentricity tends to decrease under radiation reaction, for almost the entire
inspiral. However, despite this decrease, a substantial amount of eccentricity will survive, in many cases, 
up to the point where the orbit is about to plunge (the evolution of $e$ during the inspiral 
is discussed in more detail later in this article). These orbits will probably become zoom-whirl orbits, 
especially when a rapidly spinning black hole is involved and the motion is prograde. 
Not surprisingly, the zoom-whirl behaviour will appear also when the orbit is non-equatorial (a good example is the
orbit of Fig.~\ref{fig_orbit}). Close to the separatrix we have $T_r \gg T_\theta, T_\phi $ and the body will spend 
a significant amount of time moving in a nearly circular-inclined fashion at $ r \approx r_p$.


\subsection{Hamilton-Jacobi theory in Kerr spacetime}
\label{HJ}

Despite our deep understanding of Kerr geodesic motion, certain aspects of it were 
studied only recently. For the most interesting case of generic orbits, an important issue concerns 
their periodicity properties. From this point of view, eqns.~(\ref{geod1}) are somewhat misleading as they
mix the $r$ and $\theta$ motions. As a consequence, it is not obvious how one could calculate orbital periods
$ T_r, T_\theta, T_\phi  $ from these equations. However, as we are about to describe, the full separability of the
Hamilton-Jacobi equation suggest that it is, after all, possible to decouple the $t,r,\theta,\phi $ motions by
means of an action-angle canonical representation \cite{goldstein},\cite{wolfram}. In this way, the complete periodicity of
Kerr motion is unveiled and rigorous fundamental orbital periods can be computed.     

The starting point is the `super-Hamiltonian' for a test-body \cite{MTW},
\be
{\cal H}(x^a,p_b) = \frac{1}{2}\, g^{\mu\nu}\, p_{\mu}\,p_{\nu} = -\frac{1}{2}\,\mu^2
\label{super}
\ee
where $ p^a =  dx^a/d\lambda $ is the four-momentum. Generating a canonical transformation 
$\{ x^a, p_b  \} \Rightarrow \{ \beta^a, \gamma_b\}$ with the requirement that the new
Hamiltonian vanishes (and as a consequence the new coordinates and momenta are constants) 
leads to the Hamilton-Jacobi equation,
\be
g^{\mu\nu}\,\frac{\partial S}{\partial x^\mu}\,\frac{\partial S}{\partial x^\mu} = -\mu^2
\ee
For a restricted class of coordinate systems (Boyer-Lindquist among them) this equation `miraculously' 
admits full separation of variables (see \cite{carter} for detailed analysis of this issue). The solution takes the 
form,
\be
S = \frac{1}{2} \lambda -E\,t +L_z\, \phi + S_r(r) + S_\theta (\theta)
\ee
with 
\be
 S_r(r) = \int dr\,\frac{\sqrt{R}}{\Delta} \quad \mbox{and} 
\quad S_\theta(\theta) = \int d\theta\, \sqrt{\Theta} 
\ee
Although the separation of the $t$ and $\phi$ coordinates is ensured by the stationary
and axisymmetric nature of the Kerr spacetime, the separation of the remaining $r,\theta$
coordinates is not related to any obvious symmetry\footnote{Unlike $E$ and $L_z$ which are related to Killing vectors 
$t^a$, $\phi^a$ associated with stationarity and axisymmetry (with $ E = -p_a\,t^a $, $ L_z = p_a\,\phi^a  $), the
Carter constant is related with a rank-2 Killing tensor, $ Q = K_{ab}\,p^a\,p^b $ \cite{Wald}.}.
The Carter constant $Q$ is related to the original $r-\theta$ separation constant ${\cal C}$ as $ Q = {\cal C}^2 -L_z^2 -a^2 E^2 $.  

The equations of motion follow from 
\be
p^a = g^{ab}\,\frac{\partial S}{\partial x^b} \quad \Rightarrow
\left \{\begin{array}{llll}
& p^t &=& -g^{tt}\, E + g^{t\phi}\, L_z 
\nonumber \\ \nonumber \\
& p^r &=& \pm g^{rr}\,\Delta^{-1}\,\sqrt{R}  
\nonumber \\ \nonumber \\
& p^{\theta} &=& \pm\, g^{\theta\theta}\, \sqrt{\Theta} 
\nonumber \\ \nonumber \\
& p^{\phi} &=& g^{\phi\phi}\, L_z -E\, g^{t\phi} 
\end{array} \right.
\label{geod2}
\ee
These are just the geodesic equations (\ref{geod1}). Similarly, the covariant 
momentum components are: 
$ p_t = -E,~ p_r = \pm \sqrt{R}/\Delta,~p_{\theta} = \pm \sqrt{\Theta},~p_{\phi} = L_z $.

The true periodic nature of geodesic motion in Kerr is revealed when we introduce a set of action-angle variables 
\cite{wolfram} (see \cite{goldstein} for background material). The actions are defined as 
(hereafter $i = r,\theta,\phi$)
\be
J_i = \oint \, p_a\, dx^a \Rightarrow \left \{\begin{array}{lll}
      & J_r = 2\,\int_{r_p}^{r_a}\, dr\, \sqrt{R}/\Delta  \nonumber \\ \nonumber \\
      & J_{\theta} = 2\, \int_{\theta_n}^{\theta_s}\, d\theta\, \sqrt{\Theta} \nonumber \\
        \nonumber \\
      & J_{\phi} = 2\,\pi\,L_z 
\end{array} \right.
\label{actions}
\ee
For the full four-momentum vector we can choose,
\be
\gamma_a = ( -E, J_i) 
\ee
which is to be considered as a function $\gamma_a =f_a(-\mu^2/2,E,L_z,Q)$. 
A triplet of fundamental frequencies can be defined as,
\be
\nu_i = \frac{\partial {\cal H}}{\partial J_i}
\label{freqs}
\ee
These are associated with motion with respect to $r,\theta,\phi$ coordinates.
In (\ref{freqs}) the Hamiltonian is assumed as a function of the canonical parameters 
$\{\beta^c, \gamma_a\}$. In addition there is a fourth constant, associated with the $t$ coordinate,
\be
\nu_t = -\frac{\partial{\cal H}}{\partial E}
\ee

\begin{figure}
\centerline{\includegraphics[height=6.5cm,clip]{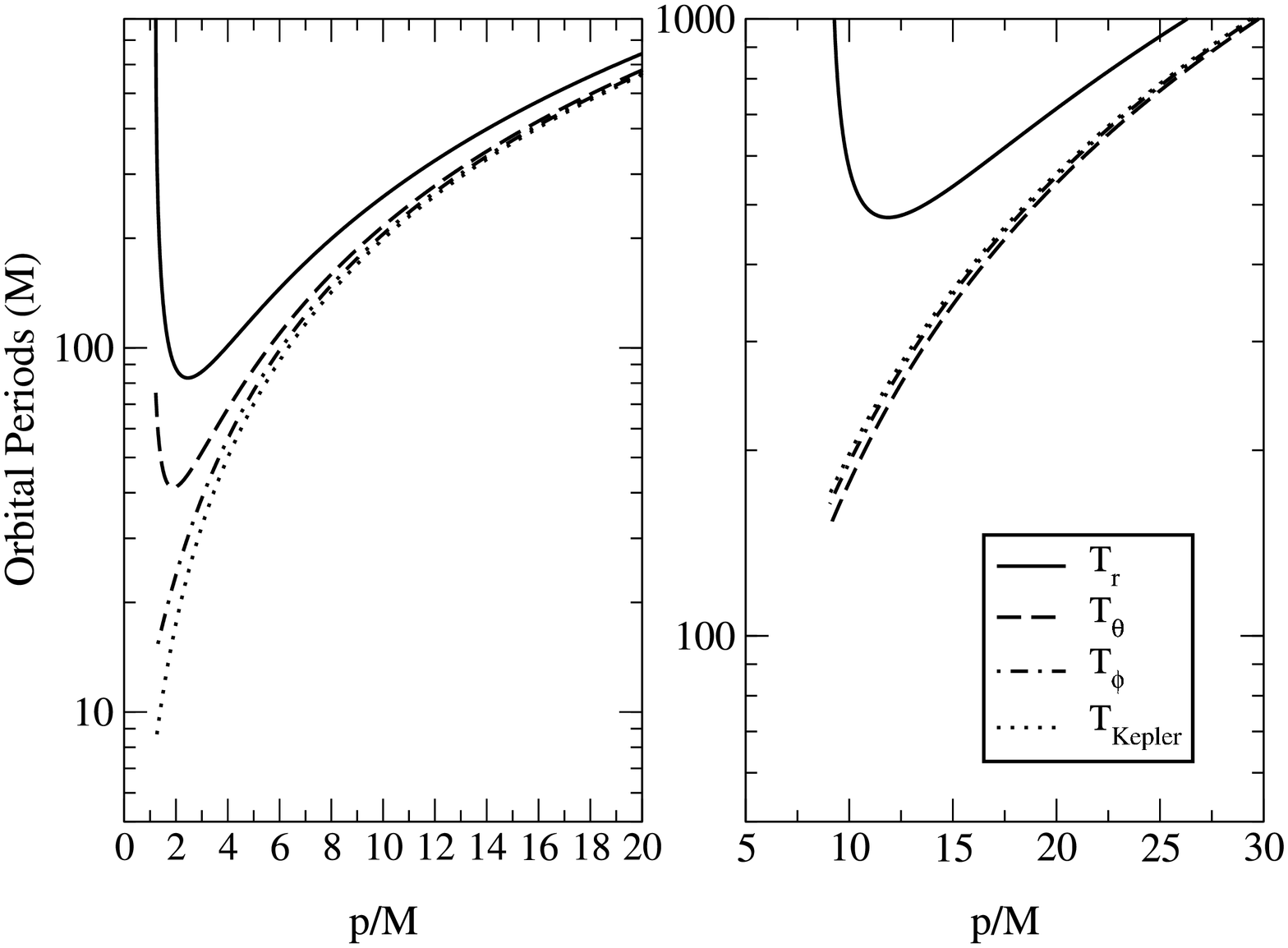}}
\caption{Fundamental Kerr periods $T_r, T_\theta, T_\phi $ for small eccentricity, small inclination orbits. The black hole spin is
$a=0.999M $ and both prograde (left panel) and retrograde orbits (right panel) are shown. Near each corresponding separatrix $T_r \to \infty $
(zoom-whirl behaviour) while $T_\theta, T_\phi  $ retain a finite value. Only for the special case where $ r_p \approx r_{+} $ 
(as in this Figure, left panel), the $\theta$-motion is `locked' by the presence of the event horizon and $T_\theta \to \infty $ too. 
For $ p \gg M$ all periods converge to the usual Keplerian period. The results shown here are typical even when the assumptions 
of small $e$, $\iota$ are dropped \cite{wolfram}.} 
\label{fig_periods}
\end{figure}

Unlike the familiar action-angle treatment of the Kepler problem \cite{goldstein}, it is not
possible to integrate and subsequently invert eqns.~(\ref{actions}) to obtain $ {\cal H}(\gamma_a)$.
Nevertheless, as shown by Schmidt \cite{wolfram}, it is still possible to obtain the desired 
derivatives $ \partial {\cal H}/\partial \gamma_a $ by employing a theorem on implicit functions
for the function $f_a$. If $D[f]$ is the Jacobian matrix of $f$ then the theorem states (provided that
$\det D[f] \neq 0$) that $ D[f]\cdot D[f^{-1}] = D[f]\cdot (D[f])^{-1} = I$. Using the fact that
${\cal H}(-E,J_i) = -\mu^2/2 $, we can then obtain
$ \partial {\cal H}/\partial E, \partial {\cal H}/\partial J_i  $ as well as
$\partial  Q/\partial E, \partial Q/\partial J_i  $. Then, the fundamental frequencies
(\ref{freqs}) are found to be,
\be
\nu_r = \frac{K(k)}{2\Lambda}, \qquad \nu_\theta = \frac{a(\mu^2 -E^2)^{1/2}\,z_{+}\,X(r_p,r_a)}{4\Lambda},
\qquad
\nu_\phi = \frac{K(k)\,Z(r_p,r_a) +L_z\,[\Pi(z_{-}^2,k) -K(k)]\,X(r_p,r_a)}{2\pi\Lambda}
\label{freqs_i}
\ee
For the fourth frequency we find,
\be
\nu_t = -\frac{\partial {\cal H}}{\partial E} = \frac{K(k)\,W(r_p,r_a) + a^2\,z_{+}^2\,E\,
[ K(k) -E(k)]\,X(r_p,r_a)}{\Lambda}
\label{freq_t}
\ee
The various integral quantities appearing in these formulae are listed in Appendix~\ref{app:integrals}.
Note that the frequencies (\ref{freqs_i}) are associated with the test body's proper time. The role of $ \nu_t $ is 
to translate them to frequencies $ \Omega_r, \Omega_\theta, \Omega_\phi$ associated with Boyer-Lindquist
time $t$ \cite{wolfram}. These are simply related as, 
\be
M\Omega_i = \nu_i/\nu_t
\label{periods}
\ee
The above frequencies are simplified by assuming small $e$ and $\iota$. Then,
\bear
\Omega_\phi &=& \frac{\Omega_{\rm K}}{1 \pm a\Omega_{\rm K}}
\\
\Omega_\theta &=& \Omega_\phi\, \left [ 1 \mp 4a\Omega_{\rm K} + 3\frac{a^2}{p^2} \right ]^{1/2}
\\
\Omega_r &=& \Omega_\phi\,\left [ 1 -\frac{6M}{p} \pm 8a\,\Omega_{\rm K} -3\frac{a^2}{p^2}  \right ]^{1/2}
\eear
where $\Omega_{\rm K} = \sqrt{M/p^3} $ is the Keplerian frequency and the upper (lower) sign refers to prograde (retrograde) motion. 
Figure~\ref{fig_periods} shows how the corresponding periods $T_i = 2\pi/\Omega_i$ vary with $p$ for a near extreme hole  $ a=0.999M$ 
for both prograde and retrograde motion. The same qualitative behaviour persists for the general periods (\ref{periods}).

The  (constant) coordinates conjugate to $\gamma_a$ are given by,
\be
\beta_a = \frac{\partial S}{\partial \gamma_a} \quad \Rightarrow \quad \beta_a + \nu_a\,\lambda =
\frac{\partial W}{\partial \gamma_a} = \frac{\partial W}{\partial \alpha_k}\, 
\frac{\partial \alpha_k}{\partial \gamma_a} = w^a
\ee
where $ \alpha_k = (-\mu^2/2,E,L_z,Q)$ and  $ W(r,\theta,\phi,\alpha_k) = S -\lambda/2 $.  
These equations define the Kerr angle variables $w_a$ and provide the frequency interpretation for $\nu_a$. 
Expanding these we obtain
\bear
\beta_t + \nu_t\, \lambda &=& -2\,\nu_t\,\left [ \frac{\partial S_r}{\partial \mu^2}  +
\frac{\partial S_\theta}{\partial \mu^2} \right ] + t -\left [ \frac{\partial S_r}{\partial E}  +
\frac{\partial S_\theta}{\partial E}   \right ]
\nonumber \\
\nonumber \\
\beta_r + \nu_r\, \lambda &=& -2\,\nu_r\,\left [ \frac{\partial S_r}{\partial \mu^2}  +
\frac{\partial S_\theta}{\partial \mu^2} \right ]  +  \frac{\partial Q}{\partial J_r} 
\left [ \frac{\partial S_r}{\partial Q}  + \frac{\partial S_\theta}{\partial Q}   \right ]
\nonumber \\
\nonumber \\
\beta_\theta + \nu_\theta\, \lambda &=& -2\,\nu_\theta\,\left [ \frac{\partial S_r}{\partial \mu^2}  +
\frac{\partial S_\theta}{\partial \mu^2} \right ]  +  \frac{\partial Q}{\partial J_\theta} 
\left [ \frac{\partial S_r}{\partial Q}  + \frac{\partial S_\theta}{\partial Q}   \right ]
\nonumber \\
\nonumber \\
\beta_\phi + \nu_\phi\, \lambda &=& -2\,\nu_\phi\,\left [ \frac{\partial S_r}{\partial \mu^2}  +
\frac{\partial S_\theta}{\partial \mu^2} \right ]  +  \frac{\partial Q}{\partial J_\phi} 
\left [ \frac{\partial S_r}{\partial Q}  + \frac{\partial S_\theta}{\partial Q}   \right ]
+ \frac{1}{2\,\pi}\, \left [ \phi + \frac{\partial S_r}{\partial L_z} + 
\frac{\partial S_\theta}{\partial L_z}   \right ]
\eear
By differentiating and combining these expressions one can arrive to the equations of motion (\ref{geod1}).
For example, from the $w_r, w_\theta$ equations we get,
\bear
r^2\,R^{-1/2}\, \dot{r} + a^2\cos^2\,\Theta^{-1/2}\,\dot{\theta} + \frac{1}{2}\, \kappa_r\,
\left [\Theta^{-1/2}\,\dot{\theta} -R^{-1/2}\,\dot{r}  \right ] &=& 1
\label{wr}
\\
\nonumber \\
r^2\, R^{-1/2}\, \dot{r} + a^2\cos^2\theta\,\Theta^{-1/2}\, \dot{\theta} + \frac{1}{2}\, 
\kappa_\theta\,\left [ \Theta^{-1/2}\,\dot{\theta} -R^{-1/2}\,\dot{\theta} \right ] &=& 1
\label{wth}
\eear 
where the overdot stands for $d/d\lambda$, and $\kappa_i = (\partial Q/\partial J_i)/\nu_i $. 
Combining these equations,
\be
(\kappa_r -\kappa_\theta)\, \left [ \Theta^{-1/2}\,\dot{\theta} - R^{-1/2}\, \dot{r} \right ] = 0
\quad \Rightarrow \quad \Theta^{-1/2}\,\dot{\theta} = R^{-1/2}\,\dot{r} = \sigma(r,\theta)
\ee
The function $\sigma(r,\theta)$ is subsequently specified by substitution in either (\ref{wr}) or (\ref{wth}). 
We get $ \sigma = (r^2 + a^2\,\cos^2\theta )^{-1} =  \Sigma^{-1} $. In this way we have arrived at the expected 
$r,\theta$ equations of motion. In a similar manner, the remaining $t,\phi$ equations of motion follow from 
the $w_t,w_\phi$ equations.


\section{Gravitational radiation from a test-body in Kerr spacetime}
\label{EMRI}


\subsection{Basic timescales}

Before venturing into our discussion of rigorous EMRI results it would be useful to
provide some basic estimates for the timescales involved. The notion of adiabaticity
in orbital evolution plays a central role in EMRI calculations. Quite simply this is a statement
about the relative magnitudes of orbital periods and the radiation reaction timescale $T_{\rm RR} $. 
For the former we can use the Keplerian period, 
\be
T_{\rm orb} \sim \frac{2\pi M}{(1-e^2)^{3/2}}\, \left ( \frac{p}{M} \right )^{3/2}
\ee
which is close to the true periods $T_r,T_\theta,T_\phi $ provided the orbit is not too close
to the separatrix (see Fig.~\ref{fig_periods}). In this latter case, $T_r$ becomes much longer than the other two
and we should choose $ T_{\rm orb} = T_r$.  

For the radiation reaction timescale we can choose among 
$ T_p = p/|\ddp|,~ T_e = e/|\de| $ or $T_{\iota} =  \iota/|\di|  $.  The first two are comparable for $p \gg M $ while the 
third is always much longer (see Section~\ref{hybrid}). As the separatrix is approached $ T_p $ becomes the shortest, and it 
makes sense to choose $T_{\rm RR} = T_p $. From the weak-field expressions of Section~\ref{hybrid} we have,
\be 
T_{\rm RR} \sim T_p = \frac{p}{|\ddp|} \sim \frac{M^2}{\mu}\,\frac{(p/M)^4}{(1-e^2)^{3/2}} 
\ee
We can easily translate these timescales into human units,
\be
T_{\rm orb} \sim 7.6\, \frac{M_6\, p_6^{3/2}}{(1-e^2)^{3/2}} \quad \mbox{min},
\qquad
T_{\rm RR} \sim  106\,\frac{M}{\mu}\, \frac{M_6\,p_6^4}{(1-e^2)^{3/2}} \quad \mbox{min}
\ee
where $M_6 = M/10^6 M_{\odot} $ and $ p_6 = p/6M $. Clearly, $ T_{\rm RR} \gg T_{\rm orb} $ for 
the relevant (for LISA)  mass ratios $ \mu/M \sim 10^{-5} -10^{-6} $. This condition only breaks down 
very close to the separatrix, $ p\to p_{\rm s} $, as $T_r \to \infty $.


\subsection{Frequency domain calculations: The Teukolsky formalism}
\label{Teuk_FD}

For about three decades, the celebrated Teukolsky formalism \cite{teuk} has been 
the prime tool for the calculation of gravitational fluxes and waveforms from test-bodies moving in 
Schwarzschild and Kerr spacetimes. The eponymous equation describes the dynamics of 
linearised radiative perturbative fields in a Kerr background. In particular, instead of dealing 
directly with metric perturbations, the Teukolsky formalism considers perturbations on the Weyl 
curvature scalars \cite{chandra}. The most relevant scalar $\psi_4$, is a result of the projection of the 
Weyl tensor on the null vectors $n^{\alpha}$, $\bar{m}^{\beta}$ which are 
members of a Newman-Penrose tetrad, that is, 
$\psi_4= -C_{\alpha\beta\gamma\delta} n^{\alpha} \bar{m}^{\beta} n^{\gamma} 
\bar{m}^{\delta} $. The feature that makes this formalism so attractive for the EMRI problem
is that the radiative fluxes (at infinity and at the horizon), as well as the two gravitational 
wave polarisations $h_{\rm +}$,$~h_{\rm x}$ can all be extracted from $\psi_4$. In a hypothetical
absence of the Teukolsky equation, one would have to compute these quantities by solving the ten coupled 
metric perturbation equations. It is a great fortune that the Kerr spacetime, the unique spacetime describing
astrophysical black holes, is a Petrov type-${\cal D}$ spacetime, which is the key property for arriving
to decoupled perturbation equations for the Weyl scalars (see Ref.~\cite{stewart} for detailed analysis). 
As the Teukolsky formalism is a well-reviewed subject (see for example \cite{chapter} for a detailed discussion in 
the context of EMRI), we shall only provide a summary and outline some basic results.

The original Teukolsky `master' perturbation equation is \cite{teuk},
\bear 
&&\left [ \frac{(r^2+a^2)}{\Delta} -a^2\,\sin^2\theta \right ]\,\frac{\partial^2\psi}{\partial t^2}
+ \frac{4Mar}{\Delta}\,\frac{\partial^2 \psi}{\partial t\,\partial \phi} + \left [ \frac{a^2}{\Delta} 
- \frac{1}{\sin^2\theta} \right ]\,\frac{\partial^2 \psi}{\partial \phi^2} -\Delta^{-s}\frac{\partial}
{\partial r} \left ( \Delta^{s+1}\,\frac{\partial \psi}{\partial r} \right ) -\frac{1}{\sin\theta}\,
\frac{\partial}{\partial\theta}\left ( \sin\theta\frac{\partial \psi}{\partial \theta} \right )
\nonumber \\
\nonumber \\
&& -2s\left [ \frac{a(r-M)}{\Delta} + \frac{i\cos\theta}{\sin^2\theta}  \right ]\,
\frac{\partial \psi}{\partial \phi} -2s\,\left [ \frac{M(r^2-a^2)}{\Delta} -r -ia\cos\theta \right ]\,
\frac{\partial \psi}{\partial t} + s[ s \cot^2\theta -1 ]\,\psi = 4\pi\,\Sigma\, {\cal T} 
\label{master}
\eear
The relevant value for the spin-parameter is $s = -2$, which makes $\psi = \rho^{-4}\psi_4 $, where     
$ \rho= (r -ia\cos\theta)^{-1}$. The source term ${\cal T} $ is made out of projections of the given energy-momentum
tensor, which for the present problem is the one for a point-particle, 
\be
T^{ab} = \frac{\mu\,u^a\,u^b}{\Sigma\sin\theta\,u^t}\,\delta(r-r(t))\,\delta(\theta-\theta(t))\,\delta(\phi-\phi(t))
\label{stress}
\ee  
along the Newman-Penrose tetrad \cite{teuk} (here  $ u^a = p^a/\mu $). This is the part of the formalism where the small body's motion
enters and is assumed to be {\em known}. Based on the adiabatic character of an EMRI system one is allowed to use the body's
geodesic motion, unless the system is examined over time intervals comparable to the radiation reaction timescale.      

Eqn.~(\ref{master}) is separable in the frequency domain by means 
of a decomposition,
\be
\psi_4 = \rho^{4}\sum_{\ell m}\int\, d\omega\,
R_{\ell m}(r,\omega) {}_{-2}S_{\ell m}^{a\omega}(\theta)\,
e^{-i\omega t+i m \varphi} \quad \mbox{and} \quad 
4\pi\,\Sigma\, {\cal T} = \sum_{\ell m}\int\, d\omega\,
T_{\ell m}(r,\omega) {}_{-2}S_{\ell m}^{a\omega}(\theta)\,
e^{-i\omega t+i m \varphi}  
\label{ps4}
\ee
The function $R_{\ell m}$ satisfies the radial Teukolsky equation,
\be
\Delta^2{d\over dr}\left({1\over \Delta}{dR_{\ell m}\over dr}
\right)
-V_T(r) R_{\ell m}=T_{\ell m} \quad \mbox{with} \quad 
V_T=-{K^2+4i(r-M)K \over \Delta}+8i\omega r+\lambda,
\label{radTeuk}
\ee
where $K=(r^2+a^2)\,\omega-ma$ and $\lambda= E_{\ell m} + a^2\omega^2 -2a m\omega$. The explicit form of 
$T_{\ell m}$ can be found in Ref.~\cite{chapter}. The spin-weighted angular functions $ _{-2}S_{\ell m}^{a\omega}(\theta)$ 
(hereafter denoted simply as $S_{\ell m}$) satisfy the following eigenvalue equation,
\be
\left [ {1 \over \sin\theta}{d \over d\theta}
 \left \{ \sin\theta {d \over d\theta} \right \}
+ a^2\omega^2 \cos^2\theta -\frac{m^2}{\sin^2\theta} + 4a\omega\cos\theta \right.
\left. + \frac{4m\cos\theta}{\sin^2\theta} -4\cot^2\theta -2 + E_{\ell m} \right ] 
S_{\ell m}=0.
\label{slm}
\ee
These are normalised as,
\be
\int_{0}^{\pi}\,d\theta\sin\theta\, | S_{\ell m} |^2 = 2\pi
\ee

A particular solution of equation (\ref{radTeuk}) can be found in terms of two independent solutions 
$R^{\rm H}_{\ell m}$, $R^{\rm \infty}_{\ell m}$ of the homogeneous equation,
\be
R_{\ell m}(r)= \frac{R^{\rm \infty}_{\ell m}(r)}{W}
\int^{r}_{r_{+}} dr'\, \frac{T_{\ell m}(r')R^{\rm H}_{\ell m}(r')}
{\Delta^{2}} +  \frac{R^{\rm H}_{\ell m}(r)}{W}
\int^{+\infty}_{r} dr'\, \frac{T_{\ell m}(r')R^{\rm \infty}_{\ell m}(r')}
{\Delta^{2}} 
\label{Rfield}
\ee
where $W$ the (constant) Wronskian $ W[ \Delta^{-1/2} R^{\rm H}_{\ell m}, \Delta^{-1/2} R^{\rm \infty}_{\ell m} ] $. 
Note that (\ref{Rfield}) is valid provided the source term $T_{\ell m}(r) $ does not extend to $r \to \infty $. 
Further discussion of this issue can be found in Section~\ref{high_ecc}.

The solutions $R^{\rm H}_{\ell m}$, $R^{\rm \infty}_{\ell m}$ are chosen as to have, respectively,  
purely `ingoing' behaviour at the horizon, and purely `outgoing' behaviour at infinity. Explicitly, 
\be
R^{\rm H}_{\ell m} \to\cases{
\Delta^2  e^{-ikr_{\ast}}\,& \cr
r^3 B^{\rm  out}e^{i\omega r_{\ast}}+
r^{-1}B^{\rm in}e^{-i\omega r_{\ast}}\,&
\cr}, \quad 
R^{\rm \infty}_{\ell m} \to\cases{ 
C^{\rm  out} e^{ik r_{\ast}}+ 
\Delta^2 C^{\rm in} e^{-ik r_{\ast}}\,& 
for $r\to r_+,$ \cr r^3  e^{i\omega r_{\ast}}\,& 
for $r\to +\infty$\cr }
\label{Rinup} 
\ee
where $k=\omega-ma/2Mr_+$, $~r_{+}= M + (M^2-a^2)^{1/2}$ is the 
outer event horizon, and $r_{\ast}$ is the usual tortoise coordinate defined by 
$dr_{\ast}/dr= (r^2+a^2)/\Delta $. From these expressions it follows that
$ W= 2i\omega B^{\rm in}$. The solution (\ref{Rfield}) describes ingoing waves at the horizon and 
outgoing waves at infinity as should be required on physical grounds.
That is,
\bear
R_{\ell m}(r\to r_+) & \to & 
{ \Delta^2 e^{-i k r_{\ast}} \over 
2i\omega B^{\rm in}}
\int^{\infty}_{r_+}dr' {T_{\ell m}(r') R^{\rm \infty}_{\ell m}(r') 
\over\Delta^{2}} 
\equiv Z^{\rm \infty}_{\ell m\omega} \Delta^2 e^{-i k r_{\ast}} 
\\
\nonumber \\
\nonumber \\
R_{\ell m}(r\to\infty) & \to &
{r^3e^{i\omega r_{\ast}} \over 2i\omega B^{\rm in}}
\int^{\infty}_{r_+}dr'{T_{\ell m}(r')  
R^{\rm H}_{\ell m}(r') 
\over\Delta^{2}} \equiv  Z_{\ell m\omega}^{\rm H}r^3e^{i\omega r_{\ast}} 
\label{asymptotics}
\eear
The source term $T_{\ell m}$ takes the form \cite{chapter}, 
\be
T_{\ell m} = \mu\int^{\infty}_{-\infty}dt\, 
e^{i\omega t-i m \varphi(t)}
\Delta^2 \Bigl [ A_{1}(r,\theta)\, \delta(r-r(t)) 
+ \left \{ A_{2}(r,\theta)\, \delta(r-r(t))\right\}_{,r}
+ \left \{ A_{3}(r,\theta)\, \delta(r-r(t)) \right\}_{,rr} 
\Bigr]_{\theta= \theta(t)} 
\label{Tlm} 
\ee
where $A_1, A_2, A_3$ are known functions of $r,\theta$. 
The complex amplitudes $Z_{\ell m\omega}^{\infty,H}$ defined in (\ref{asymptotics})
can then be written as, 
\be
Z^{\infty, H}_{\ell m\omega} =
{\mu\over2i\omega B^{\rm in}}
\int^{\infty}_{-\infty}dt\,
{\cal I}^{\infty, H}_{\ell m \omega}[r(t),\theta(t)]\,
 e^{i\omega t-i m \varphi(t)}
\label{Zs} 
\ee
where 
\be
{\cal I}^{\rm \infty, H}_{\ell m\omega} = 
\Bigl[R^{\rm \infty, H}_{\ell m}\, A_1 -{dR^{\rm \infty, H}_{\ell m} \over dr}\, A_2
 + {d^2 R^{\rm \infty, H}_{\ell m} \over dr^2} A_3 
\Bigr]_{r=r(t),\theta=\theta(t)} 
\label{Iotas}
\ee

So far, all expressions listed in this Section are valid for any bound Kerr orbit. For the special case 
of equatorial orbits, we have $\theta(t)= \pi/2$ which makes ${\cal I}^{\infty,H}_{\ell m\omega}$ 
functions of $r(t)$ only. It is straightforward to show \cite{cutler} that the quantities,
\be
\alpha^{\infty,H}(t)= {\cal I}^{\infty, H}(r(t))~e^{-im[\phi(t) -
\Omega_{\phi}t]} 
\ee
are {\it periodic} functions of time (with a period equal to the radial motion's period $T_r$). 
Consequently, they can be expanded in a Fourier series   
\be
\alpha^{\infty,H}(t)= \sum_{k=-\infty}^{+\infty}  
\alpha^{\infty,H}_{k}e^{-ik\Omega_r t} \quad \Rightarrow \quad
\alpha^{\infty,H}_{k}= \frac{1}{T_r} \int_{0}^{T_r}dt ~
\alpha^{\infty,H}(t)\, e^{ik\Omega_r t}
\label{fourier}
\ee
with $\Omega_{\rm r}= 2\pi/T_{\rm r}$.
Plugging these in (\ref{Zs}) we arrive at
\be
Z^{\infty, H}_{\ell m \omega} = \sum_{k=-\infty}^{+\infty}
Z^{\infty, H}_{\ell m k} \delta(\omega -\omega_{mk}) \;,
\label{Zlmk1}
\ee
where $\omega_{mk}= m\,\Omega_{\phi} + k\,\Omega_r $ and
\be
Z^{\infty,H}_{\ell m k}= \frac{\mu\,\Omega_r}{2i\omega_{mk} B^{in}}
\int_{0}^{T_r} dt ~ {\cal I}^{\infty, H}(r(t))~ e^{i\omega_{mk}t -im\phi(t)} \;.
\label{Zlmk2}
\ee
The last few steps can be repeated for another special case, that of circular-inclined orbits, 
i.e. $r=p,~e = 0 $. In this case, ${\cal I}^{\infty,H}_{\ell m\omega}$  are functions of 
$\theta(t)$ only, and therefore periodic with period $T_\theta$. Then we arrive at,
\be
Z^{\infty,H}_{\ell m k}= \frac{\mu\,\Omega_\theta}{2i\omega_{mk} B^{in}}
\int_{0}^{T_\theta} dt ~ {\cal I}^{\infty, H}(\theta(t))~ e^{i\omega_{mk}t -im\phi(t)} \;.
\label{Zlmk3}
\ee 
where now  $\omega_{mk} = m\,\Omega_\phi + k\,\Omega_\theta $.

Inserting either (\ref{Zlmk2}) or (\ref{Zlmk3}) back in (\ref{ps4}) one can obtain the
following expressions for $\psi_{4}$ at infinity and on the horizon, 
\be
\psi_{4} \to 
\cases{ \rho(r_{+})^{4} \sum_{\ell m k} \psi_{\ell m k}^{H} 
\,& for $r\to r_+$ 
\cr
\cr
 r^{-1} \sum_{\ell m k} \psi_{\ell m k}^{\infty}  \,&
for $r\to +\infty,$ \cr} 
\label{ps4_2} 
\ee
where
\be
\psi_{\ell m k}^{H, \infty}= Z^{\infty, H}_{\ell mk} 
S_{\ell m}(\theta) e^{-i\omega_{mk}(t -r_{\ast}) +im\varphi} 
\label{ps4_3}
\ee

Having at hand the Weyl scalar $\psi_4 $ we can immediately relate it to the two 
polarisation components $ h_{\rm +}, h_{\rm \times} $ of the transverse-traceless 
metric perturbation at $r \to \infty $ \cite{teuk},
\be
\psi_4 \approx \frac{1}{2} \left ( \frac{\partial^2 h_{\rm +}}{\partial t^2} 
- i \frac{\partial^2 h_{\rm \times}}{\partial t^2} \right ) \quad \Rightarrow \quad
h_{\rm +} -i\, h_{\rm \times} = \frac{2}{r} \sum_{\ell mk}\, \frac{Z^{\rm H}_{\ell mk}}
{\omega^{2}_{mk}}\,S_{\ell m}(\theta)\,e^{-i\omega_{mk}(t -r_{\ast}) + im\varphi}
\label{TT}
\ee
The waveform's frequency spectrum is discrete, composed of the orbital harmonics
$\omega_{mk}$.

Expressions for the gravitational wave energy and angular momentum fluxes at infinity can
be derived using the Landau-Lifschitz pseudotensor \cite{landau},
\bear
\left ( \frac{dE}{dt} \right )_{GW}^{\infty} &=& \frac{1}{16\pi} \int 
d\phi\, d\theta\, \sin\theta\,  r^2\, \left \{ \left ( \frac{\partial h_{\rm +}}{\partial t} \right )^2 +
\left ( \frac{\partial h_{\rm \times}}{\partial t} \right )^2 \right \} 
\label{LL1}
\\
\nonumber \\
\left ( \frac{dL}{dt} \right )_{GW}^{\infty} &=& -\frac{1}{16\pi}
\int d\phi\, d\theta\, \sin\theta\, r^2\, \left \{ \frac{\partial h_{\rm +}}{\partial t} \frac{\partial 
h_{\rm +}}{\partial \phi}
+ \frac{\partial h_{\rm \times}}{\partial t} \frac{\partial h_{\rm \times}}
{\partial \phi} \right \}
\label{LL2}
\eear
Time-averaging (over one orbital period $T_r$ or $ T_\theta $) these expressions we get \cite{press},
\be
\dot{E}_{GW}^{\infty} = \sum_{\ell m k}\, \frac{|Z^{\rm H}_{\ell mk}|^2}
{4\pi \omega^2_{mk}} \quad \mbox{and} \quad 
\dot{L}_{GW}^{\infty} =\sum_{\ell m k}\, \frac{m|Z^{\rm H}_{\ell mk}|^2}
{4\pi \omega^3_{mk}}
\label{GW_flux_infty}
\ee

The calculation of the respective fluxes at the black hole horizon is a subtle 
issue as expressions such as (\ref{LL1}), (\ref{LL2}) are not available. 
Despite this difficulty, Teukolsky \& Press \cite{press} were able to derive formulae for the 
horizon fluxes using the results of Hawking and Hartle \cite{hawking} on the change of the horizon surface area under
a given gravitational perturbation. The end result is, 
\be
\dot{E}_{GW}^{H} = \sum_{\ell m k} \alpha_{\ell mk} 
\frac{|Z^{\infty}_{\ell mk}|^2}{4\pi \omega^2_{mk}} 
\quad \mbox{and} \quad 
\dot{L}_{GW}^{H} = \sum_{\ell m k} \alpha_{\ell mk} 
\frac{m|Z^{\infty}_{\ell mk}|^2}{4\pi \omega^3_{mk}}  
\label{GW_flux_H}
\ee
where $\alpha_{\ell m k}$ is a complex constant (its explicit form can be found in Ref.~\cite{press}). 

For the case of generic orbits certain of the above steps cannot be taken as such. The reason is that the functions (\ref{Iotas})
are multiply-periodic and as a consequence their Fourier transform cannot be directly inverted. As recently 
discussed by Drasco \& Hughes \cite{drasco2}, this inversion is possible when the orbit is suitably re-parametrised \cite{mino}.
A detailed analysis of generic EMRI will appear soon \cite{drasco}.


\subsection{The Sasaki-Nakamura formalism}

According to the previous discussion, the calculation of gravitational 
waveforms and fluxes boils down to the computation of the complex numbers $Z_{\ell m k}^{\rm H,\infty}$ and of the 
angular functions $S_{\ell m}$. Before that, one needs to integrate the orbital equations and obtain 
$\{ r(t),\theta(t),\phi(t) \}$ and the frequencies $\Omega_r,\Omega_\theta, \Omega_\phi$. 
In addition, the homogeneous version of eqn.~(\ref{radTeuk}) has to be integrated, as $R^{\infty,H}_{\ell m}$ and
their derivatives as well as the asymptotic amplitude $B^{\rm in}$ are required.

Numerical integration of the Teukolsky equation itself proves to be somewhat problematic  
due to the long-range nature of the potential $V_T$. According to (\ref{Rinup}), the  
$B^{\rm in}$ term decays at infinity much faster than the $B^{\rm out}$ term and can only 
be extracted with very low accuracy \cite{detweiler1}. In order to tackle this problem,
Sasaki \& Nakamura developed a formalism that maps the solutions of the Teukolsky
equation to solutions of the so-called Sasaki-Nakamura equation \cite{chapter}, \cite{SN},
\be
\frac{d^2 X}{dr_{\ast}^2} -F(r)\, \frac{dX}{dr_{\ast}} - U(r)\, X= 0
\label{sneq}     
\ee
The `potentials' $ F(r), U(r)$ as well as the functions $\alpha(r),\beta(r),\eta(r) $ appearing in
subsequent formulae can be found in \cite{chapter}. The solutions of this equation are 
related to the solutions of the Teukolsky equation via the differential operator,
\be
R_{\ell m}(r)= \frac{1}{\eta} \left [ \left ( \alpha +
\frac{\beta_{,r}}{\Delta} \right ) 
\frac{\Delta X_{\ell m}}{(r^2 + a^2)^{1/2} } - \frac{\beta}{\Delta} 
\frac{d}{dr} \left ( \frac{\Delta X_{\ell m}}{(r^2 + a^2)^{1/2} }
\right ) \right ] 
\label{transf}
\ee
The key property of (\ref{sneq}) is that it encompasses a short-range potential.
This can be demonstrated more easily if we shift to the function,
\be
Y(r)= \eta^{1/2}(r) X(r) 
\ee
Then, eqn. (\ref{sneq}) transforms into the Schr\"{o}dinger-type equation,
\be
\frac{d^2 Y}{dr^2_{\ast}} + \tilde{U} Y= 0 
\label{sneq2}
\ee
with the short-range effective potential,
\be
\tilde{U} = -U -\frac{1}{4}\, F^2 + \frac{\Delta}{2\eta (r^2 + a^2)^2} \left \{
\Delta\, \eta_{,rr} - \frac{\Delta}{\eta}\, (\eta_{,r})^2 + 2M\eta_{,r} 
\left ( \frac{r^2 -a^2}{r^2 + a^2} \right ) \right \} \to   
\cases{ \omega^2  + {\cal O}(r_{\ast}^{-2})  \,& 
for $ r_{\ast} \to +\infty$ 
\cr
 k^2 + {\cal O}( e^{c r_{\ast}})  \,& for $r_{\ast} \to -\infty,$ \cr}
\label{effpot}
\ee
where $ c= (r_{+} -r_{-})/2M $ is a positive constant. As a consequence, eqn.~(\ref{sneq}) admits solutions,
\be
X^{\rm H} \to  
\cases{ {\cal A}\, e^{-ikr_{\ast}} \,& \cr
A^{\rm in}\, e^{-i\omega r_{\ast}} + A^{\rm out}\, e^{i\omega r_{\ast}} 
\, &  \cr}, \quad 
X^{\rm \infty} \to  
\cases{ D^{\rm in}\, e^{-ikr_{\ast}} + D^{\rm out}\, e^{ikr_{\ast}}  \,& for $r\to r_+$ 
\cr
{\cal D }\, e^{i\omega r_{\ast}}  \, & for $r\to +\infty.$ \cr}
\label{Xinup} 
\ee
The relation between the asymptotic amplitudes appearing in  (\ref{Rinup}) and
(\ref{Xinup}) can be deduced from (\ref{transf}),
\be
B^{\rm in} = -\frac{1}{4\omega^2} A^{\rm in}  
\label{amplitrans}
\ee
Unlike the Teukolsky equation, the Sasaki-Nakamura equation is ideal for
numerical integration, as the `ingoing/outgoing' components are of
comparable magnitude at infinity and the horizon. We can then simply find the 
desired amplitude $B^{\rm in}$ from (\ref{amplitrans}). Similarly, knowledge of the 
wavefunction $X(r)$ and its derivative at a given point immediately leads to 
the Teukolsky function $R_{\ell m}(r)$ and its derivative via the rule 
(\ref{transf}). 


\subsection{Equatorial orbits}
\label{equat}

The Teukolsky-Sasaki-Nakamura formalism has been the standard machinery for computing gravitational waveforms 
and the accompanying back-reaction for test-bodies in equatorial orbits in Schwarzschild and Kerr spacetimes 
\cite{detweiler1,szedenits,parabolic,apostolatos,cutler,shibata_ecc,tanaka,kenn98,kgdk,finn}. Once the 
$\dot{E}_{\rm GW}$, $\dot{L}_{\rm GW}$ fluxes are obtained at spatial infinity and the black hole horizon, then 
flux-balance dictates $\dE = -\dot{E}_{\rm GW}$, $\dL =  -\dot{L}_{\rm GW}$. Since $p=p(E,L_z),~e=e(E,L_z)$ are known 
functions we find,
\bear
\ddp &=& H_{\rm eq}^{-1}\, [ L_{z,e}\,\dE  - E_{,e} \dL ] 
\\
\dot{e} &=& H_{\rm eq}^{-1}\, [ -L_{z,p}\,\dE + E_{,p} \dL ]
\label{rates_eq}
\eear
where $H_{\rm eq} = E_{,p} L_{z,e} -E_{,e} L_{z,p}$.

We first focus on the waveforms. For given values of $p,e$, provided $r_p$ is sufficiently large
(something like $ \gtrsim  20 M  $), there is no significant difference between Schwarzschild and 
Kerr waveforms, since at such distances the black hole spin influences only marginally the orbital motion, 
and consequently the waveform. This similarity is no longer true in the strong field, especially when
the hole is rapidly rotating. The waveform is significantly affected by the body's high velocity, 
intense frame-dragging and orbital plane precession, as well as by backscattered `tail'
radiation.

\begin{figure}
\centerline{\includegraphics[height=7cm,clip]{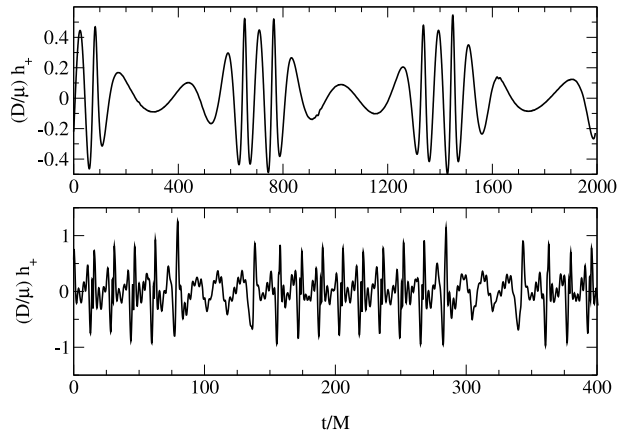}}
\caption{Examples of equatorial zoom-whirl waveforms (only $h_{+}$ component shown, with the observation point
located at $(r,\theta) = (D,90^\circ)$) for the equatorial orbits $ p =1.8M,~e =0.4,~\iota = 0^\circ $ 
(bottom panel) and $ p =10.5 M,~ e = 0.5,~\iota = 180^\circ $ (top panel). The black hole spin is $a=0.99M$.}
\label{fig_wav1}
\end{figure}

Of particular importance are the `zoom-whirl' orbits we discussed in Section~\ref{geod}
(more details can be found in \cite{kgdk}). Provided that the eccentricity at this stage is not 
too small (which is the most likely scenario, according to the results of Section~\ref{hybrid}), then 
the zoom-whirl behaviour should be taken into account for constructing waveform templates for LISA. 
A compact body in a zoom-whirl orbit will spend a considerable fraction of the orbital period in strong 
field regions  and hence will radiate strongly. It turns out that a good fraction of the averaged flux is 
radiated during the motion near the periastron. The emitted waveform is accordingly shaped
in a series of rapid `quasi-circular' oscillations separated by relatively `quiet' intervals. 
Once again, for values $a \gtrsim 0.9 M $  the zoom-whirl effect will be enhanced (assuming a prograde orbit, see
Fig.~\ref{fig_sepax}). Examples of zoom-whirl waveforms are shown in Fig.~\ref{fig_wav1}, generated by  
the equatorial Kerr Teukolsky code of Ref.~\cite{kgdk}. These waveforms correspond to two opposite extreme cases; 
prograde and retrograde orbits with $p$ very close to $p_s$ for $a=0.99M$. In agreement with Fig.~\ref{fig_sepax},
the `whirling' portion of the waveform is much more pronounced in the prograde case. Equally contrasting is the    
relative contribution of higher $\ell-$ harmonics: the retrograde waveform is essentially composed by 
$ \ell \leq \ell_{\rm max} \sim 4-5 $ while $ \ell_{\rm max} \sim 12-14 $ for the prograde waveform.   
Another important aspect (not evident in Fig.~\ref{fig_wav1}) is the waveform's directionality. 
Viewed off the equatorial plane, the waveform suffers a substantial suppression in high frequency features. 
This is because the wave's higher multipole components (which are responsible for the small-scale structure) 
are mainly `beamed' to directions close to the equatorial plane \cite{kgdk}. This effect is similar to
classical synchrotron radiation and it can be easily `explained' noting that at the limit $l=m \gg 1$ the angular
function (for $a=0$) $S_{\ell m} \sim \sin^{\ell}\theta $ and hence the radiated flux will exhibit
a $ \sin^{2\ell}\theta \approx e^{-\ell(\theta-\pi/2)/4}  $ angular dependence near the equatorial plane 
(see Refs.~\cite{breuer} for early studies in gravitational synchrotron radiation).

\begin{figure}
\centerline{\includegraphics[height=7cm,clip]{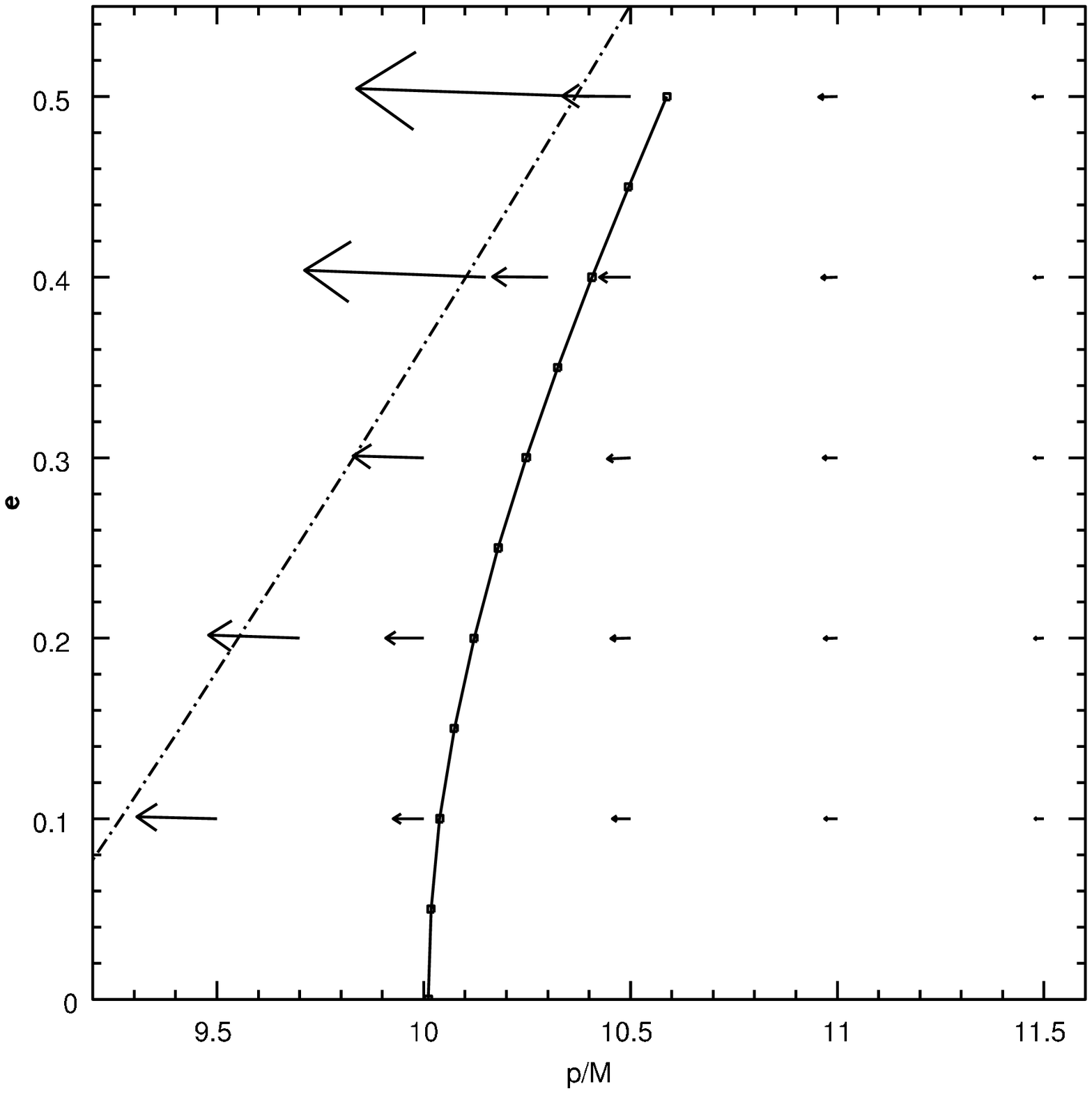}
\includegraphics[height=7cm,clip]{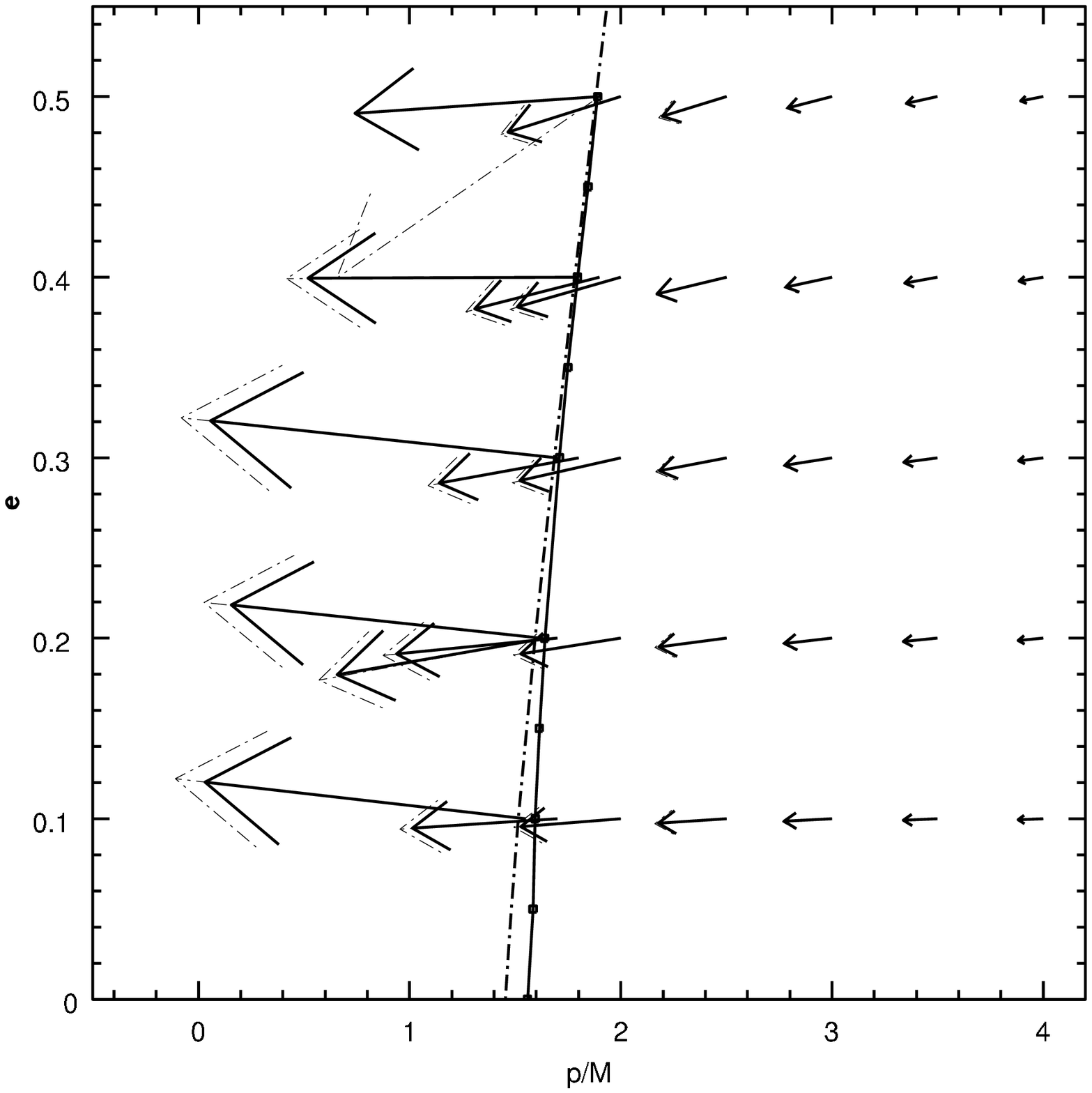}}
\caption{Radiative evolution of equatorial orbits, represented by arrows living on the $p-e$ plane. The arrows have 
components $ (M/\mu)\,[\ddp,\de] $. The right (left) panel corresponds to prograde (retrograde) motion around a 
$a =0.99 M $ Kerr hole. The separatrix and the $\de=0 $ critical curve are represented by the dashed and solid curve, 
respectively. Note the markedly different areas of the $\de > 0 $ regions.} 
\label{fig_pe_planes}
\end{figure}

Next, we discuss the evolution of equatorial orbits under radiation reaction. This topic has been the 
subject of several studies \cite{apostolatos},\cite{kenn98},\cite{cutler},\cite{kgdk}. A useful way to present the 
results is by plotting arrows $(\ddp,\de)$ for each given orbit $(p,e)$ of the $p-e$ plane, as in Fig.~\ref{fig_pe_planes}. 
For all points, we have that $\ddp <0$ which simply means that the orbit always shrinks, until the point where 
the separatrix is reached and orbit becomes a plunging one. On the other hand, the eccentricity exhibits a 
more interesting evolution. As suggested by weak-field computations \cite{pm} \cite{ryan1} (see eqn.~(\ref{deN}) 
of Section~\ref{hybrid}) we typically find $\de <0 $, that is, the orbit tends to circularise. This phenomenon 
of orbital circularization as a result of some form of dissipation is seen in many astrophysical situations, 
such as that of satellites whose orbits are decaying due to atmospheric friction. The reason is that the 
dissipating mechanism causes the body to `drop' in its potential well, the usual geometry of which ensures 
that the orbital eccentricity decreases. However, for orbits in the vicinity of the separatrix we surprisingly 
find $\de >0$.  This effect, which is confirmed by both numerical and analytical calculations, is intimately 
connected with the notion of a last stable orbit. As this orbit is approached at the end of the inspiral, 
the radial potential $R$ (defined in eqns.~\ref{geod1}) becomes shallower (as the minimum turns into a saddle 
point at plunge), and this tends to increase the eccentricity of the orbit. Shortly before plunge this 
mechanism overcomes the circularising tendency.

Using (\ref{rates_eq}) it is straightforward to show that $\de > 0$ sufficiently close to the separatrix. 
During this final stage of the inspiral the fluxes approximately obey the relation (following from 
eqns.~(\ref{GW_flux_infty}),(\ref{GW_flux_H}) as a consequence of $T_r \gg T_\phi$ which makes 
$\omega_{mk} \approx m\Omega_\phi $),
\be
\dE \approx \Omega_\phi \dL
\label{dE_circ}
\ee   
which is characteristic of a circular orbit. This relation simply means that most of the energy and 
angular momentum is radiated during the body's `whirling'  at $r \approx r_p$, during which the radius hardly 
changes and there is a single dominant frequency $\Omega_\phi  $ and its harmonics.  

Using (\ref{dE_circ}), we have near the separatrix
\be
\de \approx H_{\rm eq}^{-1}\,\dL\,[ -E_{,e} + \Omega_\phi\,L_{z,e} ] 
\ee
We always have, $\dL < 0 $ and consequently the sign of $\de$ is determined by the sign of the function
$ H_{\rm eq}^{-1}\,[ -E_{,e} + \Omega_\phi\,L_{z,e} ]  $ which encodes the body's geodesic motion close to the
separatrix. For $ p \gg M$ this quantity is positive, leading to the expected $ \de <0$ result, but sufficiently close
to the separatrix $ p \to p_s $ it takes negative values irrespective of $e$. Based on these observations, it follows that there 
must be a critical curve on the $p-e$ plane where $\de=0$. We should expect to find this curve in the vicinity of the separatrix, 
as the $\dot{e} > 0 $ effect is related to the notion of a last stable orbit. For example, for nearly circular orbits,
this critical curve is located at $p_{\rm crit} = 6.68 M $ for $a=0$ and as $a \to M$ it tends to 
`coalesce' with the separatrix at $ p_s \to  M $, for prograde motion \cite{apostolatos,kenn98,kgdk}. In the same limit, 
their separation becomes maximal for retrograde motion, $ p_{\rm crit} = 9.8 M  $. As the numerical data of 
Fig.~\ref{fig_pe_planes} demonstrate, this behaviour persists for arbitrary eccentricities: the eccentricity gain
region shrinks (expands) for prograde (retrograde) motion, as the spin increases.    

The vectors in Fig.~\ref{fig_pe_planes} were generated using $ \dE = \dE_{\infty} + \xi\,\dE_{\rm H} $
(and similarly for $\dL$) with the option of setting $ \xi = 1 $ (total flux -- dashed arrows) or $\xi =0 $ 
(horizon flux neglected--solid arrows). No obvious difference between these two choices appear in the retrograde case, 
but there is a clear deviation in the prograde case, for near-separatrix orbits. Surprisingly, inclusion of the horizon 
fluxes {\em slows down} the inspiral. Responsible for this seemingly bizarre behaviour is the phenomenon of superradiance. 
We expand further on this issue in Section~\ref{evolv} below.


\subsection{Non-equatorial orbits}
\label{nequat}

Gravitational radiation from a test-body in an circular-inclined orbit was first studied by 
Shibata \cite{shibata_circ}, more than ten years ago, by means of the Teukolsky-Sasaki-Nakamura formalism.
Shibata not only computed fluxes $\dE,\dL$ and waveforms but he also provided a first approximate expression
for the Carter constant flux $\dQ$ (more details are given in Section~\ref{Q}). 

A few years later, Kennefick \& Ori \cite{ori} were able to derive an exact expression for
$\dQ$ in terms of the components of gravitational self-force, see Appendix~\ref{app:KennOri} for
more details. The Kennefick-Ori formula relates the instantaneous fluxes $\{dE/d\tau,dL_z/d\tau,dQ/d\tau\}$,
\be
\frac{dQ}{d\tau} = {\cal H}_{,E} \frac{dE}{d\tau} + {\cal H}_{,L_z} \frac{dL_z}{d\tau} -
2\,\Delta\,\mu\,u_r\,F_{r}
\label{ko1}
\ee
The function ${\cal H}(r,E,L_z)$ is defined in Appendix~\ref{app:KennOri}.

For a circular orbit ($r=$ const, $u_r=0$) eqn.~(\ref{ko1}) yields,
\be
\dQ = {\cal H}_{,E} \dE + {\cal H}_{,L_z} \dL
\label{ko2}
\ee
for the averaged fluxes. As discussed by Kennefick \& Ori, this is a key relation for
the so-called `{\em circularity theorem}' \cite{ori},\cite{chapter},\cite{ryan2}: Kerr circular orbits remain circular 
under the influence of radiation reaction. From a practical point of view, eqn.~(\ref{ko2}) 
is an exact expression which relates $\dQ$ to the $\dE,\dL$ fluxes -- quantities that can be
calculated within the Teukolsky framework. Therefore, for circular orbits, the $\dQ$ flux 
is within reach without having to resort to any self-force computation, or any kind of approximation.     

\begin{figure}
\centerline{\includegraphics[height=9cm,clip]{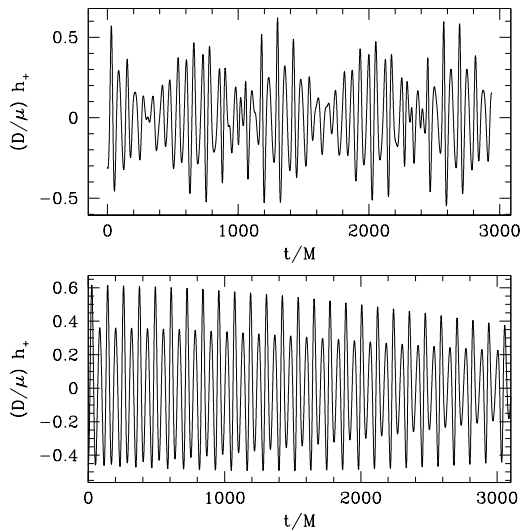}}
\caption{Waveforms from Kerr circular-inclined orbits (only $h_{+}$ component shown, with he observation point
located at $(r,\theta) = (D,90^\circ)$). The orbital parameters are $ p = 7M,~ \iota = 62.43^\circ $ and the 
black hole spin is $ a=0.95M$ (top panel) and $a=0.05M $ (bottom panel).}
\label{fig_wav2}
\end{figure}

This important result was fully exploited by Hughes \cite{scott_circ} in his study of circular-inclined EMRIs.
As in the previous case of equatorial orbits, strong field circular-inclined orbits generate the most interesting 
waveforms. A representative example (taken from Refs.~\cite{scott_circ},\cite{scott_insp}) is shown in 
Fig.~\ref{fig_wav2} (top panel) for an orbit with parameters $ p = 7M,~ \iota = 62.43^\circ $ around a rapidly spinning 
hole of $ a = 0.95M $. The dominant feature in the waveform's pattern is the modulation induced by the body's 
$\theta- $ motion (Lense-Thirring precession). The same effect operates on a much longer timescale when we set 
$a=0.05 M$, keeping the same orbital parameters, see bottom panel of Fig.~\ref{fig_wav2}. This could be 
anticipated from the fact that the precession/modulation frequency is $\Omega_\theta -\Omega_\phi \approx \pm 2aM/r^3 $,
at leading order.   

\begin{figure}
\centerline{\includegraphics[height=7cm,clip]{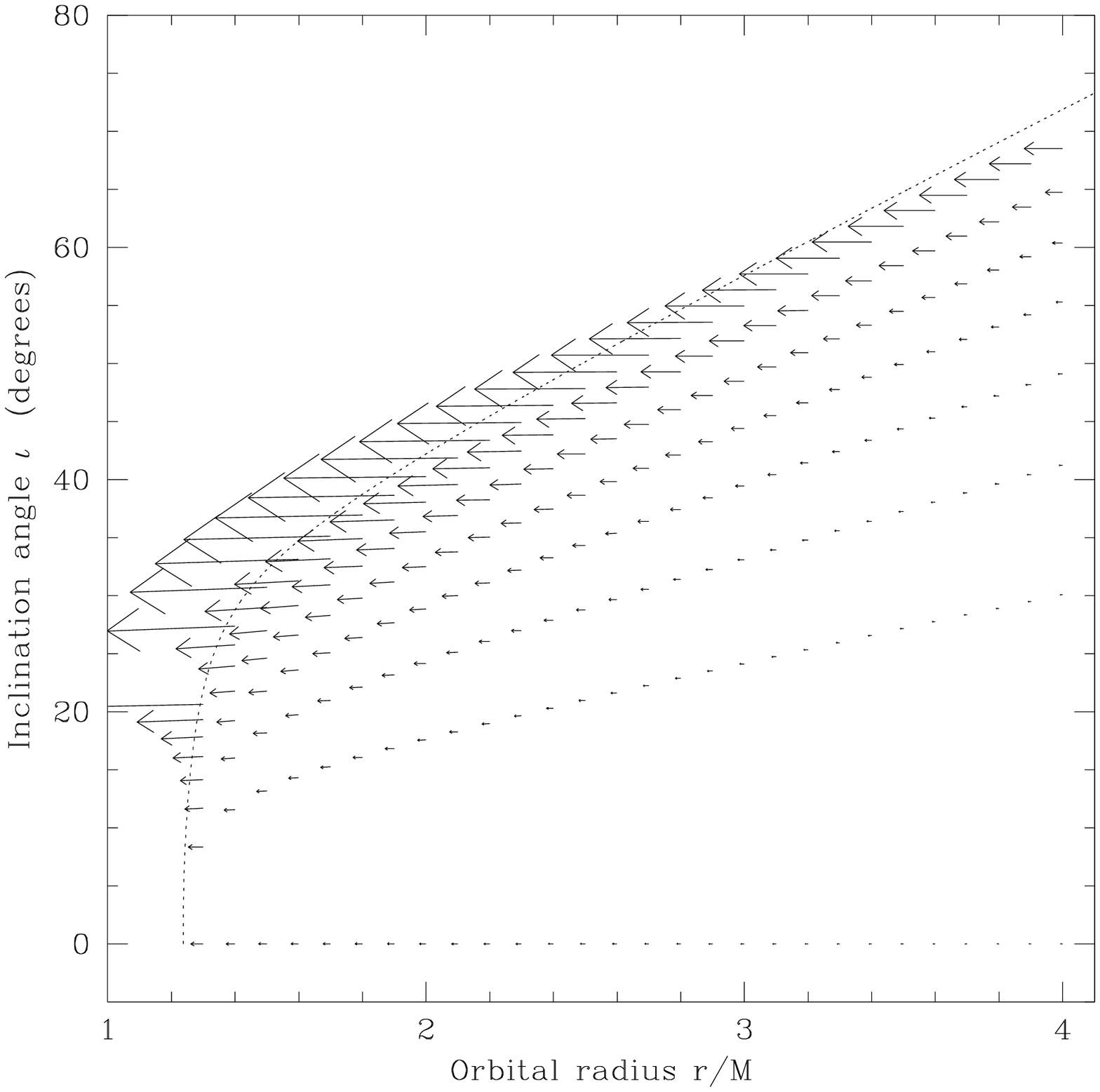}
\includegraphics[height=7cm,clip]{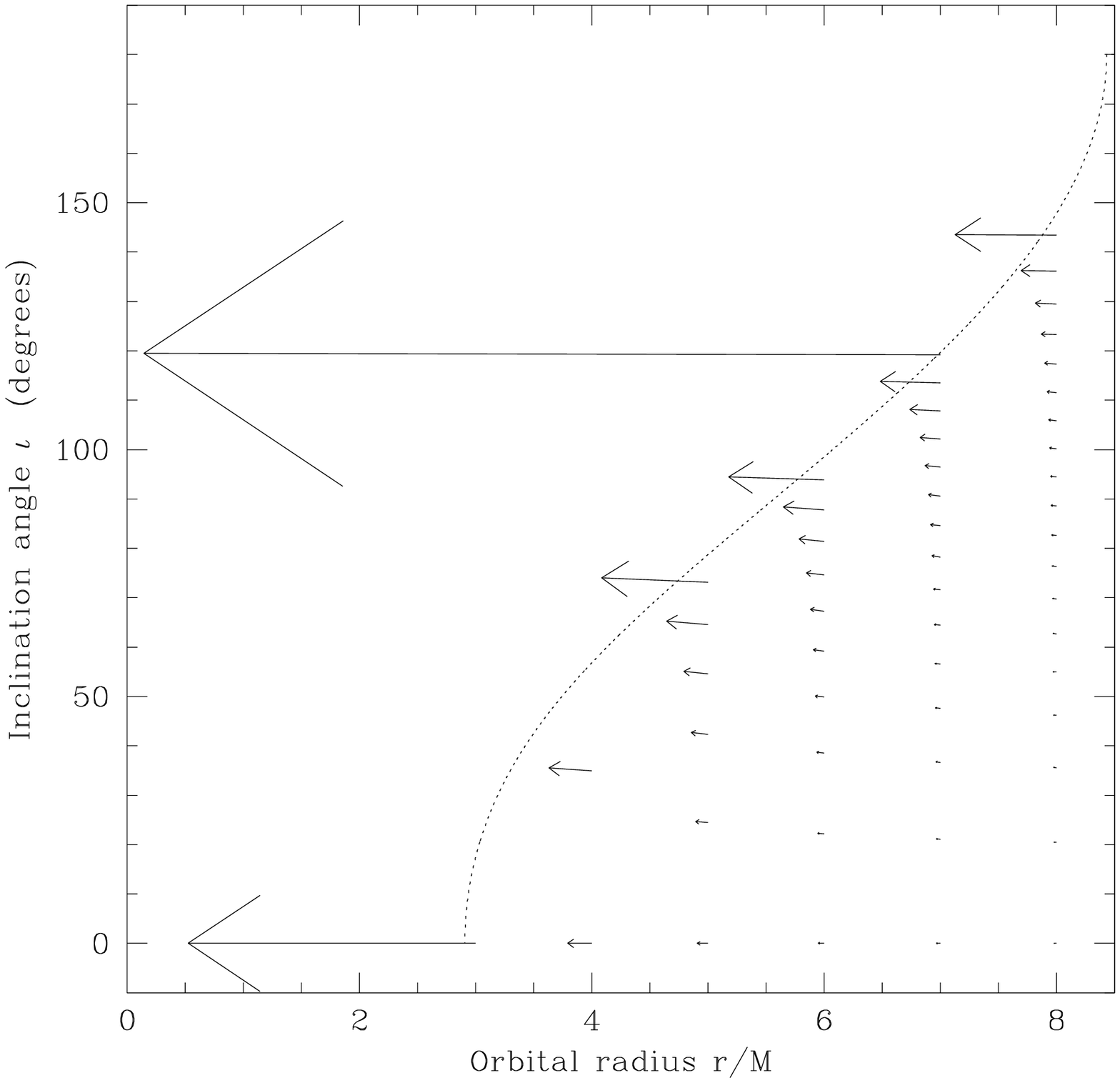}}
\caption{Radiative evolution of Kerr circular-inclined orbits, represented by arrows on the $p-\iota$ plane. 
Each arrow has components $(M/\mu)[\ddp,\di]$. The panel on the left corresponds to $a= 0.998 M$, while the panel 
on the right refers to a slower spinning hole $a=0.8M$. The dotted curves represent the separatrix for each case. 
Note the $\di < 0$ arrows near the $a=0.998M$ separatrix, and the very long arrows in the $a=0.8M$ case, 
corresponding to orbits which are about to become dynamically unstable.} 
\label{fig_pi_planes}
\end{figure}

Even more interesting are the results regarding the orbital evolution. In a fashion similar to Fig.~\ref{fig_pe_planes}, the 
rates $ \ddp,\di$ are illustrated as vectors on the $p - \iota$ plane, see Fig.~\ref{fig_pi_planes}. The relevant
expressions are,
\bear
\ddp &=& H_{\rm circ}^{-1}\, [ -L_{z,\iota}\,\dE + E_{,\iota}\,\dL ]
\\
\di &=& H_{\rm circ}^{-1}\, [ -L_{z,p}\,\dE - E_{,p}\,\dL ]
\label{dpdi}
\eear 
with $H_{\rm circ} = E_{,\iota}\,L_{z,p} -L_{z,\iota}\,E_{,p}$. It is always the case that $\ddp <0 $ 
(i.e. the orbits shrinks), but $\di$ shows a somewhat less monotonic behaviour. Provided that the black hole spin 
is $a \lesssim 0.952 M$ we have $\di > 0$ for all orbital parameters corresponding to stable circular orbits. 
Hence, the typical behaviour is a tendency of orbital angular momentum to {\em anti-align} with the black hole spin,
in agreement with the prediction of weak-field/slow-motion calculations \cite{ryan1}. As expected by symmetry 
considerations, both calculations predict $\di =0 $ when the black hole spin is turned-off.   

For rapidly spinning holes with $ a \gtrsim 0.952 M $ the above general rule is no longer true. Orbits that are 
`horizon-skimming' \cite{wilkins} (that is with $p \approx r_{+}$) evolve under radiation reaction as to 
have $\di < 0$ \cite{scott_circ}. This counter-intuitive behaviour can be explained by looking closer at the 
properties of circular orbits close to the horizon. Normally, we have the intuitive relation 
$ \partial L_z/\partial \iota < 0 $. However, horizon-skimming orbits violate this rule, i.e. they behave as 
$\partial L_z/\partial \iota > 0 $. Since $ \di = \dL\,(\partial L_z/\partial \iota)^{-1} $ and since $\dL < 0 $ always, 
it follows that $\di < 0  $ for horizon-skimming orbits \cite{scott_circ}.

The most striking feature in the radiative evolution of circular-inclined orbits is the 
small rate at which inclination changes, i.e.  $ \di/\iota \ll \ddp/p  $. In other words, during an inspiral, 
$\iota$ appears to remain almost fixed  (at least when $e=0$) under radiation backreaction, especially when the 
body is not moving close to the horizon. The data presented in the following Section will make this statement 
more quantitative. 

\subsection{Evolving the orbit}
\label{evolv}

We have so far discussed orbital evolution results in the form of $(\ddp,\de) $ and $(\ddp,\di)$ vectors.
Although quite instructive, they are, in a sense, `frozen' in time. They do not convey a clear picture
of an entire inspiral trajectory, which would start with some initial orbit and terminate at the point where 
the separatrix is crossed. Perhaps the simplest way to construct such inspiral trajectory is by pasting together 
a sequence of orbits. This should be a rather good first approximation, due to the adiabatic nature of
the evolution. Such a program has been carried out by Hughes \cite{scott_insp} for the case of circular-inclined orbits.
A similar study for equatorial-eccentric orbits is in progress and should be completed in the near future.  
The orbital evolution algorithm is the following sequence of Eulerian steps:
\bear
t_{i+1} &=& t_i + \delta t
\\
p_{i+1} &=& p_i + \ddp\,\delta t
\\
\iota_{i+1} &=& \iota_i + \di\,\delta t
\label{euler}
\eear
and the orbit's physical trajectory is described by $z^a = z_{\rm geod}^a[t,E(p_i,\iota_i),L_z(p_i,\iota_i),
Q(p_i,\iota_i)]$. The circularity theorem \cite{chapter},\cite{ori},\cite{ryan2} (Section~\ref{nequat}) 
ensures that $e=0$ throughout the inspiral. The selected time-step $\delta t$ should be consistent with the underlying 
assumption of adiabatic evolution, i.e. $\delta t \ll T_\phi, T_\theta $.  

In order to calculate the associated waveform one could in principle plug $z^a(t)$ into the Teukolsky source term, 
and follow the steps discussed in Section~\ref{Teuk_FD}. The resulting waveform would be quasi-periodic, exhibiting a familiar 
`chirping' character during inspiral.

In Ref.~\cite{scott_insp} a somewhat different prescription is given for generating the inspiral waveform.
The true periodic adiabatic waveform (as given by eqn.~\ref{TT}) becomes
\be
h_{\rm +} -i\, h_{\rm \times} = \frac{2}{r} \sum_{\ell mk}\, \frac{Z^{\rm H}_{\ell mk}}
{\omega^{2}_{mk}}\,S_{\ell m}(\theta)\,e^{i\omega_{mk}\,r_{\ast} + im\varphi}\,
e^{-i\int dt\,\omega_{mk}(t)}
\label{wav_insp}
\ee   
where $Z_{\ell m k}, S_{\ell m}(\theta)$ and $\omega_{mk} = m\Omega_\phi + k\Omega_\theta $ have
effectively become functions of time as they need to be `updated' at each timestep $t_i$. The above
two prescriptions for generating inspiral Teukolsky waveforms are not fully equivalent to each other
(in fact nobody has calculated a waveform using the former method) but, most likely, they would 
give similar results.       

\begin{figure}
\centerline{\includegraphics[height=7cm,clip]{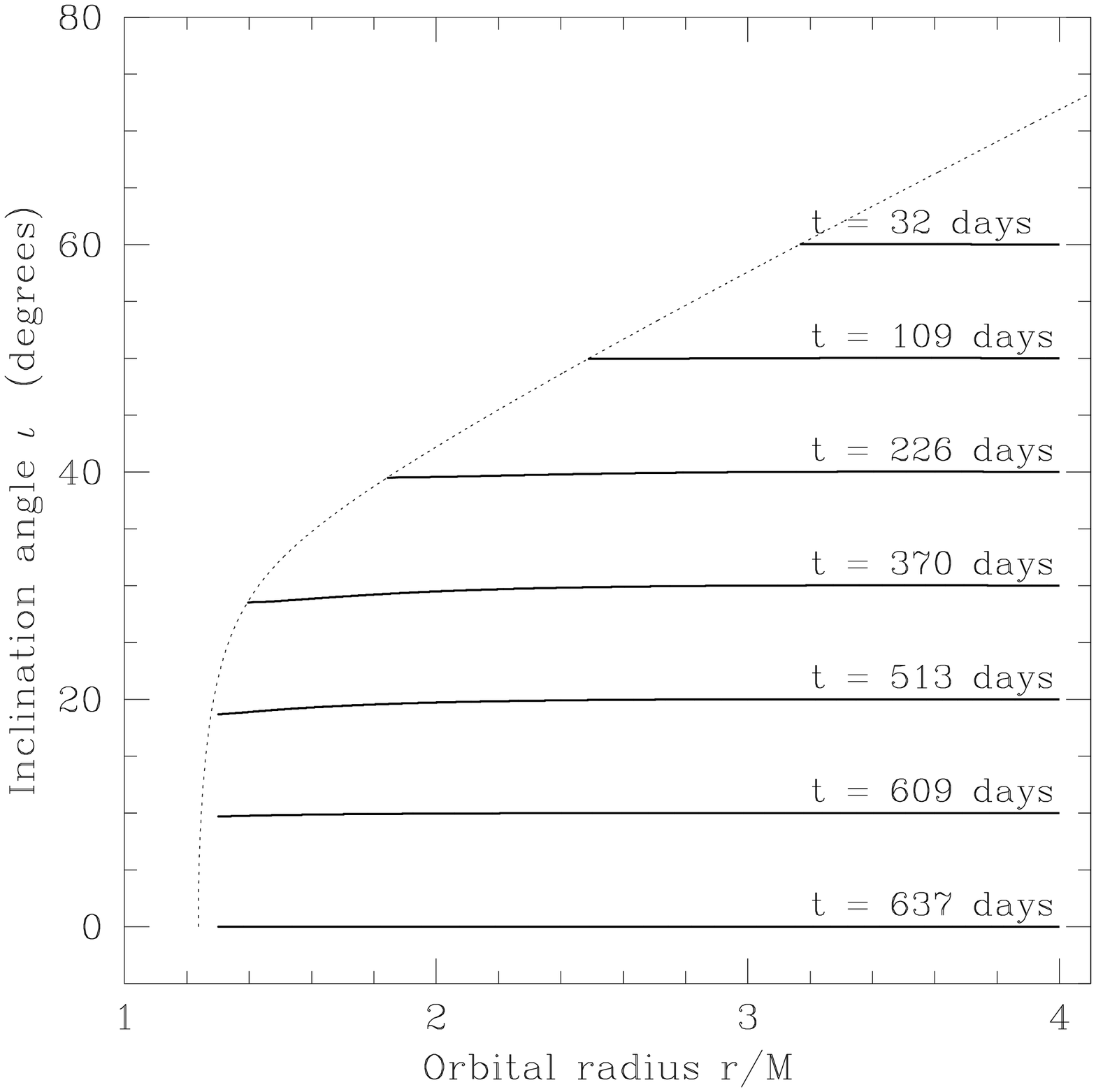}
\includegraphics[height=7cm,clip]{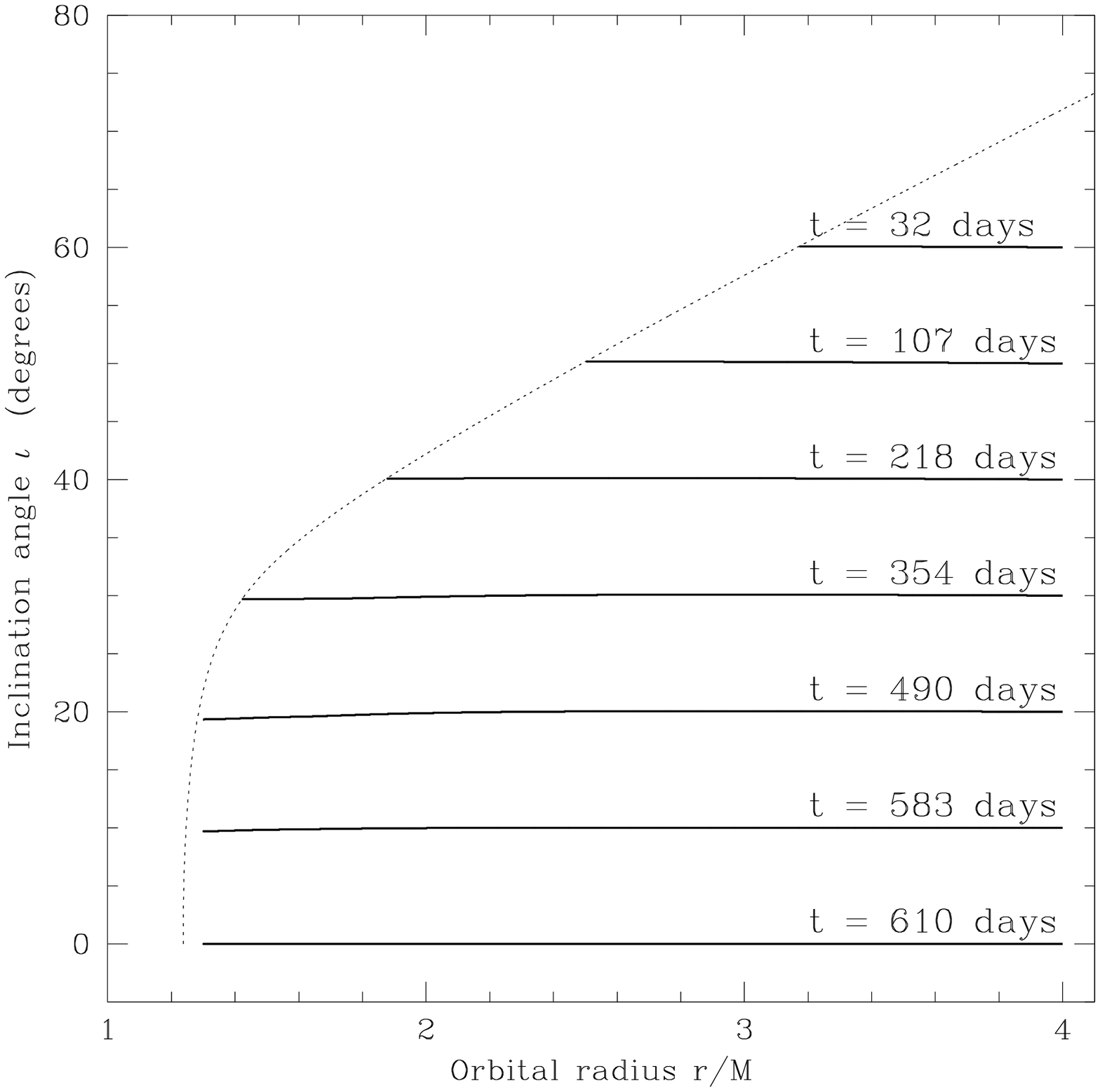}}
\caption{Teukolsky-based, Kerr circular-inclined inspirals for a $a=0.998 M$ black hole for various initial inclination 
angles. The canonical values $ M= 10^6 M_\odot $, $\mu = 1 M_\odot $ have been chosen. In generating these inspirals, 
the horizon fluxes were included (left panel) or ignored (right panel). Each curve is labelled by the total inspiral time,
starting at $p=4M$.}
\label{fig_teuk_insp}
\end{figure}

\begin{figure}
\centerline{\includegraphics[height=8cm,clip]{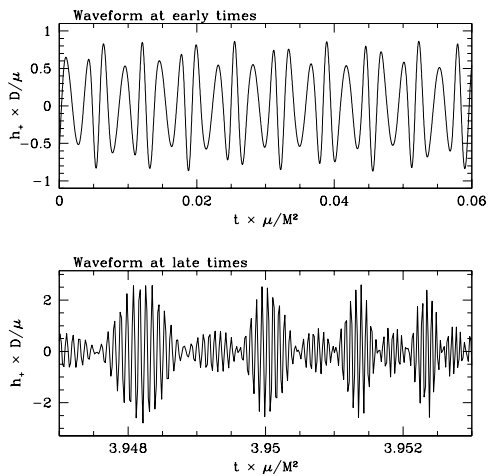}}
\caption{Teukolsky-based waveform (only $h_{+}$ component shown, with the observation point
located at $(r,\theta) = (D,90^\circ)$), corresponding to the $\iota \approx 40^\circ $ inspiral of 
Fig.~\ref{fig_teuk_insp}.}
\label{fig_wav_teuk_insp}
\end{figure}

Representative results from Hughes' study are shown in Figs.~\ref{fig_teuk_insp}~\& \ref{fig_wav_teuk_insp}. 
As anticipated from the previous discussion, even for the strong field inspiral of Fig.~\ref{fig_teuk_insp} 
the inclination angle remains almost fixed; the total accumulated change in $\iota$ is of the order $\sim 2^\circ $. 
The change in $\iota$ is hardly noticeable for inspirals that terminate at $ p \gtrsim 5 M$, as will be the case 
when the black hole is not rapidly rotating (see Fig.~5 in Ref.~\cite{scott_insp}).  
  
Perhaps, this is the most important result stemming from the study of circular inclined orbits. When combined with
Ryan's weak-field result \cite{ryan2} for $\di$ for the case of inclined and eccentric orbits (which also predicts 
that $ \di/\iota \ll \ddp/p, \de/e  $), it makes a strong case in favour of the assumption that $\iota $ hardly changes 
even for {\em generic} Kerr inspirals. This idea was suggested by Cutler as a first stab at the problem
of computing realistic generic inspirals, at least until self-force calculations are mature enough to be able to 
provide an accurate result for $\dQ$ and $\di$. In the meantime, the $\iota = const$ rule is 
a practical approximation which could facilitate the computation of Teukolsky-based generic inspirals, 
along the lines of the calculation described here. The same rule is adopted in the so-called hybrid approximation of 
Kerr inspirals, which we expose in detail in Section~\ref{hybrid}.

Another significant aspect of the inspirals of Fig.~\ref{fig_teuk_insp} is the importance of the fluxes at the 
horizon. According to the flux data from equatorial-eccentric \cite{kgdk} and circular-inclined orbits 
\cite{scott_circ}, the horizon fluxes can be as large as $\sim 10 \%$ of the fluxes radiated at infinity. 
This extreme situation was illustrated in Fig.~\ref{fig_pe_planes} for the evolution of equatorial orbits. 
Horizon fluxes of this magnitude can only be produced by bodies moving close to the horizon. In turn, such orbits are only possible 
around rapidly spinning holes, which also possess extended ergoregions. A fundamental property of an 
ergoregion is its ability to scatter back and amplify any incoming radiation field of frequency 
$ \omega < m\,\Omega_h \equiv m a/2M r_{+}$ (an excellent discussion on the many faces of superradiance and further 
references can be found in Ref.~\cite{super}). Hence, it turns out that when the horizon fluxes are significant they 
also represent superradiant back-scattered fields. In effect, the orbit {\it gains} energy and angular momentum at the 
expense of the black hole's rotational energy\footnote{In the early days of black hole perturbations, an interesting 
idea concerned the so-called `floating-orbits' \cite{floating}: bodies in strong-field Kerr orbits which would hover around the 
black hole without inspiralling, balancing their radiative losses at infinity by absorbing energy from the black hole. 
However it was soon realised that such conditions could never be realised.}. A perhaps more intuitive way to think 
of this effect is by considering the gravitational coupling of the small body with the tidal bulge that it induces 
on the horizon \cite{hawking},\cite{scott_insp},\cite{hartle} (described by the value of the Weyl scalar $\psi_0$ 
at the horizon). The bulge exerts a torque at the small body and tends to increase (decrease) its orbital 
frequency depending on the sign of $\Omega_\phi -\Omega_h $. This tidal coupling is not as exotic as it sounds,
and is observed in other celestial systems such as our own planet and the Moon.      

The data provided in Fig.~\ref{fig_teuk_insp} for the total duration of the inspiral (using canonical values
$\mu = 1 M_\odot, M = 10^6\, M_\odot $) suggest that the superradiance-induced inspiral delay can be
$\sim 5 \%$ of the total inspiral time. This effect is relevant only for rapidly spinning holes (a beautiful 
example of strong-field black hole physics that LISA will be able to probe), otherwise the horizon fluxes contribution 
quickly diminishes. 

Finally, in Fig.~\ref{fig_wav_teuk_insp} we show the waveform corresponding to two different stages of the 
inspiral of Fig.~\ref{fig_teuk_insp} (also taken from \cite{scott_insp}) with initial $\iota = 40^\circ $. 
As the orbit becomes increasingly relativistic, the orbital plane precession frequency $\Omega_\theta -\Omega_\phi$ grows and 
strongly modulates the waveform. At the same time the monotonic increase of the rotational frequency $\Omega_\phi$ shifts the 
waveform's frequency content to higher values.                

\subsection{Parabolic and high eccentricity orbits}
\label{high_ecc}

The Teukolsky solution, eqn.~(\ref{Rfield}), is not well-defined for orbits that extend to infinity, as the radial 
integrals over the source term diverge as $ r \to \infty $. In past calculations two methods have been used to
remove this divergence. In the first method \cite{szedenits}, the problem was dealt by carrying out integrations by parts, 
isolating the divergences in the resulting `surface' terms. These terms were subsequently discarded. The second method
involves the use of the Sasaki-Nakamura formalism, and in particular of the {\em inhomogeneous} equation,
\be
\frac{d^2 X}{dr_{\ast}^2} -F(r)\, \frac{dX}{dr_{\ast}} - U(r)\, X= {\cal S}(r)
\label{SN2}
\ee
The source term ${\cal S}(r)$  is related to the Teukolsky source $T_{\ell m} $ as \cite{SN},
\be
{\cal S} = \frac{\eta\,\Delta\, W(r)}{r^2\,(r^2+a^2)^{3/2}}\,\exp \left ( -i\int dr\,\frac{K}{\Delta}  \right ),
\quad \mbox{with} \quad  \frac{d^2 W}{dr^2} = \frac{r^2\, T_{\ell m}(r)}{\Delta^2}\,\exp \left ( i\int dr\,\frac{K}{\Delta}  \right ) 
\label{W}
\ee   
Solving eqn.~(\ref{SN2}) via the standard Green function method and shifting back to the Teukolsky function we arrive at,
\be
R_{\ell m}(r \to \infty) \approx \frac{r^3 e^{i\omega r_{\ast}}}{2i\omega B^{\rm in}}\, \int_{r_{+}}^{\infty} dr\,
\left ( \frac{r^2+a^2}{\Delta} \right ) \frac{X^{\rm H}}{\eta}\, {\cal S}
\label{Rfield2}
\ee
In contrast with $T_{\ell m}$, the new source term decays as $\sim r^{-5/2} $ at infinity and therefore the above integral is
finite. The price that one pays for this effective regularisation is the extra numerical integration of eqns.~(\ref{W}).  
The above method has been used in a series of papers on equatorial parabolic/eccentric \cite{parabolic},\cite{tanaka}, 
\cite{shibata_ecc} and circular-inclined orbits \cite{shibata_iota}.   

The presence of divergent integrals in the Teukolsky solution (\ref{Rfield}) does {\em not} imply a failure of the
Teukolsky equation itself. As explained by Poisson \cite{poisson}, writing the Teukolsky solution in the form (\ref{Rfield})
is legal as long as the integrals are convergent. If this is not true (as in the case of parabolic orbits), one needs to use 
the general expression,
\be
R_{\ell m}(r)= \frac{R^{\rm \infty}_{\ell m}(r)}{W} \left [ A +
\int^{r}_{r_1} dr'\, \frac{T_{\ell m}(r')R^{\rm H}_{\ell m}(r')}
{\Delta^{2}} \right ] +  \frac{R^{\rm H}_{\ell m}(r)}{W} \left [ B +
\int^{r_2}_{r} dr'\, \frac{T_{\ell m}(r')R^{\rm \infty}_{\ell m}(r')}{\Delta^{2}}  
\right ]
\ee 
where $ A,B,r_1, r_2 $ are constants. The relevant physical boundary conditions (purely outgoing/ingoing fields
at infinity/horizon respectively) can be imposed only after the integrals have been regularised. This can be achieved by integrations
by parts \cite{poisson}. The resulting integrals are finite when we take the limits $ r_2 \to \infty  $ and $r_1 \to r_{+} $ 
and the divergent surface terms are then absorbed into the constants $A,B$ which are eventually eliminated imposing the above
boundary conditions. 

Eccentric orbits formally have well-defined integrals and therefore can be described using the original solution (\ref{Rfield}). 
However, for $e$ close to unity the source integrals would grow (diverging at the $e \to 1 $ limit). This remark is relevant for 
the Teukolsky equatorial codes of Refs.~\cite{cutler},\cite{kgdk} which are based on (\ref{Rfield}). High-eccentricity tests
have revealed that even for an $ e \sim 0.9 $ orbit there is a significant error induced by the growth in the source-integral.
Therefore, for studying high-eccentricity and/or parabolic orbits one should either use the Sasaki-Nakamura solution (\ref{Rfield2}), 
or properly regularise the Teukolsky source integral \cite{szedenits}.


\subsection{Time-domain calculations}
\label{tdomain}

Computing gravitational waveforms and fluxes via the solution of the
Teukolsky equation in the frequency domain has been the `traditional' approach to the
problem for nearly thirty years, since the pioneering work of Davis {\it et al} \cite{davis} and 
Detweiler \cite{detweiler1}. This is understandable, as the full separability that the
Teukolsky equation enjoys in the frequency domain essentially reduces the problem to the integration 
of two ordinary differential equations, eqns.~(\ref{radTeuk}),(\ref{slm}). 

Working in the frequency domain indeed appears to be the appropriate computational tool for the EMRI problem
since the frequency content of the radiation field is strictly harmonic (when geodesic motion is assumed). 
However, when the orbit is significantly different than circular equatorial, the emitted waveform is the sum
of a large number of orbital frequency harmonics, especially for large eccentricity and/or inclination angle. 
In addition, strong-field motion requires the inclusion of several multipoles ($\ell_{\rm max} \sim 10 $). Orbits with
these properties are among those of relevance to LISA. For example, in the early stages of the inspiral all 
EMRI systems are expected to have $e \approx 1 $ in which case any frequency domain code would require  
substantial computational resources.  

This has motivated some authors to approach the problem from a different perspective, that is, 
working directly in the time domain \cite{tdomain1},\cite{tdomain2},\cite{khanna},\cite{martel1}. In the time domain, 
the Teukolsky equation is given by (\ref{master}). Only the $\varphi$ coordinate admits separation, hence we write
\be
\psi = \sum_{m=-\infty}^{+\infty}\,\Psi_{m}(t,r,\theta)\,e^{im\varphi}, \quad 
{\cal T} = \sum_{m=-\infty}^{+\infty}\, {\cal T}_{m}(t,r,\theta)\,e^{im\varphi}
\ee 
Then (\ref{master}) can be rewritten in the form,
\be
-\frac{\partial^2 \Phi_m}{\partial t^2} - A(r,\theta)\frac{\partial\Phi_m}{\partial t}
+ B^{2}(r,\theta)\,\frac{\partial^2\Phi_m}{\partial r_{\ast}^2} + \frac{C}{\sin\theta}\,
\frac{\partial}{\partial \theta}\left ( \sin\theta \frac{\partial \Phi_m}{\partial \theta}  \right )
- V(r,\theta)\, \Phi_m = {\cal T}_m
\label{t_teuk2}
\ee
where $\Phi_m = \Psi_m\sqrt{(r^2+a^2)\,\Delta^s} $. The functions $A,B,C,{\cal T}_m$ can be found 
in Ref.~\cite{tdomain2}.

Initial numerical investigations of the Teukolsky equation in the time domain was carried out   
by Krivan {\it et al} \cite{tcode} for the {\it homogeneous} version of equation (\ref{t_teuk2}).
Formulated as an initial value problem, these Teukolsky time-evolutions reproduced results for 
Kerr quasi-normal modes and late-time tails, already established by frequency domain calculations. 

Time evolutions of the inhomogeneous Teukolsky eqn.~(\ref{t_teuk2}) were first performed by Lopez-Aleman,
Khanna and Pullin \cite{tdomain1,tdomain2}, for equatorial Kerr orbits. The code used in these studies
is  based on the original $2+1$ homogeneous Teukolsky code of Ref.~\cite{tcode}, appropriately modified to encompass 
a test-body source term. 

The time-domain representation of a test-body source term is a particularly delicate problem. The reason is 
the presence of $\delta-$functions (and their derivatives) which are singular at the body's radial and 
angular instantaneous location (the $\delta(\phi-\phi(t))$ function can be treated analytically). 
A first cut approximation is to smear out the $\delta-$ functions by replacing them with
narrow Gaussians (with width $\sigma $ of only few grid points) \cite{tdomain1},\cite{tdomain2}, 
\be
\delta(x^a -x^a(t)) \approx \frac{1}{\sqrt{2\pi}\sigma}\,\exp[-(x^a-x^a(t))^2/2\sigma^2 ]
\label{delta}
\ee 
Typically, the radial grid is much more dense ($ \sim \mbox{few} \times 10^3 $ points) than
the $\theta-$ grid ($ \sim 50-100 $ points ).   

In a time domain evolution, fluxes and waveforms are extracted by monitoring the field at some large distance
(and close to the horizon for the horizon fluxes). The relevant flux formulae are \cite{campanelli},
 \bear
\left (\frac{dE}{dt} \right )_{\rm GW} &=& \lim_{r\to \infty}\, \left [ \frac{1}{4\pi r^6}\, 
\int d\phi\, d\theta\, \sin\theta \left | \int_{-\infty}^{t} dt^\prime\,
\psi(t^\prime,r,\theta,\phi) \right  |^2  \right ]
\\
\nonumber \\
\left (\frac{dL_{z}}{dt}  \right )_{\rm GW} &=& \lim_{r\to \infty}\, \left [ \frac{1}{4\pi r^6}\, Re \left \{
 \int d\phi\, d\theta \,\sin\theta \, \int_{-\infty}^{t} dt^\prime \partial_{\phi} \psi(t^\prime,r,\theta,\phi) \times
\int_{-\infty}^{t} dt^\prime \int_{-\infty}^{t^\prime} d\tilde{t}\,\bar{\psi} (\tilde{t},r,\theta,\phi) \right \}   \right ]
\eear 
As initial data the field is taken to be zero which is, strictly speaking, an inconsistent choice as it corresponds
to a test-body appearing from nothing. This anomaly is manifested as an artificial, propagating, radiation 
burst at the beginning of the simulation (see Fig.~\ref{fig_twav} below). Hence, one has to allow for sufficient time 
to pass in order for the computational grid to decontaminate . At the same time, the extent of the radial grid 
must be large enough as to prevent any spurious backscattering from reaching the `observation' point, 
for a predetermined time window. 

The first results obtained in \cite{tdomain1},\cite{tdomain2} are quite promising as they generally agree within 
$\sim 20\%$ with the accurate frequency domain results of \cite{kgdk},\cite{finn}, despite the crude 
representation (\ref{delta}) for the source term, and the two-dimensional nature of the problem.
A sample of flux data, taken from Ref.~\cite{tdomain2}, is provided in Table~\ref{tab_tflux1}. 
  In Fig.~\ref{fig_twav} we show a representative time domain waveform (also taken from Ref.~\cite{tdomain2}) for the
case of an equatorial eccentric orbit with $p= 4.64M, e = 0.5$. The black hole spin is $a=0.9M$ and the 
observation point is located at $r=150M,~\theta =\pi/2$.
  
\begin{figure}
\centerline{\includegraphics[height=7cm,clip]{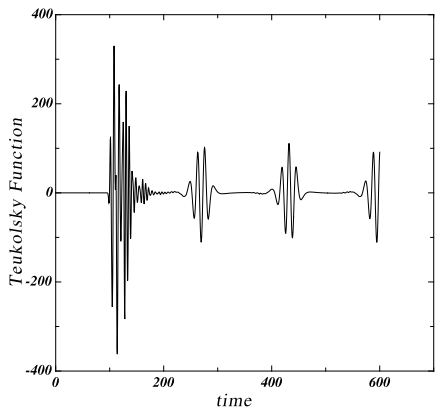}}
\caption{Time evolution of the real part of the Teukolsky function for $m=3$ and $\mu/M = 0.01 $, for a test body 
in a $p= 4.64M, e = 0.5$ equatorial orbit, around a $a=0.9M$ Kerr black hole. The observation point is located
at $r=150M,~\theta=\pi/2$. Note the strong initial burst which is an artifact resulting from imposing initial
data inconsistent with the presence of a test body.}
\label{fig_twav} 
\end{figure}

\begin{table}
\centering
\begin{tabular}{|c||c||c|c|} 
\hline
m mode & $\dE$  (Time domain) & $\dE$ (Frequency domain) & Frac. diff. \cr 
\hline \hline
$1$ &      $2.9\times10^{-10}$ &  $2.8\times10^{-10}$ & 0.036 \cr \hline
$2$ &      $1.4\times10^{-07}$ &  $1.8\times10^{-07}$ & 0.22  \cr \hline
$3$ &      $4.2\times10^{-08}$ &  $5.3\times10^{-08}$ & 0.21  \cr \hline   
\end{tabular}
\caption{Kerr time domain flux data for the same equatorial orbit as in Figure~\ref{fig_twav}.}
\label{tab_tflux1}
\end{table}

The second existing time-domain EMRI study is the one by Martel \cite{martel1}, and is focused
on bound and parabolic Schwarzschild orbits. The separation of both $\phi$ and $\theta$ dependence results in a 
$1+1$ numerical scheme. This translates to a superior computational performance and smaller
discretization-induced errors, as compared to the $2+1$ Kerr calculation.

Instead of the Teukolsky equation, Martel is evolving the inhomogeneous Zerilli and Regge-Wheeler equations 
which describe polar (even) and axial (odd) metric perturbation components, respectively. Both these 
equations have the well-known form \cite{chandra},
\be
\frac{\partial \psi_{\ell m}}{\partial t^2} -\frac{\partial \psi_{\ell m}}{\partial r_{\ast}^2} 
+ V(r)\psi_{\ell m} = S_{\ell m}
\label{Z_RW}
\ee
with a source term,
\be
S_{\ell m} = G_1(t,r)\,\delta(r-r(t)) + G_2(t,r)\,\delta^{\prime}(r-r(t))
\ee
The functions $G_1,G_2$ can be found in \cite{martel1}. The numerical method used in evolving eqns.~(\ref{Z_RW}) 
is described in \cite{martel2}, \cite{carlos} and includes a state-of-art discrete representation of the source 
term's $\delta-$function and its derivative. Martel's results are in excellent agreement 
(at the level of $\lesssim 1\%$, see Table~\ref{tab_tflux2}) with the corresponding frequency domain results, 
and not surprisingly, significantly more accurate compared to the Kerr time domain results. 

\begin{table}
\centering
\begin{tabular}{|c|c|c|c|c|}
\hline
orbit & flux& freq. domain& time domain & frac. diff. \cr
\hline \hline
$p=7.50478$ & $~\dE~$ & $3.1680 \times 10^{-4}$ & $3.1770 \times 10^{-4}$& $0.3\%$  \cr 
$e=0.188917$& $~\dL~$ & $5.9656 \times 10^{-3}$ & $5.9329 \times 10^{-3}$& $0.5\%$  \cr 
\hline\hline
$p=8.75455$ & $~\dE~$ & $2.1008\times 10^{-4}$&$ 2.1484\times 10^{-4}$& $2.3\%$ \cr 
$e=0.764124$& $~\dL~$ & $2.7503\times 10^{-3}$ & $2.7932\times 10^{-3}$& $1.6\%$  \cr \hline
\end{tabular} 
\caption{Comparison of averaged fluxes for eccentric Schwarzschild orbits obtained in the time domain
\cite{martel1}, and the frequency domain \cite{cutler}.}
\label{tab_tflux2}
\end{table}
Apart from several tables of flux data, Ref.~\cite{martel1} also presents several time-domain waveforms 
for bound and parabolic orbits ($e \leq 1$); these share the same qualitative features with the waveform 
of Fig.~\ref{fig_twav}, hence we do not present them here.  

Computing {\em averaged} fluxes in the time domain involves certain error, unlike in the frequency
domain where the averaging is performed in an exact analytic manner. For circular and small eccentricity orbits
the orbital periods are short enough (in comparison with the total duration of the time evolution)  
to allow for averaging over several orbits. In such cases the induced error is small.  
On the other hand when the eccentricity is substantial the radial orbital period becomes considerably
longer, resulting in a growth of the averaging error. This situation is evident in the sample data of
Table~\ref{tab_tflux2}. Hence, high eccentricity orbits require long time evolutions
in order to obtain high precision flux data. Meanwhile, as pointed out in \cite{martel1}, the waveform calculation
has no such problems, as no time averaging is involved.     

As we mentioned, the main goal of the time domain approach is to
be able to deal (primarily) with highly eccentric orbits, for which the standard frequency domain
formalism becomes computationally expensive. For the Schwarzschild case there is no doubt that,
indeed, the time domain approach is superior, partially because time evolving $1+1$ wave equations is easily
handled by modern computers and partially because there is an accurate representation for the singular test-body source term 
\cite{carlos}. For the more relevant Kerr problem, frequency domain calculations still lead the race, at least until Kerr time 
domain calculations reach the level of accuracy of the corresponding Schwarzchild codes. Indeed, there are ongoing programs
that aim towards that direction (for example \cite{pablo}), interfacing at the same time with self-force computations.


\section{The hybrid approximation}
\label{hybrid}

In Section~\ref{evolv} we discribed a simple algorithm for constructing the orbital inspiral
within the framework of the Teukolsky formalism. It is clear that this kind of calculation is computationally 
expensive as one needs to compute $\{ \ddp, \de, \di \}$ at each timestep. Using template banks of such 
Teukolsky-based inspiral waveforms seems as an unnecessary complication for current approximate LISA data analysis studies.
Instead, it would be highly desirable to have at hand an alternative (albeit approximate) `quick and dirty' 
scheme for computing the orbital evolution and the associated gravitational signal. 

Such a scheme is the `hybrid approximation' \cite{GHK} which is the topic of this Section. Essentially, it is a 
resurrection and extension of the older `semi-relativistic' approximation by Ruffini and Sasaki \cite{semi},
and is based on the following two-level idea:

At the first level, in calculating the radiative inspiral we replace the fluxes  $\{ \dE, \dL \} $  with approximate 
Post-Newtonian analytical expressions $\{ \dE^{PN}, \dL^{PN}\} $  (hence removing the most computationally demanding 
component of the Teukolsky calculation). For inclined orbits one has to handle the extra flux $\dQ$. This is not a problem for 
the special case of circular-inclined orbits where, as we have discussed, $\dQ$ is fully of the other two fluxes. 
However, for generic orbits, the calculation of $\dQ$ is elusive without resorting to the gravitational self-force.
It is still possible, however, to come up with certain approximations for $\dQ$. We further expand on this issue below. 
 
In calculating the rates $ \{ \ddp, \de \} $ we use the {\em exact} functional dependence on the fluxes, i.e.
\bear
\ddp &=& H^{-1} ( b_p\, \dE + c_p\, \dL + d_p\, \dQ ) \quad \Rightarrow \quad  
\ddp = H^{-1} ( b_p\, \dE^{PN} + c_p\, \dL^{PN} + d_p\, \dQ )
\nonumber \\
\de &=& H^{-1} ( b_e\,\dE + c_e\,\dL + d_e\, \dQ ) \quad \Rightarrow \quad  
\de = H^{-1} ( b_e\,\dE^{PN} + c_e\,\dL^{PN} + d_e\, \dQ )
\label{dpdedi}
\eear 
Note that the quantities $H, b_{p,e}, c_{p,e}, d_{p,e} $ encode the exact geodesic motion
and are just combinations of derivatives of $\{ E(p,e,\iota),L_{z}(p,e,\iota), Q_(p,e,\iota) \} $, 
see Appendix~\ref{app:formulae}. For $\di$ we can write the simpler expression,
\be
\di = \frac{L_z}{2\sqrt{Q} (L_z^2 +Q)}\, \left [ \dQ -\frac{2Q}{L_z}\dL \right ]
\label{di}
\ee

The inspiral $\{p(t),e(t),\iota(t) \}$ in the space of orbital elements is then 
constructed by an Eulerian algorithm, similar to the one given by eqns.~(\ref{euler}).  
The inspiral in the physical coordinate space $\{r,\theta,\phi\}$ results from the combination of
the full general relativistic geodesic equations (which are numerically integrated) and 
$\{p(t),e(t),\iota(t)\}$. With this prescription, the small body's world-line is described by
the four-vector,
\be
z^a(t) = z^a_{\rm geod}[\,t;E(t),L_{z}(t),Q(t)]
\label{traject}
\ee 
where $E(t) = E[p(t),e(t),\iota(t)]$, etc. 

At the second level of the hybrid scheme, we can compute `kludge' waveforms by assuming that the small body is
moving along the trajectory (\ref{traject}) (with the option of taking into account radiative backreaction or not) 
like a `bead on a wire' in a fictitious {\em flat} spacetime, pretending that Boyer-Lindquist coordinates are true spherical 
coordinates. Effectively, calculating the waveform becomes a problem of wave dynamics for metric perturbations in flat-spacetime 
(see \cite{MTW} for in depth discussion). Strictly speaking, this prescription is not part of a self-consistent
approximation framework; it is rather more like a `black box' designed to produce waveforms with accurate phasing. This
is the most important property of a template waveform, and is crucially sensitive to the precision of the
orbital motion. Indeed, the resulting waveforms agree well with the accurate Teukolsky waveforms. More details on the 
actual procedure and examples of kludge waveforms are given in the following Section.


\subsection{Equatorial inspirals and waveforms}
\label{hybrid_equat} 

We first apply the hybrid method to equatorial orbits. Since $\dQ= Q= \di =\iota =0  $ the rates (\ref{dpdedi}) reduce
to eqns.~(\ref{rates_eq}). A good first choice for the weak-field fluxes are the expressions derived by Ryan \cite{ryan2} for 
generic Kerr orbits. These are leading-order Post-Newtonian (PN), amended with the leading order spin term 
(which appears at 1.5 PN order), and correspond to a `Keplerian plus leading spin effect' description of the orbital motion. In fact,  
at this level of accuracy, Ryan was able to use the leading-order self force (discussed in \cite{MTW}) 
and derive a leading order result for $\dQ$ (see Section~\ref{Q}). Rewriting Ryan's fluxes in terms of our 
orbital elements yields (here we keep $\iota$ arbitrary as the same fluxes will be used for non-equatorial inspirals),
\bear
\dE &=& -\frac{32}{5}\frac{\mu^2}{M^2} 
\left ( \frac{M}{p} \right )^{5} (1-e^2)^{3/2} 
\left [ f_{1}(e)  - \frac{a}{M} \left ( \frac{M}{p} \right )^{3/2} 
f_{2}(e)\,\cos\iota \right ]
\label{dE_R}
\\
\dL &=& -\frac{32}{5}\frac{\mu^2}{M} 
\left (\frac{M}{p} \right )^{7/2}
(1-e^2)^{3/2} \left[ g_{1}(e)\, \cos\iota + \frac{a}{M} 
\left ( \frac{M}{p} \right )^{3/2}
\left[g_{2}(e) - g_{3}(e)\,\cos^2 \iota \right] \right]
\label{dL_R} 
\eear
The functions $f_{1}(e),~f_{2}(e),~g_{1}(e),~g_{2}(e),~g_{3}(e)$ are given in Appendix~\ref{app:coeffs}. 
In the limit $a=0$, eqns.~(\ref{dE_R}) and (\ref{dL_R}) reduce to the classic Peters-Mathews formulae {\cite{pm}}.

It is instructive to compare the hybrid inspirals with the inspirals generated
by approximating the entire $ \ddp, \de, \di $ formulae. At leading order we have,
\bear
\ddp &\approx& -\frac{64}{5} \frac{\mu}{M} (1-e^2)^{3/2}
\left ( \frac{M}{p} \right )^{3} \left ( 1 + \frac{7}{8}e^2  \right )
\label{dpN}
\\
\nonumber \\
\de &\approx& -\frac{304}{15} \frac{\mu}{M^2}\, e\,(1-e^2)^{3/2}
\left ( \frac{M}{p} \right )^4 \left ( 1 + \frac{121}{304} e^2 \right )
\label{deN}
\eear
These can be combined to give,
\be
p(e) = p_i  \left ( \frac{e}{e_i}  \right)^{12/19} \left [ \frac{1 + 
(121/304)\,e^2}{1 + (121/304)\,e_{i}^{2}} 
\right ]^{870/2299}
\label{peN}
\ee
where $p_i$ and $e_i$ are arbitrary initial values.

\begin{figure}
\centerline{\includegraphics[height=8cm,clip]{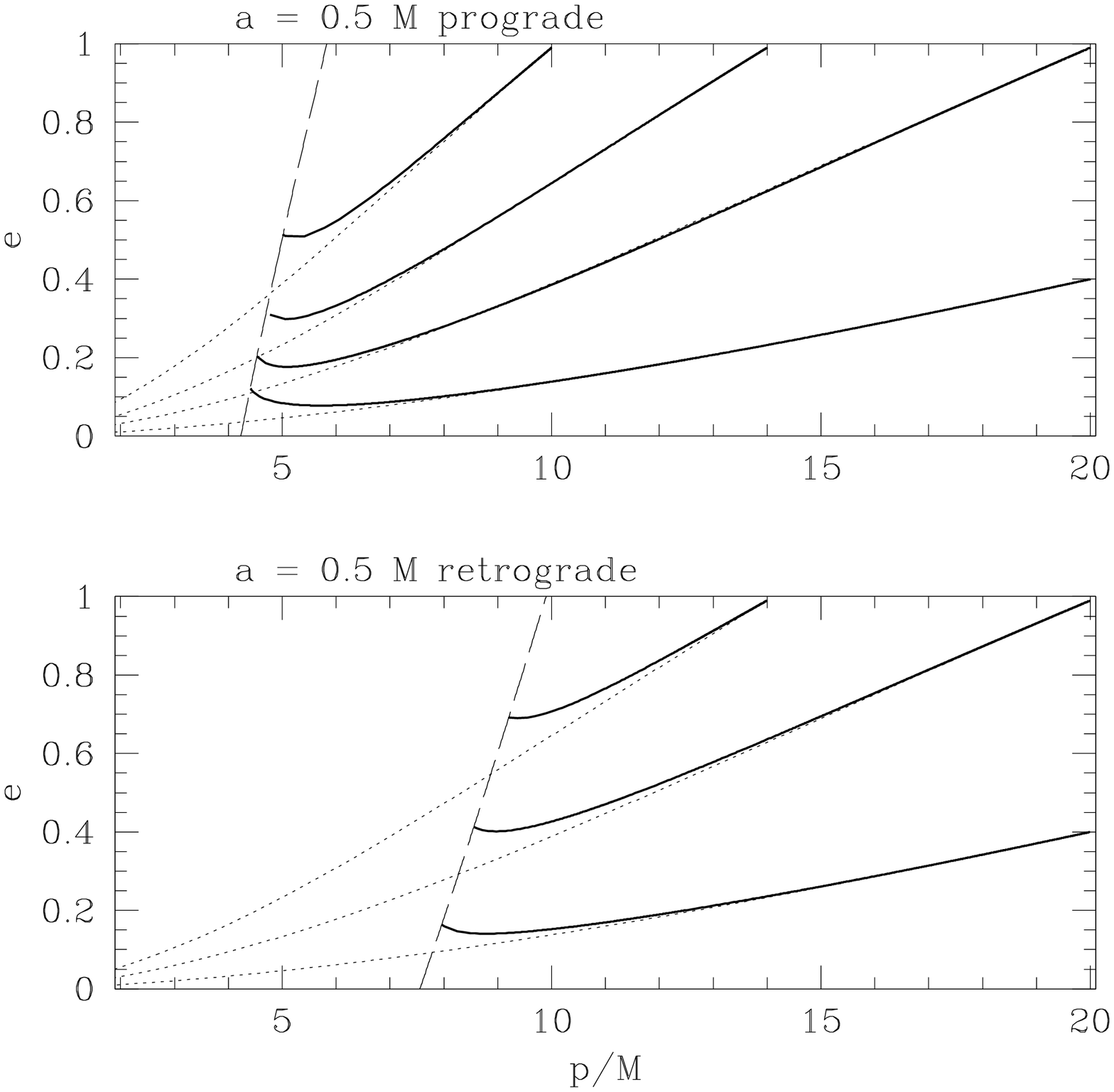}
\includegraphics[height=8cm,clip]{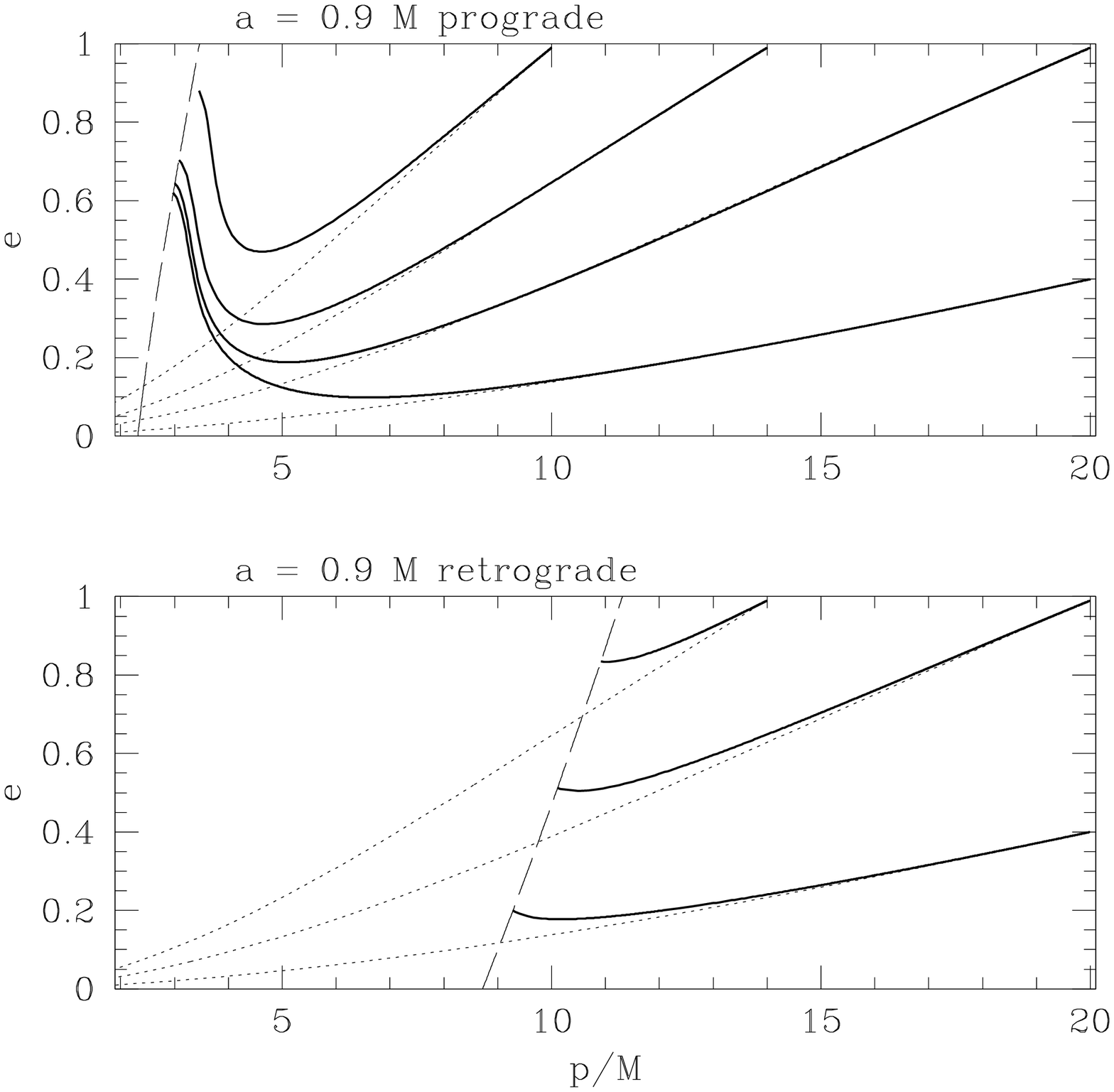}}
\caption{Examples of hybrid Kerr equatorial inspirals, into a hole with spin $a = 0.5 M$ (left panel) and 
into a hole with $a = 0.9M$ (right panel).  In each panel, the top half shows prograde inspirals and the bottom retrograde 
inspirals. In each set, the dashed line represents the separatrix separating stable from unstable orbits. The hybrid 
approximation is used to radiatively evolve orbits with initial parameters $(p_i, e_i) = (20M, 0.4)$, $(20M, 0.99)$, $(14M, 0.99)$, 
and $(10M, 0.99)$. The hybrid inspiral trajectories are shown as the solid lines in each plot (The final set is not included in 
the retrograde inspirals since the initial conditions are not stable in those cases.)  The dotted trajectories in each plot 
shows the leading-order predictions generated using eqn.~(\ref{peN}).  Note the significant qualitative difference between 
the two calculations at the vicinity of each separatrix. Note also the extremely large growth in eccentricity seen in the 
prograde inspirals for $a = 0.9M$ just before reaching the separatrix. Comparison with accurate strong-field numerical 
results shows that this growth is excessive.  Note that the time dependence of the inspiral is suppressed in this figure: most time 
is actually spent at large $p$.  The total duration of an inspiral scales with $M^2/\mu$.  The shape of a curve, however, does 
not depend on this ratio, provided that the mass ratio is extreme; these curves are universal for $\mu\ll M$.}
\label{fig_hybrid_equat}
\end{figure}

A sample of representative results for astrophysically relevant initial parameters is shown in 
Fig.~\ref{fig_hybrid_equat}.  We compare the leading-order trajectories (\ref{peN}) with the trajectories predicted by the 
hybrid scheme. The inspirals we show correspond to both prograde and retrograde orbits, for black hole spins $a = 0.5M$ 
and $a= 0.9M$. In the weak-field region $p \gg M$ the hybrid and the leading-order calculations agree well, as expected.
Crucial differences between the two methods become apparent in the strong-field regime, during the inspiral's 
final stages (this is also the epoch most relevant for LISA's observations). As is clear from 
eqn.~(\ref{deN}), the leading-order inspiral trajectory exhibits constantly decreasing eccentricity. This is not 
what the rigorous strong-field calculations, both numerical and analytical \cite{apostolatos,cutler,kenn98,kgdk}, 
predict. As we discussed in Section~\ref{equat}, there exists a region near the separatrix of stable/unstable 
orbits where $\dot{e}$ reverses sign: the eccentricity should grow near the separatrix. This feature is indeed 
present in the hybrid inspirals. Moreover, the location of the critical points where $\dot{e} = 0$ is 
in good agreement (at the order of few percent) with the numerical results of Refs.~\cite{cutler},\cite{kgdk}; 
see Ref.~\cite{GHK} for actual data. Detailed comparison of the $\{\ddp, \de\}$ values against accurate 
Teukolsky results establishes the superiority of the hybrid scheme over the leading order 
formulae (\ref{peN}). The same conclusion holds even if we consider higher order PN expansions 
for $\ddp,\de$. A sample of representative data, taken from Ref.~\cite{GHK} is provided in Table~\ref{tab_GHK}.

\begin{table}
\begin{tabular}{|c|cc|c|c|c|c|c|}
\hline\hline
$a/M$ & $p/M$ & $e$ & Calculation & $(M/\mu)\dot{p}$ & Frac.\ diff.\
in $\dot p$ & $(M^2/\mu)\dot{e}$ & Frac.\ diff.\ in $\dot e$ \\
\hline
0 & 7.505 & 0.189 & Numerical     & $-7.475\times10^{-2}$ &  ---   &
$-1.967\times10^{-3}$ &  ---   \\
 &  &  & Hybrid        & $-6.859\times10^{-2}$ & 0.0824 &
$-1.291\times10^{-3}$ & 0.3434 \\
 &  &  & Leading order & $-2.957\times10^{-2}$ & 0.6044 &
$-1.159\times10^{-3}$ & 0.4108 \\
\hline
0 & 6.9 & 0.4 & Numerical     & $-4.240\times 10^{-1} $& --- &
$ + 1.047 \times 10^{-2} $ & --- \\
 &  &  & Hybrid    &  $-3.056\times 10^{-1} $  & 0.2792    & 
$+1.506\times10^{-2}$  & -0.4384 \\
 &  &  & Leading order    & $ -3.420\times10^{-2}$   & 0.9193  &
$ -2.929\times10^{-3}$  & 1.2797 \\
\hline
0.5 & 6.5 & 0.4 & Numerical     & $-5.999\times10^{-2}$ &  ---   &
$-5.155\times10^{-3}$ &  ---   \\
 &  &  & Hybrid        & $-4.606\times10^{-2}$ & 0.2322 &
$-3.356\times10^{-3}$ & 0.3490 \\ 
 &  &  & Leading order & $-4.091\times10^{-2}$ & 0.3181 &
$-3.719\times10^{-3}$ & 0.2786 \\
\hline
0.5 & 15 & 0.4 & Numerical     & $-3.371\times10^{-3}$ &  ---   &
$-1.341\times10^{-4}$ &  ---   \\
 &  &  & Hybrid        & $-3.358\times10^{-3}$ & 0.0039 &
$-1.334\times10^{-4}$ & 0.0052 \\
 &  &  & Leading order & $-3.328\times10^{-3}$ & 0.0128 &
$-1.311\times10^{-4}$ & 0.0224 \\
\hline
0.5  & 4.8   & 0.3 & Numerical     & $-6.354\times 10^{-1}$ &  ---   &
$+1.369\times10^{-2}$ & --- \\
  &    &  & Hybrid     & $-4.858\times10^{-1} $ & 0.2354    &
$ +3.519\times10^{-2} $ & -1.5705    \\
  &    &  & Leading order & $-4.849\times10^{-2} $ & 0.9237 &
$-4.432\times10^{-3} $ & 1.3237    \\
\hline
0.9  & 5   & 0.4 & Numerical     & $-7.507\times 10^{-2}$ &  ---    &
$-9.266\times 10^{-3}$ & --- \\
  &    &  & Hybrid     & $ -4.617\times10^{-2} $ & 0.3850    &
$ -1.965\times10^{-3}$ & 0.7879  \\
  &    &  & Leading order & $ -2.698\times10^{-3} $ & 0.9641 &
$-9.732\times10^{-5}$  & 0.9895    \\
\hline
-0.99 & 10.5 & 0.4 & Numerical     & $-7.516\times10^{-2}$ &  ---   &
$-5.312\times10^{-4}$ &  ---   \\
 &  &  & Hybrid        & $-5.506\times10^{-2}$ & 0.2674 &
$-5.223\times10^{-4}$ & 0.0168 \\
 &  &  & Leading order & $-9.704\times10^{-3}$ & 0.8709 &
$-5.461\times10^{-4}$ & 0.0281 \\
\hline\hline
\end{tabular}
\caption[Table]
{Comparing the rates $\dot{p}$, $\dot{e}$ for some Kerr equatorial-eccentric orbits.  The fifth column in this 
table shows $\ddp$; the seventh column shows $\de$.  Within each section of the table, the
first row of columns five and seven contains accurate numerical data from {\cite{cutler,kgdk}}, the second row shows 
data using the hybrid scheme and the third row shows data using leading-order results. The sixth and eighth columns 
show the fractional differences between the two approximation schemes and the accurate numerical results.  In all cases but one, 
the hybrid approximation is closer to the accurate numerical calculation, sometimes substantially so.}
\label{tab_GHK}
\end{table}

Naturally, the hybrid approximation has its limitations. This can been seen, for example,
in Fig.~\ref{fig_hybrid_equat}. Three of the four cases shown appear `good' in the sense
that the trajectories appear to agree reasonably well with what we expect based on strong-field numerical analyses 
\cite{kgdk}. However, things clearly go wrong in the fourth case  ($a = 0.9M$, prograde); both the eccentricity growth 
near the separatrix and the distance of the critical curve $\dot{e}=0$ from the separatrix are excessive. This behaviour 
is not surprising as prograde orbits of rapidly rotating black holes reach rather deep into the black hole's strong field where 
the weak-field fluxes (\ref{dE_R}) and (\ref{dL_R}) cannot be trusted. As a rule of thumb, and
based on the data of Ref.~\cite{GHK} the hybrid method fails when $r_p \lesssim 5M$. This effectively constrains 
the black hole spin to $a\lesssim 0.5M$ for prograde motion. For retrograde orbits the hybrid approximation is 
favoured, since $r_p$ never comes close to the horizon, regardless of the spin.

An important prediction of the hybrid approximation is that for equatorial orbits the residual eccentricity prior to 
plunge should be substantial, in strong contrast to the prediction of the leading order formula (\ref{peN}). In many cases, 
the leading order results predict that the orbit will actually circularise prior to plunge. Because the frequency structure 
of a nearly circular inspiral is rather different from that of an inspiral with substantial eccentricity, these results 
have strong implications for the waveform models to be used in LISA's data analysis.

It is possible to further improve the performance of the hybrid scheme by simply
using more accurate fluxes for $\dE$ and $\dL$. As pointed out, Ryan's fluxes include only
the leading (1.5PN) spin-dependent term; at the same time they are fully accurate with respect to eccentricity.
Higher order (up to 2.5PN order) fluxes have been derived by Tagoshi~\cite{tagoshi_ecc}, 
under the assumption of small eccentricity (see Ref.~\cite{chapter} for a complete collection 
of PN fluxes in the extreme mass ratio limit). In practice this is not a serious restriction as 
for the majority of inspirals the eccentricity is $\lesssim 0.5 $ at the stage where the orbit resides in 
the black hole's strong field regime. At the same time, during the early stages of the inspiral we have 
$r_p \gg M $ and the higher PN corrections are small, irrespective of the eccentricity. 
However, the $e-$dependent factor in the leading PN part of the fluxes {\it is} important, but fortunately, 
this factor is fully known from the work of Peter \& Mathews \cite{pm}.    

Hence, the optimal choice for the hybrid approximation would seem to be a combination of 
Ryan's and Tagoshi's fluxes. In terms of our orbital elements, this union gives 
(full details will appear in \cite{GG}),
\bear
\dE &=& -\frac{32}{5} \frac{\mu^2}{M^2} \left(\frac{M}{p}\right)^5 
(1-e^2)^{3/2}\left [ f_1(e) -\frac{a}{M}\left(\frac{M}{p}\right)^{3/2} f_2(e) 
-\left(\frac{M}{p}\right) f_3(e) +
\pi\left(\frac{M}{p}\right)^{3/2} f_4(e) -\left(\frac{M}{p}\right)^2 f_5(e) \right.
\nonumber \\
&& \left. + \left (\frac{a}{M} \right)^2\left(\frac{M}{p}\right)^2 f_6(e) -
\pi\left(\frac{M}{p}\right)^{5/2} f_7(e)
+ \frac{a}{M}\left(\frac{M}{p}\right)^{5/2} f_8(e) \right ] ,\
\label{dE_new}
\nonumber \\
\\
\dL &=& -\frac{32}{5} \frac{\mu^2}{M} \left(\frac{M}{p}\right)^{7/2} 
(1-e^2)^{3/2}
\left [ g_1(e) -\frac{a}{M}\left(\frac{M}{p}\right)^{3/2} \{ g_{2}(e) -g_{3}(e) \} 
-\left(\frac{M}{p}\right) g_{4}(e) + \pi\left(\frac{M}{p}\right)^{3/2} g_{5}(e) \right.
\nonumber \\
&& \left. -\left(\frac{M}{p}\right)^2 g_{6}(e) 
+  \left (\frac{a}{M} \right)^2 \left(\frac{M}{p}\right)^2 g_{7}(e) -\pi\left(\frac{M}{p}\right)^{5/2} g_{8}(e)
+ \frac{a}{M} \left(\frac{M}{p}\right)^{5/2} g_{9}(e) \right ] ,\
\label{dL_new}
\eear
where the various additional coefficients are listed in Appendix~\ref{app:coeffs}. 
These fluxes include terms up to 2.5 PN order. In fact, some testing reveals that it is the 2PN fluxes 
that actually have the superior performance. This well-known result is related to the asymptotic nature of the 
PN expansion, see \cite{chapter}. 

\begin{figure}
\centerline{\includegraphics[height=7cm,clip,angle=-90]{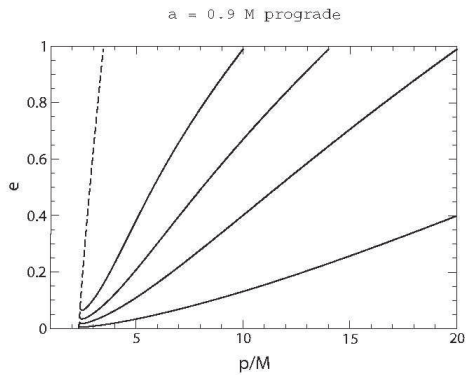}}
\caption{Improved equatorial inspirals generated with the help of the new fluxes (\ref{dE_new}),(\ref{dL_new}) additionally 
modified according to the low eccentricity fix discussed in the text. These inspirals are to be compared with
the corresponding $ a=0.9 M $ prograde inspirals of Fig.~\ref{fig_hybrid_equat}. The dashed curve represents the 
separatrix.}
\label{fig_hybrid_equat2}
\end{figure}

Another crucial improvement concerns the behaviour of the inspirals in the $e \approx 0 $ region. Although not 
particularly visible in the examples of Fig.~\ref{fig_hybrid_equat}, a closer examination of that portion of the
parameter space reveals that the hybrid approximation predicts a growth of $\de$ instead of the correct
$ \de \sim e  $ behaviour \cite{chapter},\cite{apostolatos},\cite{kenn98},\cite{ori} for nearly circular orbits.   
It is possible to rectify this problem by adding certain terms in the fluxes \cite{webpage},\cite{GG}. Since the same  
pathology also appears when $\iota \neq 0 $, we provide more details about this fix in our discussion on
generic inspirals. Modifying accordingly the fluxes (\ref{dE_new}),(\ref{dL_new}), we can generate improved hybrid equatorial
inspirals, as the ones shown in Figure~\ref{fig_hybrid_equat2}. We have chosen to display $a=0.9M$ prograde
inspirals, a case where the original hybrid calculation was clearly inaccurate. 

The new inspirals look clearly more `sane' all the way to the separatrix, devoid of the excessive eccentricity growth seen in 
Fig.~\ref{fig_hybrid_equat}. Note that the residual eccentricity prior to plunge is $ e \lesssim 0.1 $ not much different from 
what the leading order inspirals would predict. As we have mentioned elsewhere, for prograde orbits around rapidly spinning 
black holes, the $ \de >0 $ region is small and the $\de =0 $ curve is located near the separatrix, hence eccentricity 
decreases almost for the entire inspiral. 

Having discussed the first level of the hybrid approximation (construction of inspirals) we move on to the
second level, the generation of kludge waveforms \cite{kludge_paper}. We only present examples where backreaction is 
ignored, in which case it is straightforward to compare with the available Teukolsky-based equatorial waveforms of
Ref.~\cite{kgdk}. The starting point is the quadrupole formula for the flat-spacetime, trace-free metric perturbation 
$\bar{h}_{ab} = h_{ab} -\frac{1}{2} n_{ab} h$,
\be
\bar{h}^{jk} = \frac{2}{r}\,\ddot{I}^{jk}(t-r) \quad \mbox{where} \quad
I^{jk}(t) = \int dV' x'^{j}x'^{k}\, T^{00}(t,{\bf x'}), \quad (i,j,k = \{1,2,3\})
\label{quad_h}  
\ee  
is the binary's quadrupole moment and $T^{\mu\nu}$ is the small body's Minkowskian stress-energy tensor. All integrations are 
performed over the source. According to the hybrid prescription,
\be 
z^a(t) = [t,r(t),\theta(t),\phi(t) ]_{\rm BL} \Rightarrow z^a(t) = 
[t,x^1=r(t)\,\sin\theta(t)\,\cos\phi(t), x^2 = r(t)\,\sin\theta(t)\,\sin\phi(t), x^3= r(t)\cos\theta(t) ] 
\ee
From $\bar{h}^{jk} $ one can extract the transverse-traceless components $h_{+},h_{\times} $ in a standard
way \cite{MTW}. 

\begin{figure}
\centerline{\includegraphics[height=7cm,clip]{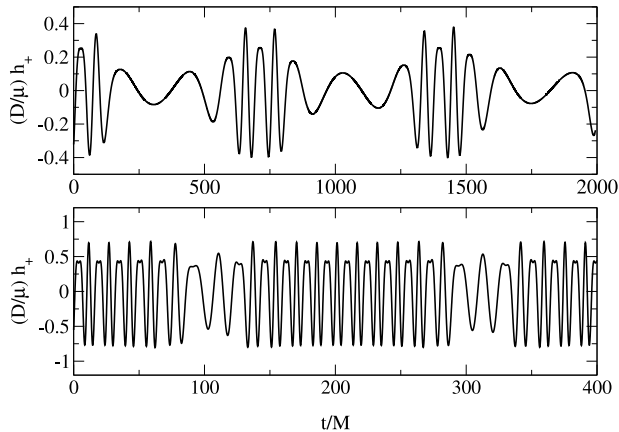}}
\caption{Sample of hybrid `kludge' waveforms for the same Kerr equatorial orbits as in Fig.~\ref{fig_wav1} for a $ a=0.99M $
Kerr hole. For the retrograde orbit $ p =10.5 M,~e = 0.5 $ (top panel) there is good agreement in both phase and amplitude 
with the analogue Teukolsky waveform. For the strong-field prograde orbit (bottom panel) $ p = 1.8M,~e=0.4 $ the two waveforms
agree well only with respect to phase. Higher multipoles and backscattering, two factors not taken into account
in the hybrid calculation, play an important role in this example.}
\label{fig_kludge1}
\end{figure}

It is straightforward to enhance the accuracy of the hybrid waveforms by including additional source multipole moments.
For example we can use the following quadrupole-octupole formula \cite{press},\cite{bekenstein},
\be
\bar{h}^{jk} = \frac2{r} \left[ \ddot{I}^{jk} - 2 n_a \ddot{S}^{ajk} +
n_a \dddot{M}^{ajk}\right]_{t-r}
\label{octu}
\ee 
where  
\be
S^{ajk} = \int dV' x'^{j}x'^{k} T^{0a}(t,{\bf x'}) \quad \mbox{and} \quad   
M^{ajk} =  \int dV' x'^a x'^j x'^k T^{00}(t,{\bf x'})
\label{moments2}
\ee
are the current-quadrupole and mass-octupole moments and $ n^i = x^i/r $. Indeed, it is possible to take into account all 
multipole moments by employing the `fast' solution found by Press \cite{press77}, but it turns out that the simpler 
quadrupole-octupole waveforms show (in most cases) little deviation from the Press waveforms. 
In Fig.~\ref{fig_kludge1} we present a small sample of kludge waveforms (for Kerr equatorial orbits) and compare them with the 
corresponding Teukolsky-based waveforms. Typically, the agreement between the two calculations is surprisingly good, despite the
simple-minded construction of the hybrid waveforms. The comparison can be quantified in terms of the overlap function between 
the waveforms; in the present case the resulting overlaps are very close to unity (see Ref.~\cite{kludge_paper} for more details). 
The use of exact geodesic motion ensures high precision phasing for the waves, even in strong-field conditions. The agreement in 
amplitude is less impressive (mainly due to the negligence of higher multipole contributions and backscattering effects) but  
this has no significant impact on the data-analysis performance of the hybrid waveforms. We need to point out that the 
high degree of phase accuracy will {\em not} be preserved when the radiative trajectory (\ref{traject}) is used (assuming the 
waveform is monitored for sufficiently long timescale $\sim T_{\rm RR}$), nevertheless, hybrid waveforms still remain valuable tools. 
A detailed quantitative discussion on the accuracy/reliability of hybrid waveforms will appear in Ref.~\cite{kludge_paper}. 

The punchline of this discussion is that hybrid waveforms have two key properties: they (i) can be very easily generated, for any 
Kerr orbit, and (ii) compare well to the existing rigorous Teukolsky-based waveforms. Hence, without much doubt, they can be used 
to produce a `leading-order' waveform template bank for LISA. Indeed, the most recent estimates on the expected EMRI event rate 
were computed with the aid of hybrid waveforms \cite{rates}.

\subsection{The elusive $\dQ $ flux}
\label{Q}

Attempting to apply the hybrid approximation to inclined orbits, one immediately faces the issue of
calculating the Carter constant flux $\dQ$. Unfortunately, unlike $\dE$ and $\dL$ 
this quantity cannot be inferred by monitoring the radiation field at infinity.
Instead, in order to compute $\dQ$, one needs knowledge of the local self-force (see eqn.~(\ref{ko3}) below). 
In fact, one of the main motivations behind the self-force program has been the calculation of this elusive flux. 
As we already mentioned, Ryan \cite{ryan2} was able to derive the following leading-order result for $\dQ$ making use 
of the known leading-order self force \cite{MTW}:  
\be
\dQ_R = -\frac{64}{5} \mu^3 \left (\frac{M}{p} \right )^3
(1-e^2)^{3/2} \sin^2\iota \left [ g_{1}(e) -\frac{a}{M}\left ( \frac{M}{p} \right)^{3/2} 
 f_{9}(e) \cos\iota \right ] 
\label{dQ_R}
\ee 
with $ f_{9}(e) = 85/8 + (211/8)e^2 + (517/64)e^4 $.       
Currently, this is the only available PN result for $\dQ$. A different approximation
emerged from the study of radiation reaction for circular-inclined Kerr orbits \cite{scott_circ},
which we reviewed in Section~\ref{nequat}. For this special orbital family,  
$\dQ$  can be expressed as a linear combination of $\dE$, $\dL$ in an exact sense; no self-force 
is required. One of the main results of Ref.~\cite{scott_insp} is that the inclination angle $\iota$ 
remains almost fixed during the radiative inspiral, even for strong-field motion.
In Appendix~\ref{app:almostsphere} (taken from Ref.~\cite{GHK}) we provide an intuitive explanation
as to why the $\iota = \mbox{const}$ rule should be a good approximation for {\it all} generic Kerr orbits. 
  
It follows from (\ref{di}) that this rule yields,
\be
\dQ_{\rm sph} = \frac{2Q}{L_z}\,\dL
\label{dQsphere}
\ee 
It is important to note that this `spherical' rule is in fact {\it exact} in a spherically
symmetric spacetime, like Schwarzschild. In such spacetime, the Carter constant is nothing
more than the projection of the angular momentum on the equatorial plane, i.e. 
$Q_{\rm sph} = L_x^2 + L_y^2 $. Then it can be shown that an inspiral in this spacetime would
proceed at exactly constant inclination angle, gravitational
waves remove the amounts of $L_x$ and $L_y$ required to hold $\iota$ fixed.      
    
Expanding now (\ref{dQsphere}) at the same order as Ryan's expression (\ref{dQ_R}) we 
get,
\be
\dQ_{\rm sph} \approx -\frac{64}{5} \mu^3 \left (\frac{M}{p} \right )^3
(1-e^2)^{3/2} \frac{\sin^2\iota}{\cos\iota} \left [ g_{2}(e)\,\cos\iota  + \frac{a}{M}
\left ( \frac{M}{p} \right)^{3/2} \{ g_{2}(e) - f_{9}(e)\,\cos^2\iota \} 
\right ] ,\
\label{dQsphere2}
\ee
At leading order this coincides with eqn.~(\ref{dQ_R}), but there is disagreement 
in the spin-dependent term. Since by construction (\ref{dQ_R}) is fully accurate
at ${\cal O}(a) $ order, we conclude that the corresponding order term in (\ref{dQsphere2})
is incomplete. The missing piece is,
\be
\delta_{Q} \equiv \dot{Q}_{R} - \dot{Q}_{\rm sph} = \frac{64}{5}\mu^3 
\left ( \frac{M}{p}  \right )^{9/2} \frac{a}{M}\,\frac{\sin^2\iota}{\cos\iota}
g_{2}(e) (1-e^2)^{3/2}
\ee
This piece is also the leading `aspherical' contribution to $\dQ$. As such, we would
expect,
\be
\dQ = \dQ_{\rm sph} + \delta_Q = \frac{2Q}{L_z}\dot{L}_z  
+ \frac{64}{5}\mu^3 \left ( \frac{M}{p}  \right )^{9/2} 
 \frac{a}{M}\,\frac{\sin^2\iota}{\cos\iota} \,g_{2}(e) (1-e^2)^{3/2}
\label{newdQ}
\ee
to be an improvement over the spherical formula (\ref{dQsphere}), see Ref.~\cite{GG}. 
Instead of keeping $\iota$ constant, this new formula predicts an increase,
\be
\di = \frac{L_z \dot{Q} -2Q\dot{L}_z}{2\sqrt{Q}(Q+L^2_z)} =
\frac{32}{5}\frac{\mu^3}{(Q+L^2_z)} \left ( \frac{M}{p} \right )^{\frac{9}{2}}
\frac{a}{M}\, \sin\iota \, (1-e^2)^{3/2} \, g_{2}(e) > 0  
\label{idot_new}
\ee 
This is a welcome feature, in qualitative agreement with the accurate Teukolsky 
results of Section~\ref{nequat}. 

Before moving on, we should mention another weak-field approximation for $\dQ$, the first in chronological order, which 
was proposed by Shibata \cite{shibata_circ} for the case of Kerr circular-inclined orbits (before the formulation of the 
circularity theorem). In terms of $L_{\pm} \equiv L_{x} \pm i L_{y} $ and their averaged fluxes $\dot{L}_{\pm}$ ,
\be
\dQ \approx \langle L_{+} \rangle\,\dot{L}_{+} + \langle L_{-} \rangle \, \dot{L}_{-} -2a^2\,
E \langle \cos\theta \rangle \,\dE
\ee
where the time average is taken over a period $T_\theta$. This expression (which nowdays is obsolete), together with 
exact Teukolsky-based fluxes $\dE$,$\dL$ allowed Shibata to compute the backreaction to the orbit (in addition to 
waveforms), with results similar to Hughes' \cite{scott_circ}.   
 
At this point we have at hand three different approximate expressions for $\dQ$, eqns.~(\ref{dQ_R}),(\ref{dQsphere}),(\ref{newdQ})
which can be incorporated in the hybrid scheme and construct generic Kerr inspirals (Section~\ref{hybrid_nequat}). 

Further insight into $\dQ$ is provided by considering the {\it exact} formula for the instantaneous evolution of 
the Carter constant as given by Kennefick \& Ori \cite{ori}, see Appendix~\ref{app:KennOri}. This time, $Q$ is expressed in the form
(\ref{Q_form2}) which leads to, 
\be
\frac{dQ}{d\tau} = {\cal G}_{,E} \frac{dE}{d\tau} + {\cal G}_{,L_z} \frac{dL_z}{d\tau} 
+ 2\,p_\theta\, F_\theta = -2a^2 E \cos^2\theta~\frac{dE}{d\tau}  
+ 2 L_z \cot^2 \theta~ \frac{dL_z}{d\tau} + 2\,p_\theta\, F_\theta
\label{ko3}
\ee
This is of course equivalent to expression (\ref{ko1}). It is an interesting exercise to derive from 
eqn.~(\ref{ko3}) the spherical rule (\ref{dQsphere}). The following argument is taken from Ref.~\cite{GG}. 
In the $a\rightarrow 0$ limit, the first term vanishes, the Carter constant 
reduces to the (square of the) projection of angular momentum on the equatorial plane, $Q= L^2 -L_{z}^{2}$ 
and the inclination angle $\iota$ becomes the true inclination of the orbital plane, since 
$\tan^2\iota=Q/L_{z}^{2}$. Taking the y-axis to lie in the orbital plane, without loss of generality, 
the Boyer-Lindquist coordinates of the particle at any point of the orbit obey the relation
\be
f(\theta,\phi) = \cos\theta - \sin\theta\,\cos\phi\,\tan\iota = 0 
\ee
Symmetry ensures that the self-force has no component perpendicular to the orbital plane and 
therefore
\be
u^a\,n_a = F^a\,n_a = 0 \qquad \Rightarrow \qquad 
\frac{F^{\theta}}{F^{\phi}} = \frac{u^{\theta}}{u^{\phi}} =
\sin^{2}\theta \,\sqrt{\tan^{2}\iota-\cot^{2}\theta} 
\ee
In this, $n_{\alpha} = \partial f/\partial x^{\alpha}$ is the normal to the orbital plane. 
The third term in equation (\ref{ko2}) thus becomes,
\be
2\,p_\theta\,F_\theta = 2\,F_\phi\,\left [ \tan^2\iota -\cot^2\theta \right ]^{1/2}\,p_\theta
= 2\,p_\phi\,F_\phi\,\left [ \tan^2\iota -\cot^2\theta \right ] 
\ee
But, $L_z = p_\phi$ and $F_{\phi} = dL_z /d\tau$ (Appendix~\ref{app:KennOri}) and (\ref{ko3}) becomes
\be
\frac{dQ}{d\tau} = 2\,(\cot^{2}\theta+\tan^{2}\iota-\cot^{2}\theta)\,L_{z}\,
\frac{dL_z}{d\tau} = 2\,\tan^{2}\iota\,L_{z}\,\dL =\frac{2\,Q\,}{L_z} \frac{dL_z}{d\tau}
\ee
which brings (\ref{ko3}) to the form (\ref{dQsphere}). The above equation is valid 
for both instantaneous and time-averaged fluxes. Turning on the spin makes 
(\ref{ko3}) depart from its spherical value and consequently $\iota$ to change. 
It is easy to deduce that some leading order terms (with respect to the spin) will be 
provided by the third term in (\ref{ko3}). This is because the leading order change in 
$F_\theta$ is linear in $a$, as can be deduced by inspecting Ryan's expressions for the 
self-force \cite{ryan1}. The first term, on the other hand, contributes only at ${\cal O}(a^2)$ 
and beyond.

One final approximate $\dQ$ for generic zoom-whirl orbits has been proposed in Ref.~\cite{GHK}.
As we discussed back in Section~\ref{equat}, equatorial zoom-whirl orbits radiate energy and angular
momentum at such rates as $ \dE \approx \Omega_\phi\,\dL $. This is a manifestation of the fact that
a large portion of the orbital period is spent at $r \approx r_p $, with the small body whirling in a 
quasi-circular path. A similar behaviour is observed for inclined zoom-whirl orbits (in which case
$T_r \gg  T_\theta, T_\phi$), hence it is natural to put forward the conjecture that that these orbits 
also radiate in a `circular' manner. That this could be true is suggested by the Kennefick-Ori
formula (\ref{ko1}). For a zoom-whirl orbit and for motion near the periastron, $r\approx
r_p$, so we should have $u^{r} \approx 0$; consequently, the unknown third term in (\ref{ko1}) 
should be negligible. Then, the conjecture is that the resulting expression for $\dQ$
\be
\dQ \approx {\cal H}_{,E} \dE + {\cal H}_{,L_z} \dL + {\cal O}(u^r)
\ee
describes the evolution of the Carter constant for all generic zoom-whirl orbits and with increasing accuracy as 
the orbit approaches the separatrix. We emphasise that this approximation should hold even for orbits deep in the 
black hole's strong-field. This conjecture could become a practical tool once a code that calculates $\dot{E}$ 
and $\dot{L}_z$ for generic orbits is developed. Furthermore, a direct comparison with (\ref{dQsphere}) 
should be a useful guide for the accuracy of the $\iota=\mbox{constant}$ rule in strong-field
situations.  Future computation of the self force will provide the ultimate test for both approximations.


\subsection{Non-equatorial inspirals and waveforms}
\label{hybrid_nequat}

\begin{figure}
\centerline{\includegraphics[height=8cm,clip]{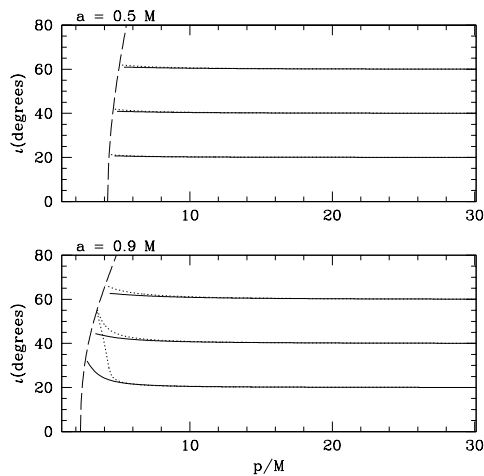}}
\caption{Circular, inclined inspiral from Ref.~\cite{GHK}. The black hole spin is $a = 0.5M$ (top) and 
spin $a = 0.9M$ (bottom). The solid lines show inspiral using the hybrid approximation; the
dotted lines show the leading order inspiral prediction.  The dashed curve represents the separatrix. }
\label{fig_insp_c}
\end{figure}

Armed with the above approximate expressions for $\dQ$ it is possible to generate inspirals
of generic Kerr orbits. These results are particularly interesting, as there are still no analogue 
Teukolsky-based results. As a warm up we first discuss circular-inclined inspirals, for which
the circularity theorem (Section~\ref{nequat}) \cite{chapter},\cite{ryan1},\cite{ori} guarantees that 
they will evolve as a sequence of circular orbits, and fixes $\dQ$. Consequently, for these orbits the 
rates (\ref{dpdedi}),(\ref{di}) become,
\bear
\ddp &=& H_{\rm circ}^{-1} ( -L_{z,\iota} \dE + E_{,\iota}
\dL )
\\
\di &=& H_{\rm circ}^{-1} ( L_{z,p} \dE - E_{,p}\dL )
\\
\de &=& e = 0 
\label{rates_c}
\eear
where $H_{\rm circ} = E_{,\iota} L_{z,p} - L_{z,\iota} E_{,p}$. The leading-order expression for 
$\ddp$, $\di$ are \cite{ryan2},
\be
\ddp = -\frac{64}{5} \frac{\mu}{M} \left ( \frac{M}{p} \right )^{3}, \qquad 
\di = \frac{244}{15} \frac{\mu}{M^2} \frac{a}{M} \left ( \frac{M}{p} \right )^{11/2} \sin\iota 
\label{leading_c}
\ee
Some examples of inclined circular inspirals are shown in Figure~\ref{fig_insp_c}, and some concrete data are provided in 
Table~\ref{tab_insp_c}. According to these results, the hybrid approximation outperforms the leading-order calculation. 
Both approximations predict that the inclination angle increases, especially close at the separatrix. However, the increase 
predicted by eqn.~(\ref{leading_c}) is far too large, particularly for rapidly spinning black holes. The inspiral predicted by the 
hybrid approximation is closer to what is seen in rigorously computed inspirals \cite{scott_circ}. Nonetheless, it too shows 
an increase in $\iota$ that is probably excessive. After all, holding $\iota$ constant produces an inspiral sequence that is 
probably closest of all to strong-field calculations and should be acceptably accurate.

\begin{table}
\begin{tabular}{|c|cc|c|c|c|c|c|}  
\hline
$a/M$ & $p/M$ & $\iota({\rm deg})$ & Calculation &
$(M/\mu)\dot{p}$ & Frac.\ diff.\ in $\dot p$ & $(M^2/\mu) \dot{\iota}$
& Frac.\ diff.\ in $\dot\iota$ \\
\hline
0.95 & 7 & 62.43 & Numerical     & $-4.657\times10^{-2}$ &  ---   &
$1.207\times10^{-4}$ &  ---   \\
     &   &       & Hybrid        & $-4.497\times10^{-2}$ & 0.0344 &
$2.639\times10^{-4}$ & 1.1864 \\
     &   &       & Leading order & $-2.750\times10^{-2}$ & 0.4095 &
$3.080\times10^{-4}$ & 1.5518 \\
\hline
0.5     & 10 & 67.56  & Numerical    & $-1.583\times10^{-2} $ & --- &
$ 1.546\times10^{-5} $ & --- \\
     &  &   &  Hybrid    & $ -1.645\times10^{-2}  $ & 0.0392 &
$ 2.043\times10^{-5} $ & 0.3215  \\
     &   &  & Leading order & $ -1.194\times10^{-2} $ & 0.2457  &
$ 2.377\times10^{-5} $ & 0.5375 \\
\hline
0.5     & 10 & 126.76   & Numerical    & $ -2.329\times10^{-2}  $ & --- &
$ 1.892\times10^{-5}  $ & --- \\
     & &  & Hybrid    & $ -2.341\times10^{-2} $ & 0.0051  &
$ 1.643\times10^{-5} $ & 0.1316  \\
     &  &  & Leading order    & $ -1.414\times10^{-2} $ & 0.3929  &
$ 2.060\times10^{-5} $ & 0.0888 \\
\hline
0.9     & 10 & 74.07 & Numerical    & $ -1.544\times10^{-2}  $ & --- &
$ 2.715\times10^{-5} $ & --- \\
     & &  & Hybrid    & $  -1.567\times10^{-2} $ & 0.0149 &
$ 3.857\times10^{-5} $ & 0.4206 \\
     & &  & Leading order    & $-1.169\times10^{-2} $ & 0.2429 &
$ 4.452\times10^{-5} $ & 0.6398 \\
\hline
0.9     & 10 & 131.57    & Numerical    & $ -3.253\times10^{-2}  $ & --- &
$ 3.887\times10^{-5}  $ & --- \\
     &  & & Hybrid    & $ -3.082\times10^{-2}  $ & 0.0526 &
$ 2.612\times10^{-5} $ & 0.3280 \\
     &  &  & Leading order    & $ -1.545\times10^{-2} $ & 0.5250 &
$ 3.464\times10^{-5} $ & 0.1088 \\
\hline
0.5     & 6 &  67.81  & Numerical    & $ -2.020\times10^{-1} $ & --- &
$ 2.094\times10^{-4}$ & --- \\
     &  & & Hybrid    & $ -1.779\times10^{-1}   $ & 0.1193 &
$ 2.992\times10^{-4}  $ & 0.4288 \\
     &  &  & Leading order    & $ -5.082\times10^{-2} $ & 0.7484 &
$ 3.954\times10^{-4} $ & 0.8882 \\
\hline
0.9     & 6  & 54.64     & Numerical    & $ -7.846\times10^{-2}  $ & --- &
$ 2.015\times10^{-4}  $ & --- \\
     &  & & Hybrid    & $ -6.950\times10^{-2} $ & 0.1142 &
$ 5.486\times10^{-4} $ & 1.7226 \\
     &  &  & Leading order    & $ -3.598\times10^{-2} $ & 0.5674 &
$ 6.268\times10^{-4} $ & 2.1107 \\
\hline
\hline
0.9     & 6 & 99.55   & Numerical    & $ -74.32 $ & --- &
$ 6.337\times10^{-4}  $ & --- \\
     &  & & Hybrid    & $ -48.02 $ & 0.3539 &
$ 5.241\times10^{-4}  $ & 0.1729 \\
     &  &  & Leading order    & $ -6.593\times 10^{-2} $ & 0.9991 &
$ 7.580\times10^{-4} $ & 0.1961 \\
\hline
\end{tabular}
\caption[Table]
{Comparing the rates $\ddp$, $\di$ for Kerr circular-inclined orbits. The accurate numerical data are
from \cite{scott_circ}, the hybrid and leading-order results refer to eqns.~(\ref{rates_c}),(\ref{leading_c}),
respectively. Typically, the hybrid scheme performs much better than the leading-order approximation when
compared to the rigorous numerical data. The only case which this is not true is for $\di$ of retrograde orbits.  
Nevertheless, this small inaccuracy has no impact on the calculation of generic inspirals, as we assume that 
$\iota=\mbox{constant}$. Note the enormous difference between the numerical and the leading order
results in the Table's final entry.  This is because that point is fairly close to the separatrix between stable 
and unstable orbits. Since the leading-order calculation has no notion of this separatrix, it is particularly 
inaccurate here.}
\label{tab_insp_c}
\end{table}

The results presented so far (equatorial, circular-inclined) have established the reliability, as well 
as limitations of the hybrid scheme. Of course, the true purpose of the scheme is the construction 
of generic inspirals; we discuss them next. Generic inspirals are governed by formulae (\ref{dpdedi}),(\ref{di}).
Before the hybrid approximation, the only available result on generic Kerr inspirals was Ryan's 
weak-field analysis \cite{ryan2}. From that work we extract the $\dE,\dL$ fluxes (\ref{dE_R}),(\ref{dL_R}).
Similarly, the natural first choice for $\dQ$ is eqn.~(\ref{dQ_R}). However, as discussed
in Ref.~\cite{GHK}, the resulting inspirals are quite disappointing, particularly in the strong field, unless the
orbit is close to the one of the above special cases. The root of the problem lies in the use of eqn.~(\ref{dQ_R}), 
which apparently is not as accurate as we would require it to be. The qualitative behaviour of the inspirals is more 
reasonable when the spherical rule (\ref{dQsphere}) is employed, which also fixes $\di=0$. 
The discussion in Appendix~\ref{app:almostsphere} gives support to the conjecture that this rule should be a good 
approximation for generic inspirals.

\begin{figure}
\includegraphics[height=8cm,clip]{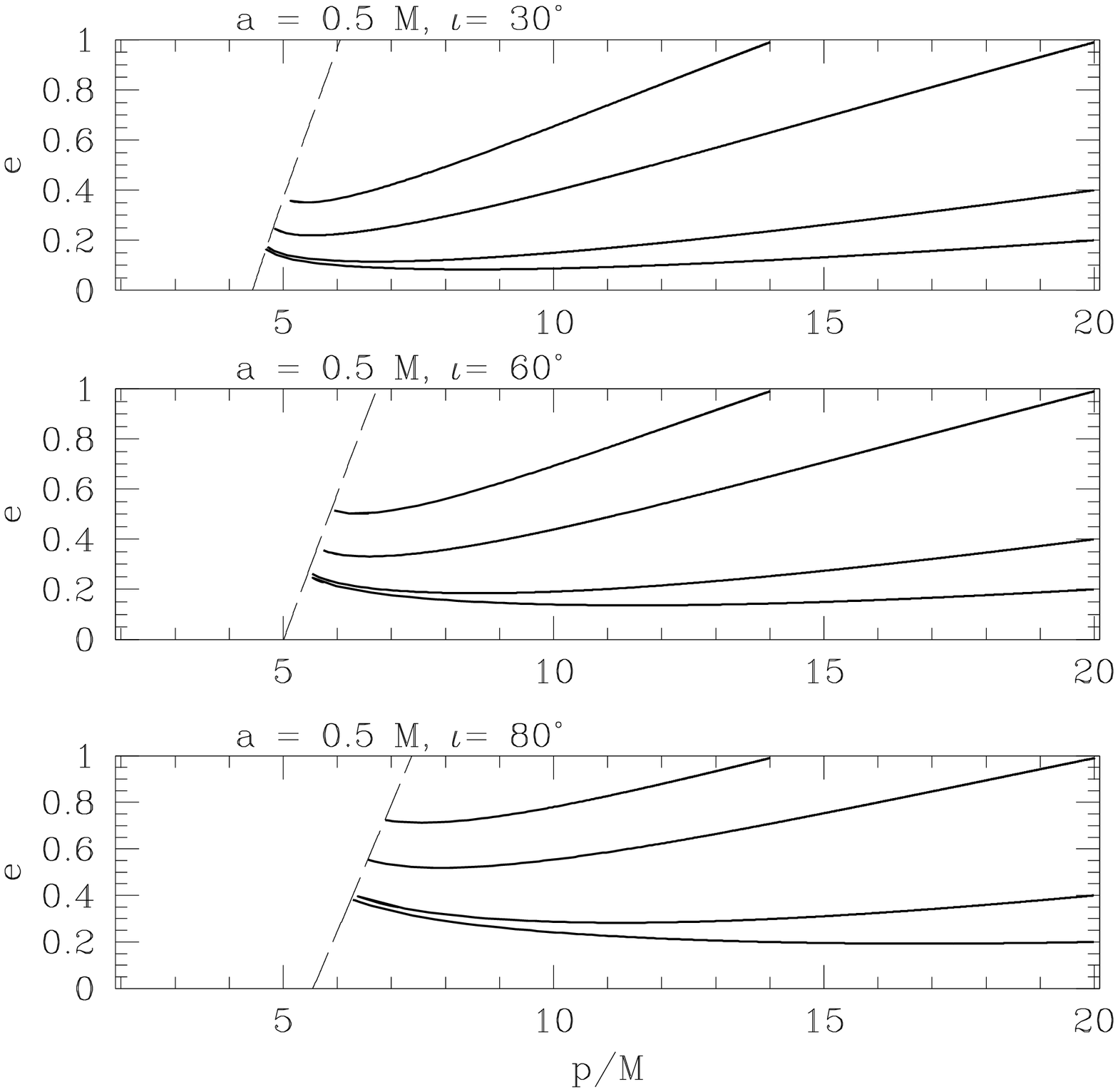}
\includegraphics[height=8cm,clip]{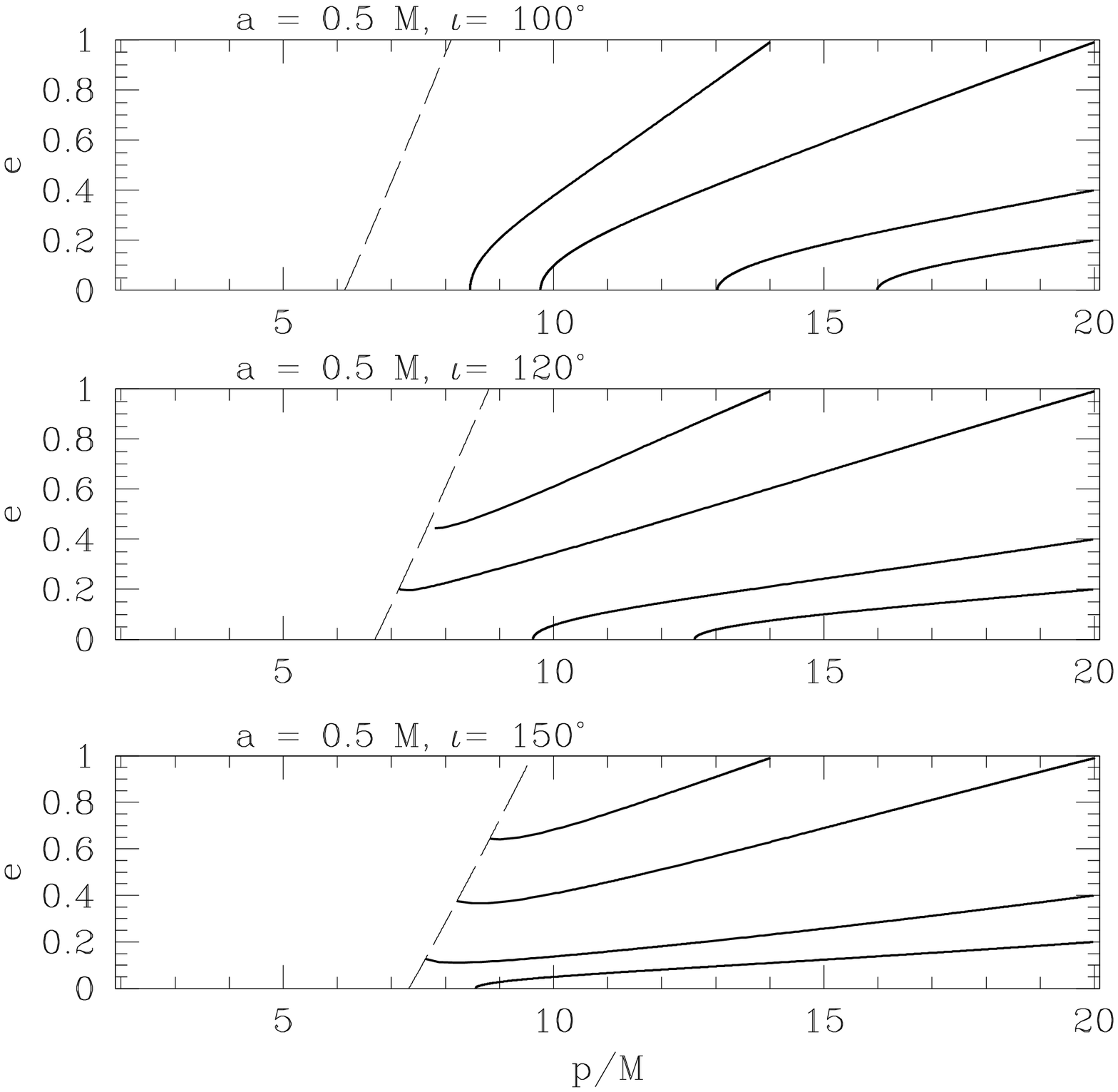}
\includegraphics[height=8cm,clip]{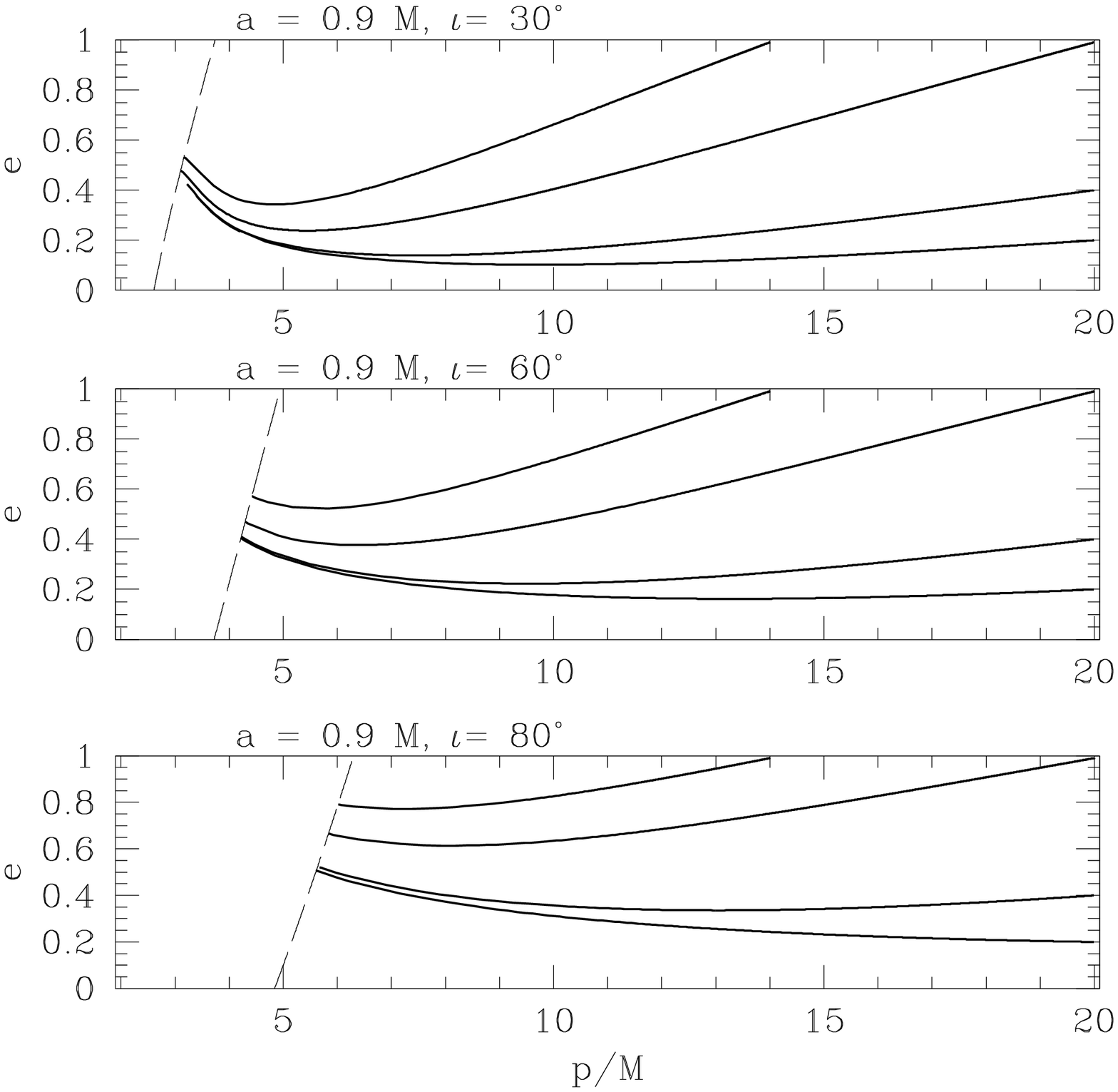}
\includegraphics[height=8cm,clip]{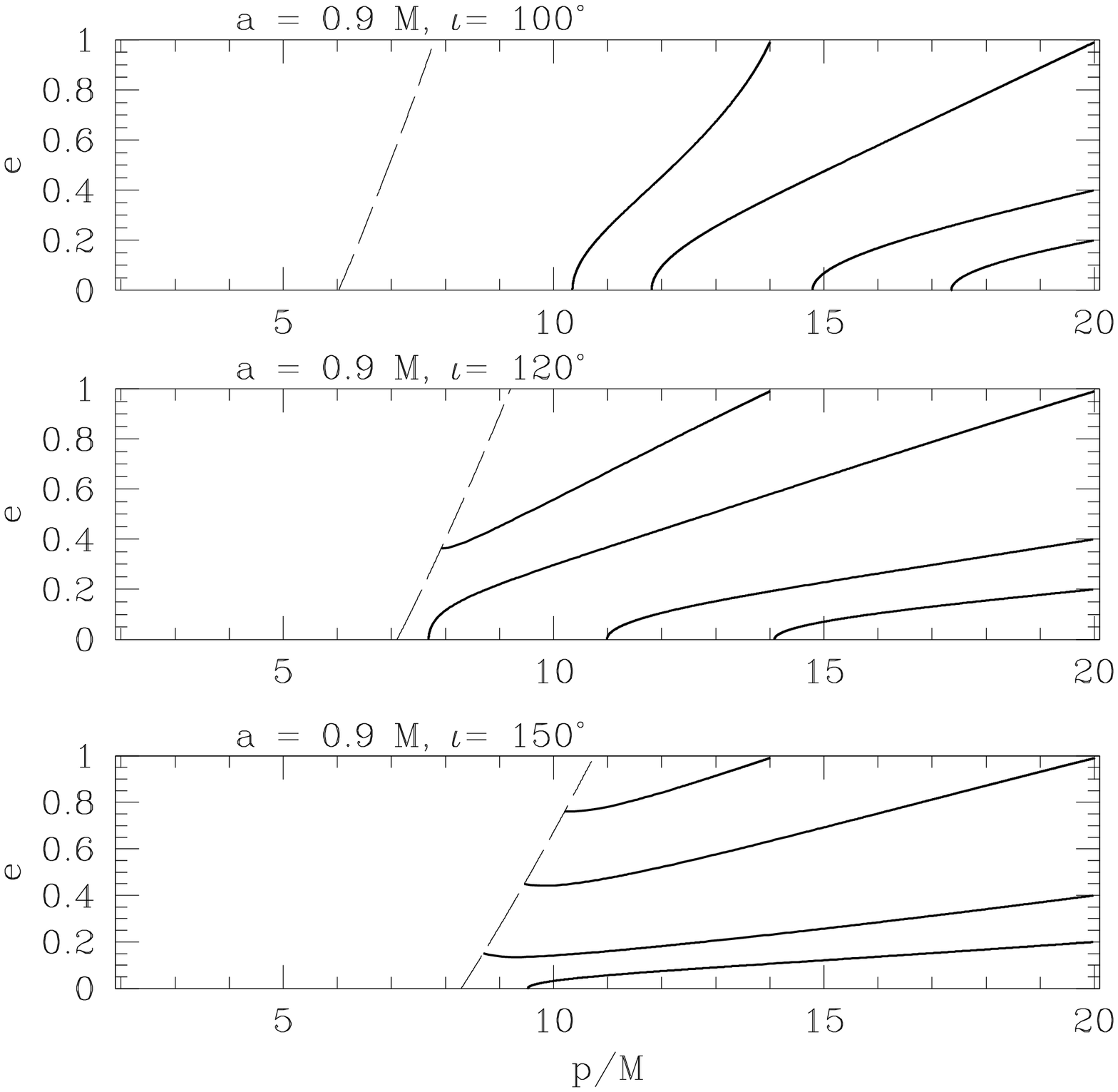}
\caption{Generic inspiral at several inclination angles into a hole with spin $a = 0.5M$ (top pair) and $a=0.9M $ 
(bottom pair).  In all plots, the dashed line represents the separatrix between stable and unstable orbits. 
As discussed in the text, the inspirals are forced to lie in surfaces of constant $\iota$.
The initial orbital parameters are $(p_i,e_i) = (20M, 0.2)$, $(20M, 0.4)$, $(20M, 0.99)$, and $(14M, 0.99)$,
and $\iota = 30^\circ$, $60^\circ$, $80^\circ$, $100^\circ$, $120^\circ$, and $150^\circ$.}  
\label{fig_insp_g1}
\end{figure}

\begin{figure}
\centerline{\includegraphics[height=9cm,angle=-90,clip]{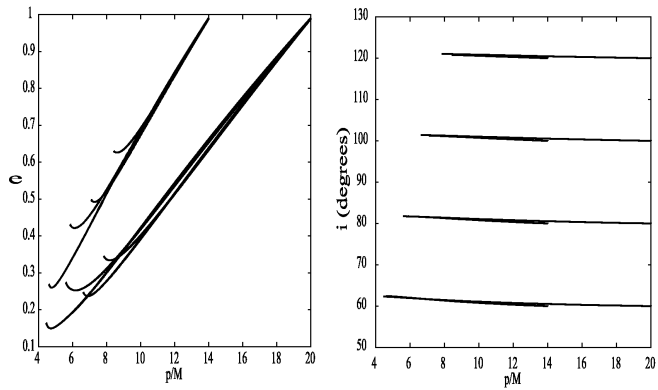}}
\caption{Improved hybrid Kerr generic inspirals, shown on the $p$-$e$ and $p$-$\iota$ planes (left and right panels 
respectively), for $a$=$0.8M$ spin. Initial orbital parameters are $e_i$=$0.99$ and $(p_i,\iota_i)$ = $(20M,60^\circ),(20M,80^\circ),
(20M,100^\circ),(20M,120^\circ),(14M,60^\circ),(14M,80^\circ),(14M,100^\circ),(14M,120^\circ)$. Each curve terminates at the
respective separatrix. These results can be compared with the earlier hybrid inspirals of Fig.~\ref{fig_insp_g1}. 
Note the significant residual eccentricity and the minor change in inclination.}
\label{fig_insp_g2}
\end{figure}

Using the fluxes (\ref{dE_R}) and (\ref{dL_R}) together with the $\di =0$ rule produces inspirals that agree with
the leading-order results when $p\gg M$, smoothly converge to the equatorial results for $\iota\to 0$ and 
$\iota \to \pi $, and that exhibit an $\dot{e} > 0$ region near the separatrix. All these are features we would
expect to be present in the true generic inspirals. Examples of generic inspirals are shown in Figs.~\ref{fig_insp_g1} 
\& \ref{fig_insp_g2}, where several interesting features can be seen. The trajectories for $\iota < 90^\circ$ are qualitatively 
similar to the equatorial, prograde trajectories shown in Fig.~\ref{fig_hybrid_equat}. In particular, each such trajectory passes 
through a critical point at which $\dot e = 0$ after which eccentricity grows. The system typically has substantial non-zero 
eccentricity when it reaches the separatrix. Also, note the excessive growth in eccentricity near the separatrix for $a = 0.9M$ 
and $\iota = 30^\circ$.  At shallow inclination angle, the separatrix is very deep in the black hole's strong field, so the 
inspiral proceeds to small $r$ before plunging.  Just as in the case of equatorial orbits for $a= 0.9M$, the weak-field flux formulae 
that we use cannot be trusted this deep into the Kerr strong field.

There are other regions of the $\{p,e,\iota\}$ space where the hybrid inspirals exhibit serious pathologies:
nearly polar orbits ($\iota \sim 90^\circ $) and nearly circular orbits ($ e\approx 0$) tend to circularise rapidly,
before reaching the separatrix. This is not a true physical effect; indeed, it is found that the behavior of inspirals exhibits a 
rather sharp transition as the inclination angle goes from slightly below $90^\circ$ to slightly above. 
For nearly circular inspirals we either observe a rapid circularisation (for $\iota > 90^\circ$) or an unphysically large radius 
where $\de$ changes sign. 

An array of improvements is available for generic inspirals \cite{GG}. First, we can use eqn.~(\ref{idot_new}) for
$\di  $. Second, we can build and utilise higher order PN fluxes, instead of
(\ref{dE_R}),(\ref{dL_R}), by combining three different PN results: Ryan's fluxes, Tagoshi's fluxes for equatorial orbits with 
small eccentricity \cite{tagoshi_ecc} and fluxes for circular-inclined orbits with small inclination, derived by Shibata {\em et al} 
\cite{shibata_iota}. The resulting fluxes can be found in Ref.~\cite{GG}.

Use of these `new' 2PN fluxes results in an overall improvement of the inspirals, especially
in the strong field regime. However, they do not ameliorate the pathological behaviour
near $\iota = 90^\circ  $ and $ e \approx 0 $. The root of the problem is found when 
we examine closer the $\de$, $\ddp$ expressions, eqns.~(\ref{dpdedi}). For the former we find
(a detailed analysis will appear in \cite{GG}), 
\be
\lim_{e \to 0} \de = \frac{1}{e}\,[ N_1(p,\iota)\,\dE + N_2(p,\iota)\, \dL + N_3(p,\iota)\, \dQ  ] + {\cal O}(e) 
\ee 
The exact fluxes would exactly eliminate the first divergent at $e=0$ term, i.e
\be
N_1(p,\iota)\,\dE(p,e=0,\iota) + N_2(p,\iota)\, \dL(p,e=0,\iota) + N_3(p,\iota)\, \dQ(p,e=0,\iota) = 0
\label{condition1}
\ee
leaving $ \de \sim e  $ which is a well known result for nearly circular orbits \cite{chapter},\cite{kenn98},\cite{ori}.
It is interesting to note that for equatorial orbits, the condition (\ref{condition1}) reduces to the familiar
$\dE = \Omega_\phi \dL $ relation. Neither Ryan's original fluxes nor the improved 2PN order fluxes 
satisfy this relation, which explains the bizarre behaviour seen for small eccentricity. Looking 
at (\ref{condition1}) a suitable fix would be,
\be
\dL^{\rm (new)} = \dL^{\rm (old)} -\left [ \frac{N_1}{N_2}\,\dE(p,e=0,\iota) + 
\frac{N_3}{N_2}\,\dQ(p,e=0,\iota) \right ]
\label{modify}
\ee     
while retaining the same $\dE,\dQ$. A similar fix for nearly polar orbits can be devised by studying the limit 
$\iota \to \pi/2 \ $ of $\de$. However, we can arrive at the same result in a more direct way by considering,
\be
\dQ = 2\,\sqrt{Q}\,\frac{\sin\iota}{\cos\iota}\,\left(\dL
+ \sqrt{Q}\,\frac{1}{\sin^{2}\iota}\,\dot{\iota} \right)
\ee
Since $\dQ$ must be well-behaved for $\iota \to \pi/2 $ it follows that,
\be
\lim_{\iota \to \pi/2} \left [ \dL +\sqrt{Q}\frac{\di}{\sin^2\iota} \right ]
\sim \cos\iota
\label{condition2}
\ee
This condition is again not exactly satisfied by our Post-Newtonian fluxes, hence
they need to be appropriately modified. As before, we can just modify $\dL$ as in (\ref{modify}).

Including all available improvements we produce some `final' generic inspirals, shown in Fig.~\ref{fig_insp_g2}.
These inspirals are well behaved throughout the $\{p,e,\iota\}$ space, exhibiting a reasonable size for the $\de > 0$
region and for the total change in $\iota $. They should be fairly good approximation of the true inspirals, but only the
comparison with future rigorous self force-based inspirals will decide if this statement is true.

\begin{figure}
\centerline{\includegraphics[keepaspectratio=true,height=2.5in,angle= 0,clip]{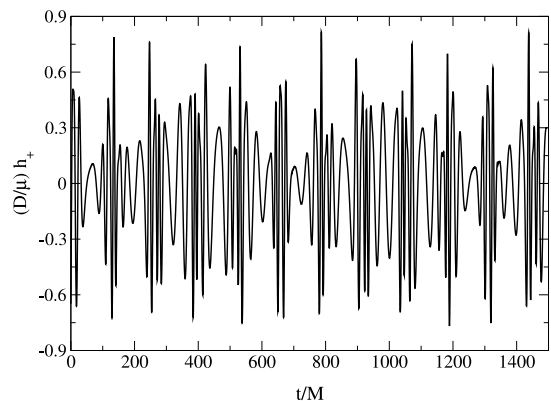}}
\caption{Kerr generic waveform calculated with the hybrid approximation, using the quadrupole-octupole
formula (\ref{octu}). The orbital parameters are $p=3.5 M,~e =0.4, ~\iota =40^\circ$ and the back hole spin $a=0.99 M $. 
Note the combined effects of zoom-whirl motion and Lense-Thirring precession in the waveform's pattern.}
\label{fig_wav_g}
\end{figure}

Our discussion on the hybrid approximation concludes with the presentation of a generic kludge waveform (see Fig.~\ref{fig_wav_g}), 
which is calculated according to the prescription given in Section~\ref{hybrid_equat}. The selected orbit is zoom-whirl, with
$p= 3.5 M$, $e =0.4$ and $\iota = 40^\circ $, for spin $a=0.99 M$. The waveform exhibits a combination of characteristics
already seen separately in equatorial and circular-inclined waveforms, and is an example of what LISA could actually
observe. 
   
\section{Mapping spacetimes with LISA}
\label{mapping}

It is universally accepted that the `dark' objects harboured in galactic nuclei are in fact black holes as
predicted by General Relativity. But it should also be acknowledged that this belief is partially
rooted in our faith to General Relativity itself, rather than direct astronomical evidence. So far,
it has not been possible to actually probe the gravitational field in the vicinity of these dark objects,
and determine whether it is described by the Kerr metric or something else \cite{MBHs_review}. Existing 
astronomical observations can only probe the massive body's asymptotic Keplerian field (there is, however, evidence 
of existence of event horizons in accordance with the ADAF accretion model \cite{adaf}). 
This is strong evidence in favor of massive black holes, but is far from being taken
as decisive proof of the Kerr metric.

LISA will offer a unique opportunity to clarify this matter. Gravitational radiation from EMRI events 
will serve as an optimal probe of black holes spacetimes, verifying their Kerr identity (the most likely 
outcome of this experiment) or identify a more exotic massive object. In order to carry out 
such a `spacetime-mapping' program, the LISA data analysis community will require the construction of waveform 
templates for EMRI around a non-Kerr object. Despite its potential importance, this program still remains in 
its infancy.

The foundations were laid out by Ryan, in a series of papers \cite{ryan3},\cite{ryan4},\cite{ryan5},\cite{ryan_boson}. The central idea 
of his analysis is to express the general stationary-axisymmetric spacetime in terms of multipole mass and current moments, 
denoted as $M_\ell$ ($\ell = 0, 2, 4, ...) $ and $S_\ell$ ($\ell = 1,3,5,... $), respectively, 
following the formalism of Geroch, Hansen, Fodor, Hoenselaers and Perjes \cite{mpoles}. The metric then can be written in the form,
\be
ds^2 = -e^{\gamma + \delta}\,dt^2 + e^{2\alpha}\,( dr^2 + r^2\,d\theta^2) 
+ e^{\gamma -\delta}\,r^2\sin\theta^2( d\phi -\omega dt)^2
\label{metric_multi}
\ee
The metric functions $\gamma,\delta,\omega,\alpha $ depend only on $r,\theta$ and each one of them
can be written as an expansion in multipole moments. For example \cite{ryan_boson},
\bear
\delta &=& \sum_{n=0}^{+\infty}\, -2\frac{M_{2n}}{r^{2n+1}}\,P_{2n}(\cos\theta) + \quad (\mbox{higher order terms})
\\
\nonumber \\
\omega &=& \sum_{n=1}^{+\infty}\, -\frac{2}{2n-1}\,\frac{S_{n-1}}{r^{2n+1}}\,\frac{P_{2n-1}^1(\cos\theta)}
{\sin\theta} + \quad (\mbox{higher order terms}) 
\label{metric_exp}
\eear
where $P_{2n},~P^{1}_{2n-1}$ are Legendre and associated Legendre polynomials, respectively. 
These expansions illustrate the lowest order appearance (with respect to $1/r$) of each multipole moment:
higher moments `weight' higher powers of $1/r$. The prescription for writing the metric components in terms
of multipole moments is given in \cite{ryan3}, \cite{mpoles} and makes use of the classic Ernst potential \cite{ernst}. 

Kerr spacetime is special in the sense that its higher multipole moments are all `locked' to the first 
two moments, the mass $M \equiv M_0$ and spin $J \equiv S_1 = a M $,
\be
M_l + i\,S_l = M(ia)^{\ell}, \quad \ell = 0, 1, 2, ...
\label{nohair}
\ee 
In essence, this statement is the famous `no-hair' theorem of General Relativity \cite{MTW}: 
the two-parameter $(M,a)$ Kerr metric is the only possible description for the exterior field of 
(uncharged) black holes with the corresponding mass and spin.  

Given the metric (\ref{metric_multi}), the task is to study geodesic motion of a test-body and the induced gravitational 
perturbation. However, neither of these issues is straightforward. The main difficulty that arises is that many 
of the special properties of Kerr spacetime are now lost. Firstly, the fact that the new spacetime 
(\ref{metric_multi}) is not necessarily of algebraic type ${\cal D}$ means that it is not {\it a priori} 
possible to carry out a program that would lead to a Teukolsky-like decoupled perturbation equation. Recall that 
decoupling of the various perturbation equations for the Weyl scalars (in the context of the Newman-Penrose 
formalism) is a property crucially related to the type ${\cal D}$ character of a spacetime 
\cite{stewart}. Secondly, even at the simple level of geodesic motion, the existence of a `third' constant,
like $Q$, is not guaranteed anymore. This is a familiar situation in the study of orbital motion in galactic potentials 
\cite{galactic} and it could well lead to undesired effects like chaotic behaviour.

These issues of principle forced Ryan to consider only special cases: (i) the gravitational wave-phase
for a circular equatorial orbit, and (ii) the orbital frequencies for circular orbits with small
eccentricity and small inclination. For the former calculation, Ryan resorted to Thorne's multipolar-expansion 
for the flux \cite{thorne} and obtained,
\be
\dE = \frac{32}{5}\left (\frac{\mu}{M}\right)^2 v^{10}\,\left [ 1 -\frac{1247}{336}\,v^2 + 
\left \{ 4\pi -\frac{11}{4}\,\frac{S_1}{M^2} \right \}\,v^3 
-\left \{ \frac{44711}{9072} -\frac{1}{16}\,\frac{S_1^2}{M^4} + \frac{2M_2}{M^3} \right \}\, v^4 
+ {\cal O}(v^5)  \right ]
\ee
where $ v = \sqrt{M/p} $.
Then the wave-phase can be written as,
\be
\Phi(t) = 2\pi\int dt\, f(t) = 2\pi \int df\,f\, \frac{dE/df}{\dE} \quad
\mbox{with} \quad f =  2\,\Omega_\phi/2\pi = \Omega_\phi/\pi
\label{phase}
\ee
For studying geodesic motion one can again write a Hamilton-Jacoby equation which, however, is only separable
with respect to $t$ and $\phi$ coordinates (a consequence of the spacetime's stationarity and axisymmetry); 
separability with respect to $r$ and $\theta$ is no longer possible (apart from the trivial case $\theta = \pi/2$). 
Then, the small body's momentum is given by:
\be
p^t = -g^{tt}\, E + g^{t\phi}\, L_z, \qquad p^{\phi} = g^{\phi\phi}\, L_z -E\, g^{t\phi}
\ee
and
\be
g_{tt}\,(p^t)^2 + 2g_{t\phi}\,p^t\,p^{\phi} + g_{\phi\phi}\,(p^{\phi})^2 + 
g_{rr}\,(p^r)^2 + g_{\theta\theta}\,(p^{\theta})^2 = -\mu^2
\label{4mom}
\ee
Following Ryan, assuming a small eccentricity - small inclination orbit allows eqn.~(\ref{4mom})
to decouple (with respect to $r$ and $\theta$) at leading order in $e,\iota$. 
In this way, it is possible to work out the fundamental orbital frequencies \cite{ryan3}, 
\bear
\Omega_r &=& \Omega_\phi\, \left [ 1- 3 v^2 +\frac{4S_1}{M^2}\,v^3 - \left (\frac{9}{2} 
-\frac{3\,M_2}{2\,M^3} \right) v^4 + {\cal O}(v^5) \right ]
\label{Om_r}
\\
\nonumber \\
\Omega_\theta &=& \Omega_\phi\, \left [ 1- \frac{2\,S_1}{M^2}\, v^3 - \frac{3M_2}{2\,M^3}\, v^4 - 
\left ( \frac{7\,S_1^2}{M^4} + \frac{3\,M_2}{M^3} \right )\,v^6 + {\cal O}(v^7) \right ] 
\label{Om_z}
\eear
and
\be
\frac{\Omega_\phi}{\mu}\,\frac{dE}{d \Omega_\phi} = -\frac{1}{3}\,v^2 + \frac{1}{2}\,v^4 
- \frac{20\,S_1}{9\,M^2}\,v^5
+ \left ( \frac{27}{8} -\frac{M_2}{M^3}  \right )\,v^6 + {\cal O}(v^7)
\label{dEdOm}
\ee
If either the phase (\ref{phase}),(\ref{dEdOm}) or the frequencies (\ref{Om_r}),(\ref{Om_z}) are measured, then it will be possible 
for LISA to extract the first few moments with sufficient accuracy, as suggested by Ryan's analysis \cite{ryan3},\cite{ryan4}.     
However, much more detailed work is required in order to construct actual templates for waveforms generated 
by bodies in generic orbits of the metric (\ref{metric_multi}).

The multipole moment formalism, despite its general and elegant character, has several serious disadvantages. 
Expanding the metric as in eqn.~(\ref{metric_exp}) is in practise an expansion in $1/r$. For any practical 
implementation this expansion has to be truncated. The induced error is not significant at large values of
$r$ but in the strong field regime several moments need to be included.

Serious difficulties emerge from the non-separability of the Hamilton-Jacobi equation and the lack of a `Teukolsky-like' 
perturbation formalism. The former somewhat complicates the description of geodesic motion, while the latter seriously 
inhibits any attempt to calculate gravitational waveforms, fluxes and back reaction to the orbit. PN results, as the ones given by 
Ryan, are unfortunately not accurate enough for constructing waveform templates for LISA. 

These problems provide sufficient motivation for approaching the problem in a different manner. 
There is some very recent work along this line \cite{bumpy},\cite{quasi} which is based on the idea of a `quasi-Kerr' metric: 
the underlying assumption is that the spacetime is not exactly described by the Kerr metric, but rather by a metric which deviates 
only {\it slightly} from Kerr. 

The actual prescriptions for building a quasi-black hole metric is markedly different in the existing studies.
In Ref.~\cite{bumpy}, Collins \& Hughes construct (as a first stab to the problem) a `bumpy' Schwarzschild 
black hole by artificially adding a certain distribution of mass $\tilde{\mu} \ll M$ around a Schwarzschild
black hole, in the form of (i) point masses at the hole's poles or (ii) an equatorial ring encircling the
black hole. In this way, the perturbed metric is axisymmetric and stationary and has a non-zero
quadrupole moment (as well as higher mass moments), whereas the Schwarzschild spacetime has only a mass
moment. Collins \& Hughes write the resulting metric in the following Weyl form (using cylindrical coordinates
$\{\rho,z,\phi \}$),
\be
ds^2 = -e^{2\psi} dt^2 + e^{2\gamma -2\psi} [ d\rho^2 + dz^2] + e^{-2\psi}\,\rho^2\, d\phi^2
\label{weyl}
\ee  
with $\psi = \psi_0 + \psi_1 $, $\gamma = \gamma_0 + \gamma_1 $. The values $\psi_0,\gamma_0$ correspond
to the exact Schwarzschild metric while $ \psi_1, \gamma_1 $ is the induced perturbation. For example, 
for $\psi_1$ we have,
\bear
\psi_1 &=& -\frac{\tilde{\mu}}{2}\,\left [ \frac{1}{\sqrt{\rho^2 + (z-b)^2}}  +
\frac{1}{\sqrt{\rho^2 + (z+b)^2}}  \right ], \qquad \mbox{point masses at poles}
\label{bump1}
\\
\psi_1 &=& -\frac{\tilde{\mu}}{2\pi}\,\int_{0}^{2\pi}\frac{d\xi}{\sqrt{\rho^2 + z^2 + b^2 -2b\rho\cos\xi}},
\qquad \mbox{equatorial ring} 
\label{bump2}
\eear
The point masses are located at $ z \pm b $ and the ring at $\rho = b$.
Knowledge of $\psi$ immediately provides $\gamma$, since they are related by the vacuum Einstein equations
(see \cite{bumpy} for relevant formulae). Both  $\psi_1, \gamma_1 $ scale as $\sim \tilde{\mu}/M  $, 
and are to be kept only to leading order in $\tilde{\mu}$. This operation is equivalent to truncation of the multipolar
structure of the metric (\ref{weyl}) to the level where only the mass and quadrupole moment are taken into 
account.

\begin{figure}
\centerline{\includegraphics[height=7cm,clip]{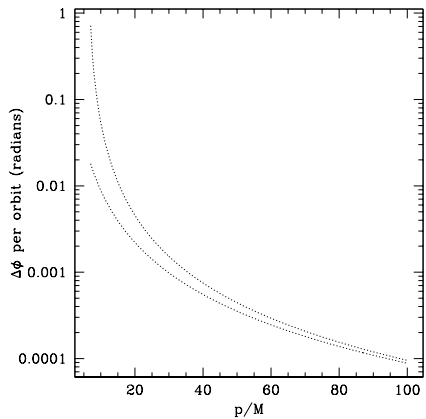}}
\caption{Periastron precession angle $\Delta\phi $ due to a combined quadrupolar distortion, eqns.~(\ref{bump1}),
(\ref{bump2}) of Schwarzschild spacetime. The top curve refer to strong-field results while the bottom curve is the
weak-field Keplerian estimate (\ref{weak_prec}).}  
\label{fig_prec}
\end{figure}

Collins \& Hughes also worked out the equatorial geodesic equations of motion for the metric (\ref{weyl}).
They find,
\be
\left (\frac{dr}{d\tau} \right )^2 = e^{-2\gamma_1}\, \left [ \frac{E}{\mu}^2 -e^{2\psi_1}\,\left ( 1-\frac{2M}{r} \right )\,
\left ( 1 + \frac{e^{2\psi_1}\,L_z^2}{\mu^2 r^2}  \right ) \right ], \quad 
\frac{d\phi}{d\tau} = e^{2\psi_1}\,\frac{L_z}{\mu r^2}, 
\quad \frac{dt}{d\tau} = e^{-2\psi_1}\,\frac{E}{\mu} \left ( 1-\frac{2M}{r} \right )^{-1}
\label{bumpy_equat}
\ee
An orbital parameter which is clearly imprinted on the waveform is the periastron advance $\Delta\phi$. 
Due to the non-zero quadrupole moment, geodesic motion in the metric (\ref{weyl}) will exhibit an additional
precessional motion in addition to the familiar Schwarzschild precession. It is easily calculated from 
eqns.~(\ref{bumpy_equat}) and is plotted in Figure~\ref{fig_prec} (for $\tilde{\mu} = 0.01M$ and $ b = 2.5M $).

In the weak-field limit ($p \gg M$) the periastron precession takes the simple form,
\be
\Delta\phi \approx  
\cases{ -3\pi\tilde{\mu}\,b^2/M p^2  \,& for point masses
\cr + 3\pi\tilde{\mu}\,b^2 / 2Mp^2 
\, & for ring \cr}
\label{weak_prec}
\ee   
According to (\ref{weak_prec}), a prolate (oblate) quadrupole deformation will induce a positive (negative) precession
in agreement with Newtonian expectations. As Fig.~\ref{fig_prec} suggests the weak-field result seriously underestimates
$\Delta\phi $ for $ p \lesssim 20 M  $. Either way, the perturbation $\psi_1,\gamma_1 $ can impart a significant 
correction to the small-body's orbit.

In their study \cite{quasi}, Glampedakis \& Babak build a quasi-Kerr spacetime in a somewhat more natural way,
by combining the exact Kerr metric with the celebrated Hartle-Thorne (HT) exterior metric \cite{HT}. 
This latter metric describes the spacetime outside {\it any} arbitrary, but slowly rotating, stationary and 
axisymmetric body. It is accurate up to ${\cal O}(J^2)  $ where $J$ is the body's spin.
The HT metric depends on three parameters: the mass $M$, spin $J$ and quadrupole moment ${\cal Q}$. 
For the special case,
\be
{\cal Q}_{\rm kerr} = -\frac{J^2}{M} 
\ee  
it reduces to the ${\cal O}(a^2)  $ Kerr metric. We can then assume an approximation
where the quadrupole moment has a slightly `wrong' value,
\be
{\cal Q} = {\cal Q}_{\rm kerr} - \epsilon\,M^3
\ee
with $\epsilon \ll 1$. Then the Hartle-Thorne metric can be written,
\be
g_{ab}^{\rm (HT)} = g_{ab}^{\rm (\alpha^2~Kerr)} + \epsilon\,h_{ab}
\label{HTmetric}
\ee
The quadrupole  term $\epsilon\,h_{ab}$ represents the {\it leading order} deviation from the Kerr metric. 
The ansatz adopted in \cite{quasi} is to replace the ${\cal O}(a^2)$ Kerr metric with the {\em full} Kerr
metric, i.e $ g_{ab}^{\rm (\alpha^2~Kerr)} \to g_{ab}^{\rm (Kerr)} $. However, prior to this operation, one more 
manipulation is required in (\ref{HTmetric}). The Kerr limit ($\epsilon \to 0$) of the original HT metric
(\cite{HT}) does not yield the Kerr metric in its Boyer-Lindquist coordinate form \cite{bardeen}.
This not surprising, as the original set of coordinates used by Hartle \& Thorne are not Boyer-Lindquist.
On the other hand, there is a strong reason for which we would like this limit to be satisfied, 
and is related to the separability of the Hamilton-Jacobi and wave equation. The Kerr limit of
the HT metric leads to {\it non-separable} dynamical equations (such as the Hamilton-Jacobi
and the scalar wave equation). These equations are separable provided that we are dealing with a 
type ${\cal D}$ spacetime (which is true for the HT metric only when $\epsilon \to 0$ ) {\it and} 
for a certain class of coordinate frames \cite{carter},\cite{stewart}. As the Boyer-Lindquist 
frame is included in this privileged class, it makes sense to express the HT metric 
in these coordinates. In fact, the relevant transformation from Boyer-Lindquist to Hartle-Thorne coordinates was 
given in the original paper \cite{HT} and it is trivially inverted, at ${\cal O}(J^2) $ accuracy \cite{quasi}.

Putting all the pieces together, the ansatz for building a quasi-Kerr metric in Boyer-Lindquist
coordinates is:
\be
g_{ab} = g_{ab}^{\rm (Kerr)} + \epsilon\,h_{ab}
\label{qmetric}
\ee        
where,
\bear
h^{tt} &=& (1-2M/r)^{-1} [\, (1-3\cos^2\theta)\,{\cal F}_1(r)] , \qquad 
h^{rr} = (1-2M/r) [\, (1-3\cos^2\theta)\,{\cal F}_1(r)]  
\\
\nonumber \\
h^{\theta\theta} &=&  -\frac{1}{r^2}\, [\, (1-3\cos^2\theta)\, {\cal F}_{2}(r)], \qquad
h^{\phi\phi} =  -\frac{1}{r^2\sin^2\theta}\,[\, (1-3\cos^2\theta)\, {\cal F}_{2}(r)] 
\\
\nonumber \\
h^{t\phi} &=& 0
\eear
The radial functions ${\cal F}_{1,2}(r) $ are written explicitly in Appendix~\ref{app:HT_functions}.

Given the metric (\ref{qmetric}), it is straightforward to obtain equations of motion for {\it equatorial} orbits,
\bear
p^t &=& p^t_{\rm (Kerr)} -\epsilon\, \frac{r\,E\, {\cal F}_1}{r-2M}
\\
\nonumber \\
p^r &=& p^r_{\rm (Kerr)} + \epsilon\,\frac{R_2}{2\,r^2 \sqrt{R_\circ}},
\qquad \mbox{with} \qquad R_2 = -r\,(r-2M)\, [ \mu^2 r^2\, {\cal F}_1 + L_z^2\,({\cal F}_1 -{\cal F}_2) ]
\nonumber \\
&& \hspace{4cm} \mbox{and} \qquad R_\circ = r^4\,E^2 -r(r-2M)\,[ \mu^2 r^2 + L_z^2]
\\
\nonumber \\
p^\theta &=& 0, \qquad p^\phi = p^\phi_{\rm  (Kerr)} -\epsilon\, \frac{L_z {\cal F}_2}{r^2} 
\label{quasi_equat}
\eear
where $p^a_{\rm (Kerr)} $ is the pure Kerr four-momentum (eqns.~(\ref{geod1})).
We have to point out that if we were merely interested in equatorial motion only, then shifting
from Hartle-Thorne to Boyer-Lindquist coordinates is an unnecessary step, as the Hamilton
Jacobi equation is trivially separable when $\theta = \pi/2 $. However, the coordinate
transformation is required in the general case in order to ensure that the Hamilton-Jacobi
equation is separable in the $\epsilon \to 0 $ limit. Then, one can still obtain approximate equations   
of motion for generic orbits by means of canonical perturbation techniques (see \cite{goldstein} for treatment
of the perturbed Kepler problem); these equations will appear in Ref.~\cite{quasi}.

\begin{figure}
\centerline{\includegraphics[height=7cm,clip]{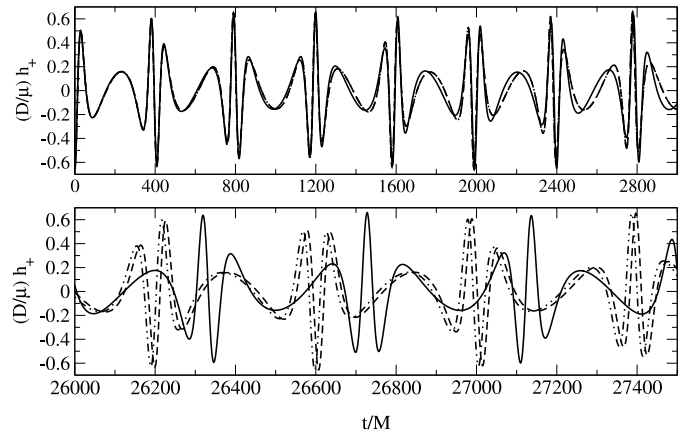}}
\caption{Quasi-Kerr (solid curves) versus Kerr waveforms for the orbit $ p = 10M,~e=0.5,~\iota = 0^\circ $. The
black hole spin is $ a=0.5M $ and the Kerr deviation parameter $\epsilon = 0.1$. In calculating the Kerr
waveforms we have either ignored or included the radiative orbital evolution (setting $\mu/M =10^{-5}$ ), 
dashed and dashed-dotted curves respectively. To demonstrate how promptly the quasi-Kerr waveform dephases we show 
two different time windows: an early adiabatic stage and a much later ($t \sim T_{\rm RR} $) stage where backreaction 
effects have become noticeable. Note that the quasi-Kerr waveform already deviates from its Kerr counterpart after
only few orbits, where the two Kerr waveforms are essentially indistinguishable.}
\label{fig_quasi_wav}
\end{figure}

\begin{figure}
\centerline{\includegraphics[height=6cm,clip]{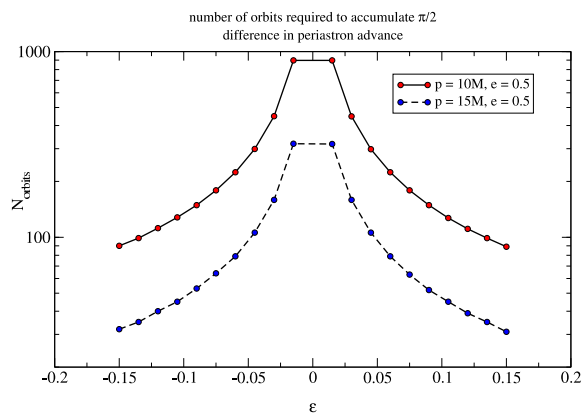}}
\caption{Number of orbits required to accumulate $\Delta\phi = \pi/2 $ periastron advance for equatorial orbits in the 
quasi-Kerr metric (\ref{qmetric}) as a function of $\epsilon$. The spin is $a=0.5M$ and the orbital paramaters are:
$p=10 M, e=0.5 $ (solid curve), $ p=15M,e=0.5  $ (dashed curve).}
\label{fig_prec2}
\end{figure}

The main goal of the quasi-Kerr framework is to facilitate the computation of strong-field 
fluxes and waveforms. So far, the only relevant results have been obtained by applying the hybrid 
approximation for the metric (\ref{qmetric}), in combination with (\ref{quasi_equat}) \cite{quasi}
(indeed, it would be straightforward to repeat the same calculation for the metrics (\ref{metric_multi}),
(\ref{weyl}), always assuming equatorial motion). Some preliminary results are shown in 
Figs.~\ref{fig_quasi_wav} \& \ref{fig_prec2} where we compare quasi-Kerr versus exact Kerr waveforms and
quantify the difference of the periastron shift. These results suggest that (say) a $\sim 10 \% $ 
deviation in the quadrupole moment would lead to significant changes in the orbit and waveform. 
Possible non-Kerr character of black hole spacetimes is a factor that should be taken into account in building
data analysis tools for LISA\footnote{In this context, an important issue that needs to be addressed is whether a Kerr
waveform with a different set of parameters could match the quasi-Kerr waveform.}.

The computation of rigorous fluxes, waveforms and backreaction to the orbit is the major challenge that lies ahead 
for the quasi-Kerr program. The metric (\ref{qmetric}) has the clear advantage of being arbitrarily close to type ${\cal D}$.
The hope is that this property {\it may} eventually allow the formulation of a `quasi-Teukolsky' equation for the 
perturbed Weyl scalars, in which case, one will be able to obtain results in a fashion similar to the one described 
in Section ~\ref{EMRI}. In the unfortunate case where this scenario will not flourish, we may be forced to deal with 
direct metric perturbations. These are described by ten coupled equations (plus four gauge constraints) which will 
require solution in the time-domain (as a `2+1' problem). Clearly, this is a much more complicated task compared to the solution 
(either in the time or frequency domain) of the single Teukolsky equation. It is interesting to note that recent approaches 
to the calculation of the gravitational self-force, is via time-evolution of black hole metric perturbations 
(see other contributions in the present Volume). Existing calculations have been focused on the Schwarzschild case; generalisation 
to Kerr is at early stages. It is likely that the experience gained by these calculations could be extremely useful for the 
quasi-Kerr problem.


\section{Recoiling black holes}
\label{recoils}

Gravitational radiation generated by an inspiralling and subsequently coalescing binary system is not only a carrier 
of energy and angular momentum but also of linear momentum. As a result, the binary's center of mass experiences a 
`recoil' kick \cite{bekenstein}. This recoil effect is expected to be of potential astrophysical importance in situations 
where massive black holes are involved. At the top of the scale we have the inspiral and final merger of supermassive 
black holes, as the aftermath of galactic mergers. At a smaller scale, there may be intermediate mass black holes 
merging at the centers of globular clusters. 

In particular, it is important to obtain an estimate for the final net recoil velocity of a 
merged black hole binary, and compare it to the escape velocity of the host stellar system.  
In this problem one has to consider a binary composed of comparable mass black holes.
The general formula for the averaged rate at which linear momentum is removed from the system 
(which is the opposite of the amount carried by gravitational waves) is \cite{MTW},
\be
\dot{P}^k = -\frac{r^2}{16\pi} \int\, d\Omega\, n^k\, \langle  \dot{h}_{+} + \dot{h}_{\times}  \rangle
\ee
where $ n^k $ is the radial unit vector. Unlike the energy and angular momentum fluxes which first appear at the 
level of the source's mass quadrupole moment, linear momentum is a higher order effect and requires combination of 
quadrupole mass and current moments together with mass octupole moment. This leading order formula was first 
derived by Bekenstein \cite{bekenstein},
\be
\dot{P}^k = -\frac{2}{63}\, \langle \frac{d^3{\cal I}^{ij}}{dt^3} \, \frac{d^4{\cal M}^{ijk}}{dt^4} \rangle 
-\frac{16}{45}\, \langle \epsilon^{kij}\,  \frac{d^3{\cal I}^{il}}{dt^3}\, \frac{d^3{\cal S}^{jl}}{dt^3}  
\rangle  
\label{mom_lead}
\ee    
where ${\cal I}^{ij},{\cal M}^{ijk},{\cal S}^{ij} $ the symmetric and trace-free mass parts of
$ I^{ij}, S^{ijk}, M^{ijk} $, see Ref.~\cite{thorne} for details. 

A first recoil result was obtained by Fitchett \cite{fitchett}, using (\ref{mom_lead}) under the assumption 
of Keplerian motion for the binary. For circular orbits Fitchett's result is,
\be
v_F \approx 1480\,\frac{f(q)}{f_{\rm max}}\, 
\left ( \frac{2G(M_1 +M_2)/c^2}{r_{\rm term}} \right )^4 \mbox{km/s}
\label{recoil_F}
\ee       
where $q = M_1/M_2 \leq 1 $ is the binary's mass ratio, $r_{\rm term}$ is a `cut-off' radius where
gravitational wave emission terminates, and $f(q) = q^2(1-q^2)/(1+q)^5 $. According to (\ref{recoil_F}),
for the expected  $r_{\rm term} \sim \mbox{few} \times M $, the recoil velocity can be 
$v_F \sim 10 -100 $ km/s. Typical escape velocities span a range $\sim 10-10^3 $ km/s,
from globular clusters to galaxies. In view of these numbers, Fitchett's result clearly establishes 
the astrophysical importance of the recoil effect and provides motivation for more detailed work.   
  
More detailed work means taking into account strong-field effects in the motion of the 
coalescing black holes since, after all, the momentum flux grows with decreasing separation for the
two black holes. However, this is a venture into the thorny area of fully nonlinear General Relativity
hence more suitable for hard-core numerical relativity (see \cite{num_rel} for estimates on binary black hole 
recoil). Fortunately, some valuable input can be provided by much simpler perturbative 
methods. Although black hole perturbation theory (such as the Teukolsky formalism) assumes an extreme mass ratio
$q \ll 1$, results can be extrapolated to (say) $q \sim 0.5 $ with a modest error (with the same reasoning 
we can also neglect the spin of the lighter black hole). In fact, the recent study by Volonteri {\em et al} \cite{volonteri}
suggests that low-redshift  ( $z \lesssim 1 $) black hole mergers predominantly occur with $ q \lesssim 0.1 $.

Within the Teukolsky framework the first relevant calculations were performed by Fitchett \& Detweiler
\cite{detweiler_reco} for circular Schwarzschild orbits. Linear momentum fluxes were also computed for bound and plunging 
parabolic orbits in Kerr and Schwarzschild spacetimes \cite{parabolic}.   

In terms of the amplitudes $Z_{\ell m}$ appearing in the expansion for $\psi_4$,
\be
\psi_4 = \frac{1}{r} \sum_{\ell m} Z_{\ell m}\, S_{\ell m} e^{im\varphi -i\omega_m (t- r_\ast)}
\ee
the linear momentum flux is \cite{detweiler_reco},
\bear
\dot{P}_x &=& \frac{1}{4\Omega_{\phi}^{2}}\, \sum_{\ell,\ell^\prime,m} \left [ \int_{0}^{\pi} d\theta\sin^2\theta\,
\frac{Z_{\ell m}}{m}\,S_{\ell m} \left \{ \frac{\bar{Z}_{\ell^\prime m-1}}{m-1}\,S_{\ell^\prime m-1}\,
e^{i\Omega_\phi (t-r_{\ast})}  + \frac{\bar{Z}_{\ell^\prime m+1}}{m+1}\,S_{\ell^\prime m+1}\,
e^{-i\Omega_\phi (t-r_{\ast})} \right \}  \right ]
\\
\dot{P}_y &=& \frac{i}{4\Omega_{\phi}^{2}}\, \sum_{\ell,\ell^\prime,m} \left [ \int_{0}^{\pi} d\theta\sin^2\theta\,
\frac{Z_{\ell m}}{m}\,S_{\ell m} \left \{ \frac{\bar{Z}_{\ell^\prime m-1}}{m-1}\,S_{\ell^\prime m-1}\,
e^{i\Omega_\phi (t-r_{\ast})}  - \frac{\bar{Z}_{\ell^\prime m+1}}{m+1}\,S_{\ell^\prime m+1}\,
e^{-i\Omega_\phi (t-r_{\ast})} \right \}  \right ]
\\
\dot{P}_z &=& 0
\label{dPxyz}
\eear
Note that these expressions are valid only for circular equatorial orbits. The total recoil velocity simply follows by 
integrating (\ref{dPxyz}) from some initial time $t_0$ (when the binary separation is large and $v_{x,y}(t_0)=0 $),
\be
v_{x,y} = \frac{1}{M_1+M_2}\, \int_{t_0}^t\, dt\, \dot{P}_{x,y}
\label{v_reco}
\ee
with $\dot{P}_{x,y} $ calculated for a sequence of quasi-circular orbits.
The most recent study on recoilling black holes is the one by Favata {\it et al} \cite{favata} 
who focused on Kerr circular equatorial orbits. In a realistic scenario there is no {\em a priori}
reason to expect the orbit to be `equatorial' (for a moderate mass ratio system, the term equatorial
should be interpreted as alignment of the total orbital angular momentum and the spin of the heavier
black hole). Still, one can gain useful information by examining the two extreme opposite cases of equatorial prograde 
and retrograde motion.

\begin{figure}
\centerline{\includegraphics[height=8cm,clip]{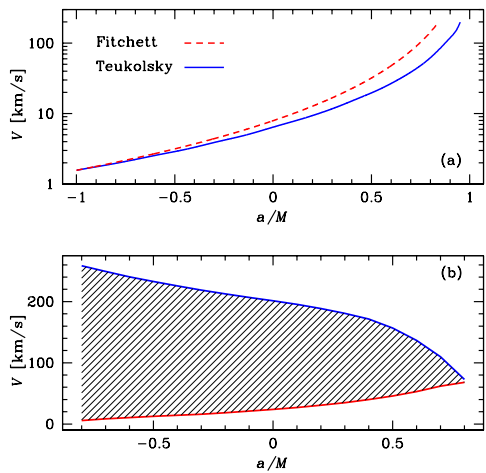}}
\caption{Recoil velocity as a function of black hole spin $a/M$ for $q \approx 0.1$.  
(a) Recoil velocity up to the separatrix.  The solid curve is the Teukolsky-based result while the dashed curve shows
the Newtonian recoil prediction eqn.~(\ref{recoil_F}), which is substantially higher for $ a \gtrsim 0.5 M $ prograde inspirals.
(b) Upper and lower limits for the total recoil. The shaded region represents the uncertainty in the final kick velocity.
The upper limit curve is generated by assuming a continuation of the power-law behaviour $\dot{P}_{\rm tot} \sim p^{-\alpha} $ 
that fits the inspiral down to the separatrix. On the other hand, the lower-limit curve is generated with the help 
of the hybrid scheme.}
\label{fig_recoil}
\end{figure}

Estimating the recoil from eqns.~(\ref{dPxyz}),(\ref{v_reco}) is an accurate procedure (apart from the error 
introduced by the finite mass ratio) as long as the orbit is stable, $ p > p_s $. Results on the recoil velocity, 
as computed by Favata {\em et al}, are shown in Figure~\ref{fig_recoil}. According to this data, Fitchett's leading 
order formula is in fairly good agreement with the Teukolsky-based results. Some noticable deviation appears for 
prograde motion around $a \gtrsim 0.5 M $ Kerr holes. The orbit ventures deep inside the hole's 
strong field rendering the leading-order result (\ref{recoil_F}) unreliable. 

The major uncertainty in the recoil calculation comes from the plunging part of the inspiral. Indeed, this part  
could well provide the dominant contribution to the final recoil velocity, especially when the motion
is retrograde (in which case the separatrix is beyond $ 6 M$). In contrast, prograde orbits with $a \approx M $
receive little contribution from the plunge since $ p_s \approx r_{+} $.  

\begin{figure}
\centerline{\includegraphics[height=6cm,clip]{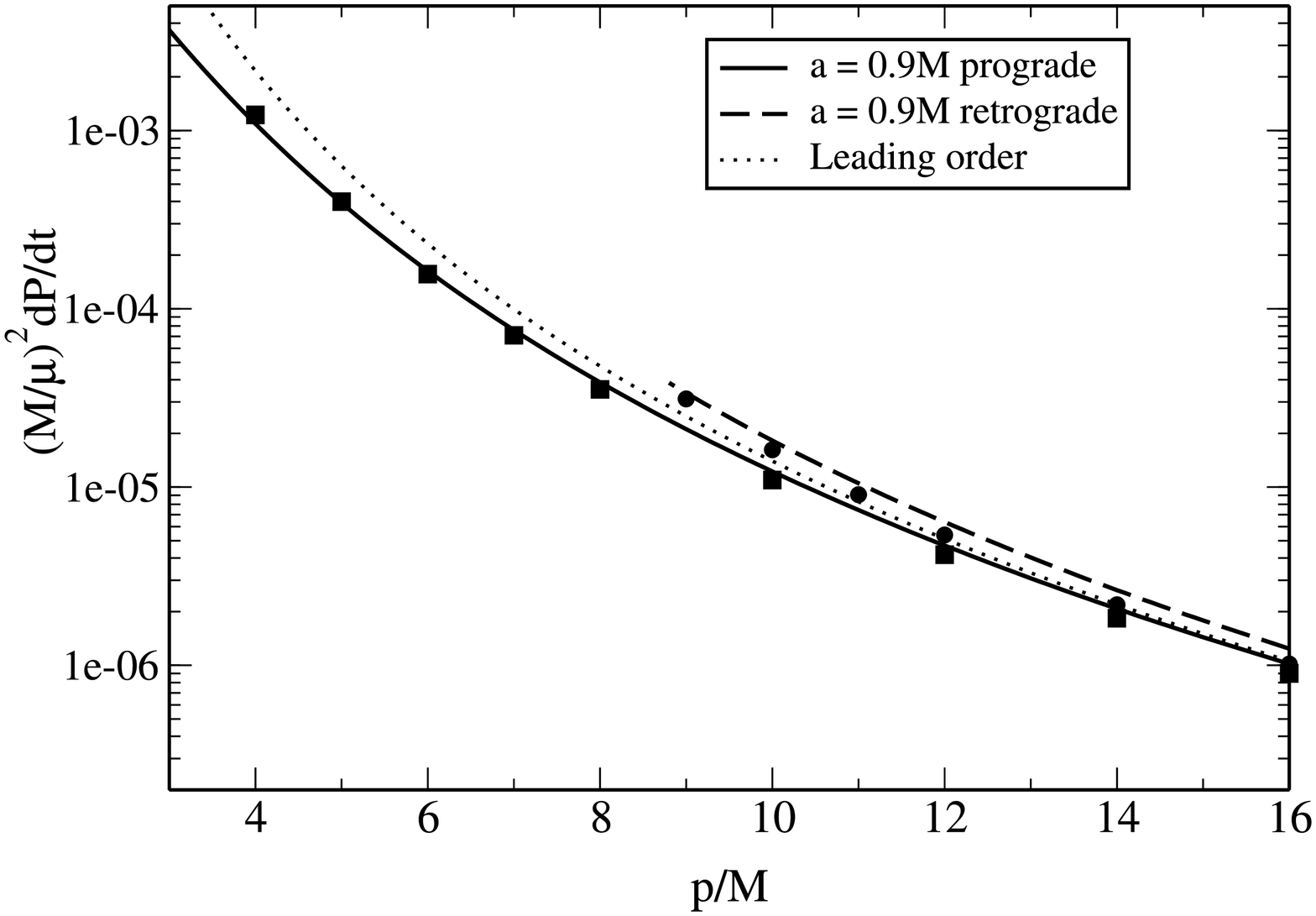}}
\caption{Linear momentum flux at infinity for Kerr circular-equatorial orbits. We compare results obtained by
the hybrid approximation (solid and dashed curves for prograde and retrograde motion respectively) with
accurate Teukolsky-based results (box and circle symbols) and the leading order prediction (dotted curve)
as given by eqn.~(\ref{recoil_F}). The black hole spin is set to $a=0.9M$.}
\label{fig_recoil2}
\end{figure}

Somewhat surprisingly, there is no available Teukolsky-based analysis for this kind of orbits. In Ref.~\cite{favata} two different 
approximations are employed in order to obtain some rough estimate for the contribution of the plunge portion of the 
inspiral. Firstly, a power-law behaviour $\dot{P} \sim p^{-\alpha}  $ is assumed, as a continuation of the similar behaviour 
during the adiabatic quasi-circular inspiral (see for example Fig.~\ref{fig_recoil2}, taken from Ref.~\cite{favata}).
The resulting recoil velocity is designated as `upper limit' in Fig.~\ref{fig_recoil}. Secondly, the hybrid approximation 
(Section~\ref{hybrid}) can be employed by using the exact plunging geodesic trajectory in the leading-order formula 
(\ref{mom_lead}). In this way, the `lower-limit' recoil curve of Fig.~\ref{fig_recoil} is generated. The hybrid result should not 
be such a bad approximation given the overall accuracy of this calculation. This is what Fig.~\ref{fig_recoil2} suggests, where 
we compare $\dot{P}_{\rm hybrid} $ against $\dot{P}_{\rm Teuk} $ for $ p > p_s $. In any case, what we learn with confidence 
from Fig.~\ref{fig_recoil} is that for a merging binary with the heavier member rapidly spinning and the lighter one in prograde motion, 
the resulting recoil velocity is $ \sim 50-100 $ km/s {\em smaller} than the Newtonian expectation.

\section{Concluding remarks}
\label{conclusions}

We have presented a variety of material on EMRIs, from rigorous Teukolsky-based results (both in the frequency and time domains) 
for special orbits (equatorial-eccentric and circular-inclined) to approximate results for generic orbits (hybrid approximation). 
Guided by these studies we have a fairly detailed knowledge of the gravitational dynamics of the system and of the associated waveform. 
However, there is still plenty of distance until the final goal of developing realistic waveforms for LISA is reached. Presumably, 
this program will be fulfilled in the near future by the self-force framework. But even within the limits of the standard 
flux-balance/Teukolsky formalism there is much work to do. A first priority is the computation of rigorous waveforms
and fluxes from a test-body in a generic orbit, without worrying about the orbital evolution and $\dQ$. Fortunately, this analysis
is expected to appear soon \cite{drasco}. Another important task is the construction of Teukolsky-based inspirals,
as the ones discussed in Section~\ref{evolv}. We already have the tools for producing equatorial-eccentric inspirals,
and it will even be possible to `fake' generic inspirals by using rigorous $\dE,\dL  $ fluxes and some approximation
for $\dQ$ (as the ones described in Section~\ref{Q}).   

Another layer of complexity can be added regarding the modelling of the small body itself. The assumed `point-particle' 
is in reality another Kerr black hole (or some other compact body) with intrinsic properties beyond the mass $\mu$. 
Modelling the small body, one could take one more step and include the spin. Formal equations of motion for a spinning test-body have 
been known for a long time \cite{spinning}, but only recently there were actual calculations for the particular case of Kerr spacetime 
\cite{hartl}. The coupling of the body's spin with the background curvature results in a non-geodesic motion with extra spin degrees of 
freedom and loss of one integral of motion (the Carter constant). Additional work is required in order to understand the motion of a 
spinning body under the influence of radiation reaction (plus the backreaction to its spin), within both the adiabatic flux-balance and 
the self-force framework.

If we are prepared to accommodate the idea that LISA might actually discover non-Kerr massive objects, then we must be prepared to
study EMRIs around non-Kerr metrics. The material of Section~\ref{mapping} is just a first modest step towards that direction.
The actual computation of `realistic' waveforms/orbital evolution could develop to a daunting task, likely to involve direct time-evolution
of metric perturbations. Also it is not clear yet if by including only the leading-order quadrupole moment deviation is sufficient
for actually telling the difference between a true quasi-Kerr object and a Kerr black hole with slightly different set of $M,a$. If this 
turns out to be the case, and additional multipole moments are required, we may then have to revitalise (and perhaps modify) Ryan's 
multipolar formalism.


\acknowledgements

I am grateful to Judith Adams and Carlos Lousto for their kind invitation to write this
review article and for repeatedly stretching the submission deadline on my behalf. I am also grateful to 
my collaborators Stas Babak, Jonathan Gair, Scott Hughes and Daniel Kennefick for their vital contributions 
to this article. I acknowledge support from PPARC Grant PPA/G/S/2002/00038.    


\appendix

\section{}
\label{app:almostsphere}

The occurrence of a `third' orbital constant $Q$ in axisymmetric gravitational fields is not an exclusive 
feature of general relativity.  For example, it is familiar from Newtonian celestial mechanics applied 
to orbital motion in galactic gravitational potentials (see, for example \cite{galactic}, where the third
constant is denoted $I$).  The departure of $Q$ from $L_x^2 + L_y^2$ can then be attributed to the `asphericity' 
of the potential.  If such a potential does not deviate very much from sphericity, $L^2$ (the square of the 
total angular momentum) turns out to be almost constant, so that $Q$ should be, after all, nearly $L^2 - L_z^2$.

\begin{figure}
\centerline{\includegraphics[height=9cm,clip]{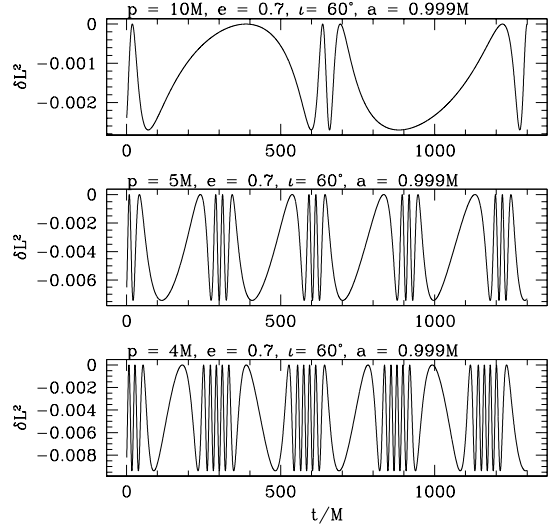}}
\caption{Examining our notion of ``total angular momentum'' for
strong-field Kerr black hole orbits.  Each panel compares the angular
momentum squared $L^2\equiv p_\theta^2 + (\sin\theta)^{-2} p_\phi^2$
to $Q + L_z^2$: the quantity plotted is $\delta L^2\equiv L^2/(Q +
L_z^2) - 1$.  The top panel shows these quantities over an orbit with
$p = 10M$, the center panel the quantities over an orbit with $p =
5M$, and the bottom over an orbit $p = 4 M$.  In all cases, the orbits
have eccentricity $e = 0.7$, inclination $\iota = 60^\circ$, and are
about a hole with spin $a = 0.999M$.  Even deep in the strong field,
$L^2$ differs very little from $Q + L_z^2$ --- the greatest deviation
in this sample is about $1\%$.  Since $Q + L_z^2$ is a constant by
definition, this also shows that $L^2$ is approximately conserved over
the orbit.}
\label{fig5}
\end{figure}

It is straightforward to check whether this behaviour of $L^2$ occurs in Kerr spacetime. The definition we use 
for $L^2$ is identical to that used in Schwarzschild spacetime,
\begin{equation}
L^2 = p_{\theta}^2 + (\sin\theta)^{-2} p_{\phi}^2\;,
\label{Ltot}
\end{equation}
For the Carter constant $Q$ we have,
\begin{equation}
Q = p_{\theta}^2 + (\sin\theta)^{-2} p_{\phi}^2 -p_{\phi}^2 +
a^2\cos^2\theta(\mu^2 - E^2)\;.
\label{Qeq}
\end{equation}
Combining these two expressions gives
\begin{equation}
Q = L^2 - L_{z}^2 + a^2 \cos^2\theta(\mu^2 - E^2)\;.
\label{QnL}
\end{equation}
In other words, $Q$ can be interpreted as the projection of the total angular momentum on the equatorial plane, 
apart from the `aspherical' term $a^2\cos^2\theta(\mu^2 - E^2)$.  This interpretation makes sense when
the aspherical term is small, that is, when $a\ll M$ (slow rotation) and/or $E\approx 1$ (weak-field orbits).  
In practice, we find that this term is often significantly smaller than the preceding terms even for motion in 
strong-field regions of rapidly rotating holes.  We illustrate this in Fig.\ {\ref{fig5}}, showing how the
quantity $\delta L^2 \equiv L^2/(Q + L_z^2) - 1$ varies with time for a variety of generic orbits around a 
rapidly spinning hole.

Examining Fig.\ {\ref{fig5}}, we see that $L^2$ deviates very little from $Q + L_z^2$ even when the small body 
is deep in the black hole's strong field; in this sample, the difference is no more than about
$1\%$.  This shows that interpreting $Q$ as a squared projection of angular momentum into the equatorial plane 
is sensible.  Because $Q +L_z^2$ is a constant quantity, this figure also demonstrates that
$L^2$ is nearly constant.  This is exactly what we expect for motion in an axisymmetric potential that is almost spherical.  
These pieces of evidence suggest that the Kerr spacetime is not as aspherical as we might have expected, at least for 
the purposes of this argument, lending credence to our suggestion that the `$\iota =\mbox{constant}$' assumption should 
be reliable.

\section{}
\label{app:integrals}

Here we list the integrals appearing in the expressions for the Kerr fundamental
frequencies, eqns.~(\ref{freqs_i}),(\ref{freq_t}). We have 
\be
\Lambda = K(k)\,Y(r_p,r_a) + a^2\,z_{+}^2\,[K(k) -E(k)]\,X(r_p,r_a) 
\ee
with
\bear
K(k) &=& \int_{0}^{\pi/2}\frac{d\psi}{\sqrt{1-k\,\sin^2\psi}}, \qquad
E(k) = \int_{0}^{\pi/2}d\psi\, \sqrt{1-k\,\sin^2\psi}
\\
\nonumber \\
\Pi(z_{-}^2,k) &=& \int_{0}^{\pi/2}\,\frac{d\psi}{(1-z_{-}^2\sin^2\psi)\,
\sqrt{1-k\,\sin^2\psi}} 
\eear
and
\bear
X(r_p,r_a) &=& \int_{r_p}^{r_a}\, \frac{dr}{\sqrt{R}}, \qquad 
Y(r_p,r_a) = \int_{r_p}^{r_a}\, \frac{r^2 dr}{\sqrt{R}}
\\
\nonumber \\
Z(r_p,r_a) &=& \int_{r_p}^{r_a}\, dr\frac{r [L_z\,r -2M\,(L_z-aE)]}{\Delta\,\sqrt{R}},
\qquad 
W(r_p,r_a) = \int_{r_p}^{r_a}\, dr \frac{r[r^3 +a^2\,(E/\mu)\,r -2a\,(L_z -aE)]}{\Delta\,\sqrt{R}}
\eear


\section{}
\label{app:KennOri}

In this Appendix we provide a brief discussion on the Kennefick-Ori formula for the Carter constant 
flux $\dQ$, see Ref.~\cite{ori}. The starting point is to express all three constants $C=\{E,L_z,Q\}  $ in a 
functional form $C=C(x^b,p_a) $. In particular, for $E,L_z$ we simply have,
\be
E = -p_t, \quad L_z = p_\phi
\ee   
while for $Q$ we have a choice between the two (equivalent) forms,
\bear
Q &=& p_\theta^2 + \cos^2\theta\, \left [ a^2 (\mu^2 -p_t^2) + \frac{p_\phi^2}{\sin^2\theta}  \right ]
\equiv p_\theta^2 + {\cal G}(\theta,E,L_z)
\label{Q_form2}
\\
Q &=& \Delta^{-1}\left [ p_t (r^2 + a^2) + a p_\phi  \right ]^2 -(p_\phi +a p_t)^2 -r^2 -\Delta\, p_r^2 \equiv {\cal H}(r,E,L_z) -\Delta\, p_r^2
\label{Q_form1}
\eear

Defining the self-force as $F_a = D\,p_a/D\tau $, Kennefick \& Ori derive,
\be
\frac{dC}{d\tau} = \frac{\partial C}{\partial p_a}\, F_a
\ee
for the instantaneous rates. Clearly, $F_t = -dE/d\tau $ and $F_\phi = dL_z/d\tau $. For $\dQ$ we obtain the
twin expressions,
\bear
\dQ &=& {\cal H}_{,E}\,\frac{dE}{d\tau} + {\cal H}_{,L_z}\,\frac{dL_z}{d\tau}  -2\,\Delta\,p_r F_r
\label{ko_r}
\\
\dQ &=& {\cal G}_{,E}\,\frac{dE}{d\tau} + {\cal G}_{,L_z}\,\frac{dL_z}{d\tau} + 2\,p_\theta\,F_\theta
\eear



\section{}
\label{app:formulae}

This appendix contains explicit expressions (in terms of $E$, $L_z$,
$Q$ and their derivatives) for the various functions appearing in the
formulae (\ref{dpdedi}) for the rates $\ddp$, $\de$, $\di$.
First,
\be
H = Q_{,p}\, E_{,e}\, L_{z,\iota} -Q_{,p}\, E_{,\iota}\, L_{z,e}
- Q_{,e}\,E_{,p}\,L_{z,\iota}
+ Q_{,e}\, E_{,\iota}\, L_{z,p} + Q_{,\iota}\, E_{,p}\, L_{z,e} 
- Q_{,\iota}\, E_{,e}\, L_{z,p}
\label{Hfunc}
\ee
and
\bear  
b_p &=& Q_{,\iota} L_{z,e} -Q_{,e} L_{z,\iota}, \qquad b_e = L_{z,\iota} Q_{,p} -Q_{,\iota} L_{z,p}
\\
\nonumber \\
c_p &=& E_{,\iota}Q_{,e} -E_{,e} Q_{,\iota}, ~~\qquad c_e =  Q_{,\iota} E_{,p} -E_{,\iota} Q_{,p}
\\
\nonumber \\
d_p &=& E_{,e} L_{z,\iota} -E_{,\iota} L_{z,e}, \qquad d_e = E_{,\iota} L_{z,p} -E_{,p} L_{z,\iota}
\eear


\section{}
\label{app:coeffs}

The functions $f_{1}(e) - f_{8}(e)$ and $g_{1}(e) - g_{9}(e)$ appearing in the flux-formulae 
(\ref{dE_R}),(\ref{dL_R}),(\ref{dE_new}),(\ref{dL_new}) are:
\bear
f_{1}(e) &=& 1+ \frac{73}{24}e^2 + \frac{37}{96} e^4 ,
\qquad
f_{2}(e) =  \frac{73}{12} + \frac{823}{24}e^2 + 
\frac{949}{32} e^4 + \frac{491}{192} e^6 ,
\qquad f_3(e) = \frac{1247}{336} + \frac{9181}{672} e^2 ,\quad\quad 
\nonumber \\
\nonumber \\
f_4(e)  &=& 4 + \frac{1375}{48} e^2, \quad\quad\quad\quad~
f_5(e) = \frac{44711}{9072} + \frac{172157}{2592} e^2 ,
\quad\quad\quad\quad\quad\quad\quad 
f_6(e) = \frac{33}{16} + \frac{359}{32} e^2  ,\
\nonumber \\
\nonumber \\
f_7(e) &=& \frac{8191}{672} + \frac{44531}{336} e^2 ,  
\quad\quad
f_8(e) = \frac{3749}{336} - \frac{5143}{168} e^2   
\eear
and
\bear
g_1(e) &=& 1 + \frac{7}{8} e^2 ,   
\quad\quad\quad\quad\quad
g_{2}(e) = \frac{61}{24} + \frac{63}{8} e^2 + \frac{95}{64} e^4, \qquad 
g_{3}(e) = \frac{61}{8} + \frac{91}{4} e^2 + \frac{461}{64} e^4
\nonumber \\
\nonumber \\
g_{4}(e) &=& \frac{1247}{336} + \frac{425}{336} e^2  ,\quad\quad~
g_{5}(e) = 4 + \frac{97}{8} e^2 , \qquad\qquad\qquad
g_{6}(e) = \frac{44711}{9072} + \frac{302893}{6048} e^2 
\nonumber \\
\nonumber \\
g_{7}(e) &=& \frac{33}{16} + \frac{95}{16} e^2 , \qquad\qquad
g_{8}(e) = \frac{8191}{672} + \frac{48361}{1344} e^2 ,\quad\quad\quad
g_{9}(e) = \frac{417}{56} - \frac{37241}{672} e^2 
\eear


\section{}
\label{app:HT_functions}
The radial functions ${\cal F}_{1,2}(r) $ appearing in the quasi-Kerr metric (\ref{qmetric}) are:
\bear
{\cal F}_{1}(r) &=& -\frac{5(r-M)}{8Mr(r-2M)}\,(2M^2 + 6Mr -3r^2) -\frac{15 r(r-2M)}{16 M^2}\,\ln\left ( \frac{r}{r-2M} \right )
\\
\nonumber \\
{\cal F}_{2}(r) &=& \frac{5}{8Mr}\,(2M^2 -3Mr -3r^2) + \frac{15}{16M^2} (r^2 -2M^2)\ln \left ( \frac{r}{r-2M}\right )  
\eear


\end{document}